\documentclass[prb,aps,floatfix,floats,superscriptaddress,notitlepage]{revtex4-1}
\usepackage{epsfig}
\usepackage{enumerate}
\usepackage{amsmath}
\usepackage{amssymb}
\usepackage{color}
\usepackage{theorem}
\usepackage{braket}
\usepackage{mathtools}
\usepackage{tensor}
\usepackage{tikz}
\usetikzlibrary{decorations.markings}
\DeclarePairedDelimiter{\abs}{\lvert}{\rvert}
\renewcommand{\Im}{\mathfrak{Im}}
\renewcommand{\Re}{\mathfrak{Re}}
\newcommand{\ZZ}{\mathbb{Z}}
\newcommand{\D}{\mathcal{D}}
\newcommand{\C}{\mathcal{C}}
\newcommand{\oO}{\mathcal{O}}
\newcommand{\N}{\mathcal{N}}
\newcommand{\K}{\mathcal{K}}
\newcommand{\J}{\mathcal{J}}

\newcommand{\ri}{\textrm{i}}
\newcommand{\be}{\begin{equation}}
\newcommand{\ee}{\end{equation}}
\newcommand{\bea}{\begin{eqnarray}}
\newcommand{\eea}{\end{eqnarray}}

\def\nn{\nonumber\\}
\def\fr#1{(\ref{#1})}

\def\tht{{\tilde\theta}}

\def\ri{{\rm i}}

\def\C{\mathcal C}
\def\cC{\mathcal C}

\def\S{\mathcal S}

\def\D{\mathcal D}

\def\cG{\mathcal G}
\def\R{\mathcal R}

\def\K{\mathcal K}
\def\J{\mathcal J}
\def\T{\mathcal T}

\def\CC{\mathbb C}
\def\ZZ{\mathbb Z}

\def\RR{\mathbb R}
\def\M{\mathcal M}
\def\B{\mathcal B}
\def\A{\mathcal A}
\def\repS{{|\Phi_{\tilde{s}}\rangle_{\rm NS}}}

\theoremstyle{plain}

\allowdisplaybreaks[4]
\begin{document}
\title{Quantum quench in the sine-Gordon model}
\author{Bruno Bertini}
\affiliation{
The Rudolf Peierls Centre for Theoretical Physics, Oxford
University, Oxford OX1 3NP, UK}
\author{Dirk Schuricht}
\affiliation{
Institute for Theoretical Physics,
Center for Extreme Matter and Emergent Phenomena, 
Utrecht University, Leuvenlaan 4,
3584 CE Utrecht, The Netherlands}
\author{Fabian H. L. Essler}
\affiliation{
The Rudolf Peierls Centre for Theoretical Physics, Oxford
University, Oxford OX1 3NP, UK}
\begin{abstract}
We consider the time evolution in the repulsive sine-Gordon quantum
field theory after the system is prepared in a particular class of
initial states. We focus on the time dependence of the one-point
function of the semi-local operator
$\exp\big(\ri\beta\Phi(x)/2\big)$. By using two different methods
based on form-factor expansions, we show that this expectation value
decays to zero exponentially, and we determine the decay rate by
analytical means. Our methods generalize to other correlation
functions and integrable models.  
\end{abstract}
\maketitle

\section{Introduction}
The last decade has witnessed great advances in the understanding of
the non-equilibrium dynamics of isolated many-particle quantum
systems. For example, it has become possible to generate
one-dimensional quantum gases and study their time evolution after a
quantum quench, i.e. a sudden change in the system
parameters\cite{kww-06,expquenches}. The quantum Newton's
cradle\cite{kww-06} in particular emphasized the importance of 
conservation laws and motivated the study of quantum quenches in
integrable models. One key result of this work has been that the
stationary properties of observables at late times after the quench
can be described in terms of a generalized Gibbs
ensemble (GGE)\cite{rigol}, which has by now been fully established for
several systems \cite{gge1,CEF}, while strong evidence in favour of the GGE
has been found for many others
\cite{gge2,fioretto2010,ggeTE,iucci}. Generalisations of the GGE to
initial states in which the probability distributions of conserved
charges are not infinitely sharp in the thermodynamic limit are
discussed in Ref.~[\onlinecite{andrei}].

A recent focus of attention has been the problem of determining
expectation values of local observables in the stationary state after
a quantum quench in \emph{interacting} integrable models, and the
construction of appropriate GGEs for these cases\cite{gge3,FCEC,Nardis14a}.  

The problem of describing the time evolution of local observables in
integrable models after a quantum quench is harder and less well
developed. So far analytic results have been obtained only for
certain quenches to conformal field theories \cite{CC07} and models
that can be mapped to free fermions or bosons
\cite{ggeTE,fioretto2010,CEF,LL,CE_PRL13,SE:2012,iucci}. A logical
next step is then to develop techniques for describing the time
evolution after quenches to interacting integrable models. Here we
address this problem for a paradigmatic integrable quantum field
theory, the quantum sine-Gordon model 
\be
{\cal H}_{\rm SG}=\frac{v}{16\pi}\int dx\left[\big(\partial_x\Phi\big)^2+
\frac{1}{v^2}\big(\partial_t\Phi\big)^2\right]-\lambda\int dx\,\cos(\beta\Phi).
\label{H_SG}
\ee
We are interested in the situation where the system is prepared in a
particular kind of initial state $|\Psi_0\rangle$, which is not an
eigenstate of ${\cal H}_{\rm SG}$. Here we focus on the case where
$|\Psi_0\rangle$ corresponds to an integrable boundary condition in
the sine-Gordon model \cite{GZ}. This choice of initial state is
motivated by the situation encountered for quenches to conformal field
theories\cite{CC07}, which can be mapped to a boundary problem with
appropriately defined boundary state. Our protocol of imposing a
particular initial state differs from the usual quench setup, where
one prepares the system in the ground state of a given Hamiltonian and
then changes one or more system parameters instantaneously. We discuss
the realizability of integrable boundary states in standard quench
protocols in Section~\ref{sec:initialstate}. 
Given our initial state, we consider unitary time evolution
\be
|\Psi_t\rangle=e^{-\ri{\cal H}_{\rm SG}t}|\Psi_0\rangle,
\ee
and our goal is to determine expectation values of the kind
\be
F^\alpha_{\Psi_0}(t)=\frac{\langle\Psi_t|e^{\ri\alpha\Phi(x)}|\Psi_t\rangle}
{\langle\Psi_t|\Psi_t\rangle}.
\ee
We do so by two complementary approaches. The first is based on the
``representative state'' formalism proposed in
Ref.~[\onlinecite{CE_PRL13}]. The second is a generalisation of the
low-density expansion introduced for studying quenches in the
transverse field Ising chain\cite{CEF} and the Ising field
theory\cite{SE:2012}. 

Quantum quenches to the massive regime $\beta<1$ of the sine-Gordon model
were considered by Iucci and Cazalilla\cite{iucci}. They focussed on
two special cases, namely quenches from the massless regime to the
free-fermion point $\beta=1/\sqrt{2}$ and in the semiclassical regime
$\beta\to 0$. These cases are characterized by no or weak interactions
respectively, and as a result are amenable to treatment by simpler
methods than the ones developed in our work. The solvability of the
model at the free-fermion point was also employed by Foster et
al.~\cite{Foster} to investigate the amplification of initial density
inhomogeneities. 

Gritsev et al.\cite{gritsev08b} considered the time evolution of the
\emph{local} operator $e^{\ri\beta\Phi}$ after preparing the system in
an integrable boundary state similar to ours. However, they focussed
on the attractive regime $\beta<1/\sqrt{2}$ where, in addition to
solitons and antisolitons, breather bound states exist. By
employing a form-factor expansion and evaluating the first few terms,
they calculated the power spectrum of the vertex operator
$e^{\ri\beta\Phi}$ and showed that it possesses sharp 
resonances corresponding to the creation of breather states and
discussed implications for experiments on split one-dimensional Bose
condensates. 
For the approach Ref.~[\onlinecite{gritsev08b}] to apply, higher order
terms in the Lehmann expansion must be negligible. This is generally
not the case\cite{CEF,SE:2012}: for semi-local operators of the kind
considered here these terms are in fact divergent at late
times. Our main purpose is to develop methods for determining the
late-time behaviour in this much more complicated case.

For $\beta>1$ the cosine perturbation is irrelevant at low
energies and quenches in this very different regime have been
considered in Refs.~[\onlinecite{irrelevant}]. 

The remainder of this article is organized as follows: In the next
section we provide some context for our work by summarising
applications of the sine-Gordon model to magnetic solids and systems
of trapped, ultra-cold atoms. This is followed by a brief review of
some well-known facts related to the integrability of the sine-Gordon
model in section~\ref{sec:integrFT}. We then introduce the notion of
``integrable'' quantum quenches, and discuss various issues relating
to particular kind of initial state we consider. In
Section~\ref{sec:RSA} we determine the time evolution of the one-point
function $F_{\Psi_0}^{\beta/2}(t)$ using a method based on the
recently proposed   ``representative state"
approach\cite{CE_PRL13}. In Section~\ref{sec:LCE} we apply a
complementary method based on a linked-cluster expansion, which was
originally developed for quenches in the transverse-field Ising
model\cite{CEF,SE:2012}. 
Both approaches show that the one-point function $F_{\Psi_0}^{\beta/2}(t)$
decays exponentially in time, see Eq.~\eqref{eq:finalresult}, with a calculable
decay time $\tau$. We conclude with a discussion of our results and 
an outlook for further investigations in Section~\ref{sec:conclusions}. 
Technical details encountered in the course of our analysis
are presented in a number of appendices. 

\section{Realisations of the sine-Gordon model}
The sine-Gordon model is known to emerge as the low-energy description
in a variety of contexts, both in solids and systems of trapped,
ultra-cold atoms.

\subsection{Solids}
The sine-Gordon model is obtained as the low-energy limit of a variety
of quantum spin chain models, see
e.g. Ref.~[\onlinecite{review}]. Perhaps the best experimental
realisation in this context is in field-induced gap Heisenberg magnets
like CuBenzoate\cite{FIG_E,FIG_T}. The underlying lattice model in
these systems is of the form 
\be
H=J\sum_{j=1}^L\Bigl[S^x_jS^x_{j+1}+S^y_jS^y_{j+1}+\Delta
S^z_jS^z_{j+1}\Bigr]+h_u\sum_{j=1}^LS^z_j+h_s\sum_{j=1}^L (-1)^j S^x_j.
\ee
In the thermodynamic limit the low-energy sector is described by a quantum sine-Gordon model \fr{H_SG}, but typically $\beta<\frac{1}{2}$ so that the sine-Gordon model is in the attractive regime. The repulsive regime can be realized as the low-energy limit of the spin-1/2 XYZ chain 
\be
H=J\sum_{j=1}^L\Bigl[(1+\gamma)S^x_jS^x_{j+1}+(1-\gamma)S^y_jS^y_{j+1}+\Delta S^z_jS^z_{j+1}\Bigr],
\label{HXYZ}
\ee
where we take 
\be
J>0\ ,\quad 0<\Delta<1\ ,\quad \gamma\ll 1.
\ee
In this regime we may regard \fr{HXYZ} as a perturbation of the spin-1/2 XXZ chain by the relevant operator
$\sum_j[S^x_jS^x_{j+1}-S^y_jS^y_{j+1}]$. In the low-energy limit this gives a sine-Gordon model \fr{H_SG} with
\be
-\Delta=\cos(\pi\beta^2)\ ,\ \frac{1}{2}<\beta^2<1,\quad
v=\frac{J}{2}\frac{\sin(\beta^2)}{1-\beta^2}\ ,\quad
\lambda\propto \gamma.
\ee
The transverse spin operators are bosonized as follows
\bea
S^+_j&=&S^x_j+iS^y_j\sim {\cal C}(-1)^j:e^{\ri\frac{\beta}{2}\Phi(x)}:+{\cal A}
\left[:e^{\ri\frac{\beta}{2}\Phi(x)+\frac{\ri}{2\beta}\Theta}:+
:e^{\ri\frac{\beta}{2}\Phi(x)-\frac{\ri}{2\beta}\Theta}:\right]+\dots\nn
&\equiv& J^+(x)+(-1)^j n^+(x)\ ,\quad x=ja_0.
\eea
In this realisation of the sine-Gordon model the operator $e^{\ri\frac{\beta}{2}\Phi(x)}$ thus corresponds to the staggered magnetisation. The bosonized form of the longitudinal spin operator is
\be
S^z_j\sim\frac{a_0}{4\pi\beta}\partial_x\Theta(x)-{\cal B}(-1)^j
\sin\left(\frac{\Theta(x)}{2\beta}\right)+\ldots
\ee
We note that the constants ${\cal A}$, ${\cal B}$ and ${\cal C}$ are known\cite{lukyanov}. In the XYZ realisation the Bose fields are compactified 
\be
\Theta(x)=\Theta(x)+4\pi\beta\ ,\qquad
\Phi(x)=\Phi(x)+\frac{4\pi}{\beta}.
\label{compactification}
\ee
This implies that the only local vertex operators are of the form
\be
{\cal O}_{n,m}=e^{\ri\frac{n}{2\beta}\Theta(x)+\ri\frac{\beta m}{2}\Phi(x)}.
\label{vertop}
\ee
Defining the topological charge as
\be
\label{topcharge}
Q=\frac{\beta}{2\pi}\int_{-\infty}^\infty dx\ \partial_x\Phi(x),
\ee
we see that ${\cal O}_{n,m}$ carry topological charge $2n$.

\subsection{Systems of ultra-cold trapped atoms}
Other realisations of the sine-Gordon model can be found in systems of
interacting one dimensional bosons\cite{Giamarchi}. 
For example, one may consider a single species of bosons in a periodic
potential
\be
H=\frac{1}{2m_0}\int dx |\partial_x\psi|^2+\frac{1}{2}\int dx\ dx'\
V(x-x')\rho(x)\rho(x')+\int dx\ [V_L(x)-\mu]\rho(x),
\label{Hboson}
\ee
were $\psi(x)$ is a complex scalar field, $V(x)$ describes
density-density interactions between bosons, $\mu$ a chemical
potential, and $V_L(x)$ a periodic lattice potential. Models like
\fr{Hboson} can be realized in systems of trapped, ultra-cold atoms,
where $V_L(x)$ accounts for the optical lattice\cite{BBZ}. 
 ``Bosonizing the boson''\cite{bosonisation}  
\bea
\psi^\dagger(x)&\sim&\sqrt{\rho_0}\sum_{p\in\mathbb{Z}}
e^{2\ri p[\pi\rho_0x-\sqrt{\frac{K}{8}}\Phi(x)]}e^{-\frac{\ri}{\sqrt{8K}}\Theta(x)}\ ,\nn
\rho(x)&\sim&\rho_0-\sqrt{\frac{K}{8\pi^2}}\partial_x\Phi(x)
+\rho_0\sum_{p\neq 0}e^{2\ri p[\pi\rho_0x-\sqrt{\frac{K}{8}}\Phi(x)]},
\eea
where $\rho_0$ is the average density, leads to a low-energy
description\cite{Giamarchi} in terms of a sine-Gordon model
\fr{H_SG}. Here the $\cos(\beta\Phi)$ originates  from the periodic
potential $V_L(x)$, and the Luttinger parameter $K$ characterizes the
interactions. In order to access the repulsive regime of the
sine-Gordon model ($\beta^2>1/2$), the density-density interaction
$V(x)$ should be sufficiently long-ranged. Of particular interest for
our work is the case where the periodic potential is such that there
is on average one boson for every two minima of $V_L(x)$, i.e.  
\be
V_L(x)\sim V_L\cos(4\pi\rho_0x),
\ee
which leads to a sine-Gordon model with $\beta^2=2K$. As long as $K<1/2$, the leading oscillating term in the density is
\be
\rho_{\rm
  osc}(x)\sim\cos\bigg(2\pi\rho_0x-\frac{\beta}{2}\Phi(x)\bigg)
+{\rm higher\ harmonics}.
\ee
Hence in this realisation of the sine-Gordon model our results pertain to the leading oscillatory behaviour of the boson density. We note that the Bose fields again are compactified according to \fr{compactification}.

A second realisation of the sine-Gordon model with ultra-cold atoms is
in terms of coupled one-dimensional condensates\cite{gritsev07}. The
microscopic Hamiltonian is taken to be
\bea
H&=&\sum_{j=1,2}\int dx\left[\frac{1}{2m_0}\bigl(\partial_x\Psi^\dagger_j(x)\bigr)
\bigl(\partial_x\Psi_j(x)\bigr)+g\rho_j^2(x)\right]-t_\perp\int dx
\left(\Psi^\dagger_1(x)\Psi_2(x)+{\rm h.c.}\right),
\eea
where $\rho_j(x)=\Psi^\dagger_j(x)\Psi_j(x)$. Bosonising\cite{bosonisation} and then transforming to total and relative phases 
\be
\Phi_\pm(x)=\sqrt{\frac{1}{8K}}
\left[\frac{\Phi_1(x)\pm\Phi_2(x)}{\sqrt{2}}\right],\
\Theta_\pm(x)=\sqrt{\frac{K}{8\pi^2}}
\left[\frac{\Theta_1(x)\pm\Theta_2(x)}{\sqrt{2}}\right],
\ee
one finds a low-energy effective Hamiltonian of the form
\bea
H_+&=&\frac{v}{16\pi}\int dx\left[\big(\partial_x\Phi_+\big)^2+
\big(\partial_x\Theta_+\big)^2\right]\ ,\nn
H_-&=&\frac{v}{16\pi}\int dx\left[\big(\partial_x\Phi_-\big)^2+
\big(\partial_x\Theta_-\big)^2\right]-\lambda\int dx\ 
\cos\left(\frac{\Phi_-}{\sqrt{4K}}\right).
\eea
Here $\lambda\propto t_\perp$ and the parameters $v$ and $K$ are
functions of $g$, $m_0$ and the density $\rho_0$ of the
condensates. 

\section{Sine-Gordon model as an integrable quantum field theory}\label{sec:integrFT} 
In the regime of interest to us here, $1/\sqrt{2}\le\beta<1$,
the elementary excitations in the sine-Gordon model
are massive solitons and antisolitons possessing a relativistic
dispersion relation. In the attractive regime $\beta<1/\sqrt{2}$
there solitons and antisolitons can form bound states known as
breathers. At the Luther-Emery point $\beta=1/\sqrt{2}$ the model is
equivalent to a free massive Dirac theory~\cite{LutherEmery74}, thus
the quench dynamics can be investigated by simpler
methods\cite{iucci} than the ones developed here.

A basis of eigenstates of the sine-Gordon model in the repulsive
regime is conveniently constructed by employing the
Faddeev-Zamolodchikov creation and annihilation operators\cite{FZ}
$Z^\dagger_a(\theta)$, $Z_a(\theta)$, where the index $a=\pm$
corresponds to the creation and annihilation of solitons and
antisolitons respectively. They are taken to fulfil the 
algebra
\bea
Z_{a_1}(\theta_1)Z_{a_2}(\theta_2)&=&
S_{a_1a_2}^{b_1b_2}(\theta_1-\theta_2)Z_{b_2}(\theta_2)Z_{b_1}(\theta_1),\nn
Z^\dagger_{a_1}(\theta_1)Z^\dagger_{a_2}(\theta_2)&=&
S_{a_1a_2}^{b_1b_2}(\theta_1-\theta_2)Z^\dagger_{b_2}(\theta_2)Z^\dagger_{b_1}(\theta_1),\nn
Z_{a_1}(\theta_1)Z^\dagger_{a_2}(\theta_2)&=&2\pi\delta(\theta_1-\theta_2)
\delta_{a_1,a_2}+
S_{a_2b_1}^{b_2a_1}(\theta_2-\theta_1)Z^\dagger_{b_2}(\theta_2)Z_{b_1}(\theta_1),
\eea
where $S_{ab}^{cd}(\theta)$ is the two-particle scattering matrix of the sine-Gordon model. For $\beta^2>\frac{1}{2}$ it is given by
\bea
S^{++}_{++}(\theta)&=&S^{--}_{--}(\theta)=S_0(\theta)=
-\exp\left[\ri\int_0^\infty\frac{dt}{t}
\sin\left(\frac{t\theta}{\pi\xi}\right)\frac{\sinh\big(\frac{\xi-1}{2\xi}t\big)}
{\sinh\big(\frac{t}{2}\big)\cosh\big(\frac{t}{2\xi}\big)}\right],\nn
S^{+-}_{+-}(\theta)&=&S^{-+}_{-+}(\theta)
= S_T(\theta)S_0(\theta),\quad S_T(\theta)=-\frac{\sinh\big(\frac{\theta}{\xi}\big)}
{\sinh\big(\frac{\theta-\ri\pi}{\xi}\big)},\nn
S^{+-}_{-+}(\theta)&=&S^{-+}_{+-}(\theta)= S_R(\theta)S_0(\theta),\quad
S_R(\theta)=-\frac{\ri\sin\big(\frac{\pi}{\xi}\big)}
{\sinh\big(\frac{\theta-\ri\pi}{\xi}\big)},
\label{Smatrix}
\eea
where we have defined
\be
\xi=\frac{\beta^2}{1-\beta^2}.
\ee
The S-matrix fulfils the Yang-Baxter equations, crossing and unitarity relations
\bea
S^{b_2b_3}_{a_2a_3}(\theta_2-\theta_3)S_{a_1b_3}^{b_1c_3}(\theta_1-\theta_3)S_{b_1b_2}^{c_1c_2}(\theta_1-\theta_2)&=&S_{a_1a_2}^{b_1b_2}(\theta_1-\theta_2)S_{b_1a_3}^{c_1b_3}(\theta_1-\theta_3)S_{b_2b_3}^{c_2c_3}(\theta_2-\theta_3), 
\label{AA:4}\\
S_{a_1a_2}^{c_1c_2}(\theta)S_{c_1c_2}^{b_1b_2}(-\theta)&=&
\delta_{a_1}^{b_1}\delta_{a_2}^{b_2},
\label{eq:Sunitarity}\\
S_{ab}^{cd}(\ri\pi-\theta)&=&S_{\bar{c}b}^{\bar{a}d}(\theta)
=S_{a\bar{d}}^{c\bar{b}}(\theta),\quad \bar{a}=-a.
\label{eq:Scrossing}
\eea
Further relations that will prove useful in the following are
\begin{equation}
\bigl(S_{ab}^{cd}(\theta)\bigr)^*=S_{ab}^{cd}(-\theta)\quad\text{for}\quad\theta\in\mathbb{R},\quad
S_{ab}^{cd}(\theta)=S_{ba}^{dc}(\theta)=S^{ab}_{cd}(\theta)=
S_{\bar{a}\bar{b}}^{\bar{c}\bar{d}}(\theta).
\label{eq:Srelations}
\end{equation}
The zero temperature ground state $|0\rangle$ is defined by
\be
Z_a(\theta)|0\rangle=0,
\ee
and a basis of eigenstates is obtained by acting with creation operators
\be
|\theta_1,\ldots,\theta_N\rangle_{a_1\ldots a_N}=Z^\dagger_{a_1}(\theta_1)\ldots
Z^\dagger_{a_N}(\theta_N)|0\rangle.
\label{scatteringstates}
\ee
Energy and momentum of the states \fr{scatteringstates} are given by
\be
E=\Delta\sum_{j=1}^N\cosh\theta_j,\quad
P=\frac{\Delta}{v}\sum_{j=1}^N\sinh\theta_j.
\label{energymomentum}
\ee
The topological charge operator $Q$ defined in \fr{topcharge} acts on the basis \fr{scatteringstates} as 
\be
Q \ket{\theta_1,\ldots,\theta_N}_{a_1\ldots a_N}=\left\{\sum_{i=1}^{N}a_i\right\}|\theta_1,\ldots,\theta_N\rangle_{a_1\ldots a_N}.
\label{tcscatteringstates}
\ee
Importantly, matrix elements of local operators 
\be
{f^{\cal{O}}}_{a_1\dots a_N}^{b_1\dots b_M}
(\theta_1',\dots,\theta_M'|\theta_1,\ldots,\theta_N)
={}_{b_1\dots b_M}\langle\theta_1',\dots,\theta_M'|{\cal
  O}(0)|\theta_1,\ldots,\theta_N\rangle_{a_1\dots a_N}
\ee
can be calculated by means of the form-factor bootstrap approach \cite{Smirnov92book,FFbootstrap,Lukyanov97,Delfino04,LZ}. The basic axioms underlying this approach are summarized in Appendix \ref{sec:FFaxioms}.

\section{``Integrable'' quantum quenches}\label{sec:initialstate}
As shown by Calabrese and Cardy\cite{CC07}, the calculation of
expectation values of local operators after a quench can be mapped to
a corresponding problem in boundary quantum field upon analytic
continuation to imaginary time. In the case where the post-quench
Hamiltonian is integrable, there is then a particular class of
boundary conditions, and concomitantly initial states, namely those
compatible with integrability\cite{GZ}. Given the relative simplicity
of such states, they constitute a natural starting point for quenches
to interacting integrable models\cite{fioretto2010}. A characteristic
feature of these states is that they can be cast in the form\cite{GZ} 
\be
|\Psi_0\rangle=\exp\left(\int_0^\infty \frac{d\theta}{2\pi} 
K^{ab}(\theta)Z^\dagger_a(-\theta)Z^\dagger_b(\theta)\right)|0\rangle.
\label{psi0}
\ee
Here the matrix $K^{ab}(\theta)$ is obtained from a solution of the reflection equations and fulfils 
\bea
K^{a_1c_1}(\theta_1)K^{c_2c_3}(\theta_2)
S^{a_2c_4}_{c_2c_1}(\theta_1+\theta_2)
S^{b_2b_1}_{c_3c_4}(\theta_1-\theta_2)&=&
K^{c_1b_1}(\theta_1)K^{c_2c_3}(\theta_2)
S^{b_2c_4}_{c_3c_1}(\theta_1+\theta_2)
S^{a_2a_1}_{c_2c_4}(\theta_1-\theta_2),\label{BYBE}\\
K^{ab}(\theta)&=&S^{ab}_{cd}(2\theta)K^{dc}(-\theta).
\label{BUE}
\eea
For our purposes it will be useful to exploit the fact that if
$K^{ab}(\theta)$ is a solution to \fr{BYBE}, other solutions can be
obtained by multiplication with an even function $g(\theta)$ 
\be
\widetilde{K}^{ab}(\theta)= K^{ab}(\theta) g(\theta).
\label{Ktilde}
\ee
Introducing a function $g(\theta)$ is crucial in the quench context in
order to obtain well-defined results (see the discussion below). 
A particular choice of $g(\theta)$ advocated in
Ref.~[\onlinecite{fioretto2010}] in light of the corresponding
situation for conformal field theories\cite{CC07} is 
\be
g(\theta)=e^{-2\tau_0\Delta\cosh\theta},
\label{extrapolationtime}
\ee
which defines an ``extrapolation time'' $\tau_0$.
An immediate question is whether boundary states like \fr{psi0} make
for physically meaningful initial states after quenches to a
sine-Gordon model. A particular choice of $K^{ab}(\theta)$ corresponds
to fixed boundary conditions 
\be
\Phi(t=0,x)=0.
\ee
These can be realized by quenching the parameter $\lambda$ in
\fr{H_SG} from infinity to a finite value at time $t=0$, i.e. 
\be
\lambda\Big|_{t<0}=\infty\longrightarrow \lambda\Big|_{t>0}=\lambda_+.
\label{eq:lambdainftyinitial}
\ee
The problem with such a quench is that it corresponds to an initial
state after the quench with infinite energy density, which is clearly
undesirable. This problem is reflected in the fact that for
integrable boundary conditions one typically has 
\be
\lim_{\theta\to\infty}\sum_{a,b}|K^{ab}(\theta)|^2={\rm const},
\ee
which essentially corresponds to having a finite density of
excitations even for infinite-energy particles. The simplest way to
suppress the presence of such excitations in the initial state is to
introduce a function $g(\theta)$, and consider an initial state of the
form \fr{psi0} with $K$ replaced by $\widetilde{K}$. The particular
choice \fr{extrapolationtime} is too restrictive for quenches to
integrable massive QFTs\cite{spyros2}, and an important question is
what kind of initial conditions can be described by an appropriate
choice of $\widetilde{K}$\cite{spyros2,spyros}. One 
might hope that an approximate realisation of such an initial state
could be obtained in a quantum quench  
\be
\lambda\big|_{t<0}=\lambda_-\longrightarrow \lambda\big|_{t>0}=\lambda_+,
\ee
where $\lambda_\pm$ are both large but finite, and their difference is
small. This is clearly an important issue worthy of further
investigation, but is beyond the scope of our present work. 

In the following we will simply set aside the question of how our
initial state can be prepared and impose it to be of the form
\fr{psi0}, where $K^{ab}(\theta)$ is a solution of equations \eqref{BYBE} and \eqref{BUE}. We will furthermore use the freedom
\fr{Ktilde} to be able to consider $K^{ab}(\theta)$, and hence the
quench, to be small in the sense of Refs.~[\onlinecite{CEF}]. 
In Sections \ref{sec:LCK4} and \ref{sec:LCalpha} we will make the
additional simplifying assumption that\cite{note1}
\begin{equation}
K^{aa}(\theta)=0,
\label{eq:DirichletK}
\end{equation}
which is for example the case for the initial condition $\Phi(t=0,x)=0$.

\section{``Representative State'' Approach}\label{sec:RSA}
In Ref.~[\onlinecite{CE_PRL13}] a novel approach to calculating
expectation values of local operators after quantum quenches to
integrable models was proposed. It was subsequently applied to
determine the infinite-time behaviour of certain local observables in
an interaction quench from free to interacting bosons in the
Lieb-Liniger model\cite{Nardis14a} and very recently the XXZ
chain\cite{QAXXZ}, and is referred to as
``quench-action'' approach in these works. Using thermodynamic
arguments it was argued in Ref.~[\onlinecite{CE_PRL13}] that  
\bea
\lim_{L\to\infty}
\frac{{}_L\langle\Psi_t|{\cal
    O}(x)|\Psi_t\rangle_L}{{}_L\langle\Psi_t|\Psi_t\rangle_L}&=&
\lim_{L\to\infty}
\frac{{}_L\langle\Psi_0|{\cal
    O}(t,x)|\Psi_0\rangle_L}{{}_L\langle\Psi_0|\Psi_0\rangle_L}\nn
&=&\lim_{L\to\infty}\frac{1}{2}\left[
\frac{{}_L\langle\Psi_0|{\cal
    O}(t,x)|\Phi\rangle_L}{{}_L\langle\Psi_0|\Phi\rangle_L}
+\frac{{}_L\langle\Phi|{\cal
    O}(t,x)|\Psi_0\rangle_L}{{}_L\langle\Phi|\Psi_0\rangle_L}
\right],
\label{repstate}
\eea
where $|\Psi_0\rangle_L$ is the initial state in a large, finite
volume $L$, and $|\Phi\rangle_L$ is a \emph{representative state}
characterized by the requirements that 
\begin{enumerate}
\item{} $|\Phi\rangle_L$ is an eigenstate with eigenvalue $E_\Phi$ of
  the Hamiltonian in a large, finite volume.  
\item{} The expectation values of the macroscopically many (local)
integrals of motion $I^{(n)}$ inherent to quantum integrable models
\cite{korepinbook} are the same in the initial and the representative
state
\bea
\label{intofmotion}
\lim_{L\to\infty}\frac{1}{L}\frac{{}_L\langle\Phi|I^{(n)}|\Phi\rangle_L}{{}_L\langle\Phi|\Phi\rangle_L}=
\lim_{L\to\infty}\frac{1}{L}\frac{{}_L\langle\Psi_0|I^{(n)}|\Psi_0\rangle_L}{{}_L\langle\Psi_0|\Psi_0\rangle_L},\quad n=1,2,\ldots\,.
\eea
\end{enumerate}
In the representative state approach there are several main steps in determining the time evolution of local operators after quenches to integrable models:
\begin{itemize}
\item{}Construct a basis $\{|\theta_1,\ldots,\theta_N\rangle_L^{s}|s=1,\ldots, 2^N\}$ of $N$-particle energy eigenstates in the finite volume.
\item{}Construct finite-volume matrix elements of (semi-) local
operators ${\cal O}$ with respect to these eigenstates.
\item{}Given an initial state $|\Psi_0\rangle$, determine a
  corresponding representative state $|\Phi\rangle_L$. 
\item{} For a given operator ${\cal O}(x)$, work out the time evolution using a Lehmann representation of \fr{repstate} in terms of energy eigenstates, i.e. work out matrix elements of the form
\bea
{}_L\langle\Psi_0|{\cal O}(t,x)|\Phi\rangle_L&=&\sum_{N\geq 0}
\sum_{\{|\theta_1,\ldots,\theta_N\rangle_L^s\}}\tensor*[^{}_{L}]{\braket{\Psi_0|\theta_1,\ldots,\theta_N}}{^s_L}
\tensor*[^{s}_{L}]{\braket{\theta_N,\ldots,\theta_1|{\cal O}(t,x)|\Phi}}{^{}_L}\nn
&=&\sum_{N\geq 0}\sum_{\{|\theta_1,\ldots,\theta_N\rangle_L^s\}}\tensor*[^{}_{L}]{\braket{\Psi_0|\theta_1,\ldots,\theta_N}}{^s_L}
\tensor*[^{s}_{L}]{\braket{\theta_N,\ldots,\theta_1|{\cal O}(0,0)|\Phi}}{^{}_L}\
e^{\ri t(E_\Phi-\sum_{j=1}^N\Delta\cosh\theta_j)}.
\eea
\end{itemize}
We note that ultimately we are only interested in the limit $L\to\infty$ and use finite $L$ only as a regularization. As a result we only take into account the dominant finite-size corrections.

\subsection{Basis of energy eigenstates at large, finite volume}
Our starting point for constructing energy eigenstates in the finite volume are the infinite-volume scattering states \fr{scatteringstates}. The idea is to consider appropriate linear combinations of scattering states, and then to impose boundary conditions on these, see e.g. Ref.~[\onlinecite{fehtak11}]. To this end,  we introduce a \emph{transfer matrix} acting on scattering states as follows  
\begin{equation}
\label{REA:1}
\T(\lambda|\{\theta_k\})\ket{\theta_1,\dots,\theta_N}_{a_1\dots a_N}=\T(\lambda|\{\theta_k\})^{b_1\dots b_N}_{a_1\dots a_N}\ket{\theta_1,\dots,\theta_N}_{b_1\dots b_N}\,,
\end{equation}
where  
\begin{equation}
\label{REA:2}
\T(\lambda|\{\theta_k\})^{b_1\dots b_N}_{a_1\dots a_N}=\S_{c_N\,\,a_1}^{c_1b_1}(\lambda-\theta_1)\cdots\S_{c_{N-1} a_N}^{c_N\,\,b_N}(\lambda-\theta_N)\,.
\end{equation} 
As the S-matrix is a solution of the Yang-Baxter equation, the transfer matrices form a commuting family
\be
\Bigl[\T(\lambda|\{\theta_k\}),\T(\mu|\{\theta_k\})\Bigr]=0.
\ee
Hence the transfer matrices can be diagonalized simultaneously
\be
\T(\lambda|\{\theta_k\})\ket{\theta_1,\dots,\theta_N}^s=
\Lambda^{s}(\lambda|\{\theta_k\})\ket{\theta_1,\dots,\theta_N}^s\ ,\quad
s=1, \dots, 2^{N}.
\label{eigenstransfer}
\ee
We refer to the labels $s$ as \emph{polarisations}. Details of the
construction of transfer-matrix eigenstates and explicit expressions
for the eigenvalues $\Lambda^s(\lambda|\{\theta_k\})$ are given in
Appendix \ref{app:QISM}. By construction energy and momentum of the
states (\ref{eigenstransfer}) are still given by formulae
(\ref{energymomentum}). The basis transformation between scattering
states and transfer matrix eigenstates can be cast in the form 
\bea
\label{REA:3a}
\ket{\theta_1,\dots,\theta_N}^s&=&\sum_{a_1\dots a_N}\Psi^{s}_{a_1\cdots a_N}(\{\theta_k\})\ket{\theta_1,\dots,\theta_N}_{a_1\dots a_N}\,,\\
\label{REA:3b}
\ket{\theta_1,\dots,\theta_N}_{a_1\dots a_N}&=&\sum_{s}{\Psi}^{s}_{a_1\cdots a_N}(\{\theta_k\})^*\ket{\theta_1,\dots,\theta_N}^s\,,
\eea
where the amplitudes $\{\Psi^{s}_{a_1\cdots
  a_N}(\{\theta_k\})\}_{s=1,\dots,2^{N}}$ satisfy  
\bea
\T(\lambda|\{\theta_k\})_{a_1\cdots a_N}^{b_1\cdots b_N}\Psi^{s}_{a_1\cdots a_N}(\{\theta_k\})&=&\Lambda^{s}(\lambda|\{\theta_k\})\Psi^{s}_{b_1\cdots b_N}(\{\theta_k\})\,,\label{REA:4}\\
\sum_{a_1\dots a_N}\Psi^{s}_{a_1\cdots a_N}(\{\theta_k\})^*\Psi^{r}_{a_1\cdots a_N}(\{\theta_k\})&=&\delta_{rs}\,,\label{REA:5}\\
\sum_{s}\Psi^{s}_{a_1\cdots a_N}(\{\theta_k\})^*\Psi^{s}_{b_1\cdots
  b_N}(\{\theta_k\})&=&\prod_{j=1}^N\delta_{a_j, b_j}\,.\label{REA:6} 
\eea
The last two equations ensure that
$\{\ket{\theta_1,\dots,\theta_N}^s\}_{s=1, \dots, 2^{N}}$ is an
orthonormal and complete set. Since the topological charge operator
$Q$ commutes with the transfer matrix we can choose $\{\Psi^{s}_{a_1\cdots
  a_N}(\{\theta_k\})\}_{s=1,\dots,2^{N}}$ such that $Q$ is diagonal 
in the basis $\{\ket{\theta_1,\dots,\theta_N}^s\}_{s=1, \dots, 2^{N}}$ 
\be
Q\ket{\theta_1,\dots,\theta_N}^s=Q(s)\ket{\theta_1,\dots,\theta_N}^s\,,\quad 
Q(s)\in\left\{N,N-2,\dots,-N\right\}.
\ee
  
We are now in a position to consider a large, finite volume $L$ and impose periodic boundary conditions on the transfer matrix eigenstates
\begin{equation}
\label{REA:7}
e^{i \ell \sinh \theta_i}\ket{\theta_1,\dots,\theta_N}^s_L
=\T^{-1}(\theta_i|\{\theta_k\})
\ket{\theta_1,\dots,\theta_N}^s_L
\,,\qquad i=1,\dots,N,
\end{equation}  
where we have introduced a dimensionless length $\ell=L\Delta/v$. Equations \fr{REA:7} lead to quantization conditions for the rapidities in the finite volume and are known as Bethe-Yang equations\cite{fehtak11}
\be
\label{REA:8}
e^{i \ell \sinh
  \theta_i}\Lambda^{s}(\theta_i|\{\theta_k\})=1\,,\qquad 
i=1,\dots,N,\quad s=1,\dots,2^{N}.
\ee
For practical purposes the logarithmic version of \fr{REA:8} is more convenient 
\bea
2\pi I_i^{s}=
\ell\sinh{\theta_i}-\ri\log{\Lambda^{s}(\theta_i|\{\theta_k\})}
\equiv Q_i^{s}(\theta_1,\dots,\theta_N)
\,,\quad i=1,\dots,N,\quad
s=1,\dots,2^{N}.
\label{REA:9}
\eea
Here $I_i^{s}$ are integer or half-odd integer numbers that uniquely
specify a given solution and concomitantly the corresponding transfer
matrix eigenstate. By virtue of these facts one may use the integers
(rather than the rapidities) as labels for constructing a basis of
eigenstates\cite{pozsgaytakacs}. The latter is given by the states 
\begin{equation}
\ket{\{I_1,\dots,I_N\}}^s_L\ ,\quad
I_1< \cdots < I_N\ ,
\label{REA:10}
\end{equation}
where we impose normalisation conditions
\begin{equation}
\tensor*[^{r}_{L}]{\braket{\{J_M,\dots,J_1\}|\{I_1,\dots,I_N\}}}{^s_L}=\delta_{M N}\delta_{r s}\delta_{J_1 I_1}\cdots\delta_{J_N I_N}\,.
\label{REA:12}
\end{equation}
In a given sector specified by $N$ and $s$ the Jacobian matrix of the (invertible) mapping between rapidities $\{\theta_k\}_{k=1,\dots,N}$ and integers $\{I_k\}_{k=1,\dots,N}$ is given by  
\be
\J_N^s(\theta_1,\dots,\theta_N)_{ij}=\partial_{
  \theta_j}Q_i^s(\theta_1,\dots,\theta_N)\, .
\ee 
The \emph{$N$-particle density of states with polarisation $s$} is defined as the Jacobian 
\be
\rho_N^s(\theta_1,\dots,\theta_N)=\det \J_N^s(\theta_1,\dots,\theta_N)\,.
\label{denstate}
\ee 

\subsection{2-folded sine-Gordon model}
When quantising the theory in the finite volume, care has to be taken to account for the absence of spontaneous symmetry breaking of the discrete symmetry 
\be
\Phi(x)\longrightarrow\Phi(x)+\frac{2\pi n}{\beta}\ ,\ n\in\mathbb{Z}
\label{symmetry}
\ee
of the sine-Gordon Hamiltonian. In the applications of the sine-Gordon model we have in mind, the Bose field is compactified on a ring with radius
\be
R=\frac{4\pi}{\beta}.
\ee 
This implies that on a classical level there are two vacuum states $\ket{0}$ and $\ket{1}$, with corresponding field expectation values
\be
\braket{n|\Phi(x,t)|n}=\frac{2\pi}{\beta}n,\qquad n=0,1.
\ee
This theory is known as the 2-folded sine-Gordon model SG($\beta$,2),
and has been analysed in some detail in
Ref.~[\onlinecite{takacskfolded}]. We now review some relevant
results. On the quantum level, the two ground states are mapped one
into one other by the transformation
\begin{equation}
\label{REA:14}
T:\Phi(x)\,\longrightarrow\,\Phi(x)-\frac{2\pi}{\beta},
\end{equation}
which is a symmetry of the Hamiltonian. The linear combinations of $\ket{0}$ and $\ket{1}$ that are eigenvectors of $T$ are given by  
\begin{equation}
\label{REA:21}
\ket{0}_{\textrm{R},\textrm{NS}}=\frac{\ket{0}\pm\ket{1}}{\sqrt{2}}.
\end{equation}
In the infinite volume, one can construct scattering states over each
of the two ground states. They are denoted by 
\be
\label{REA:22}
\ket{\theta_1,\dots,\theta_N}^{n}_{a_1\dots a_N}
\ee
where $a_i=\pm 1$ are topological charge quantum numbers and $n=0,1$
label the two ground states. In order to define the SG($\beta$,2)
model in a finite volume, one considers eigenstates of the
``shift''-operator $T$ defined above  
\begin{align}
& \ket{\theta_1,\dots,\theta_N}^{\rm R}_{a_1\dots
    a_N}=\frac{1}{\sqrt{2}}\Big\{\ket{\theta_1,\dots,\theta_N}^{0}_{a_1\dots a_N}+\ket{\theta_1,\dots,\theta_N}^{1}_{a_1\dots a_N}\Big\}\,,\\
& \ket{\theta_1,\dots,\theta_N}^{\rm NS}_{a_1\dots a_N}=\frac{1}{\sqrt{2}}\Big\{\ket{\theta_1,\dots,\theta_N}^{0}_{a_1\dots a_N}-\ket{\theta_1,\dots,\theta_N}^{1}_{a_1\dots a_N}\Big\}\,.\label{NSR}
\end{align}
In a finite volume, periodic boundary conditions select only states for which the set 
$\{a_i\}$ is subject to the constraint 
\be
\sum_{i=1}^{N}a_i\equiv0\,\textrm{mod}\,2.
\label{REA:23}
\ee
Condition (\ref{REA:23}) is equivalent to requiring that the number $N$ of
particles be even. States involving an odd number of (anti) solitons
are incompatible with the boundary conditions and are not part of the
Hilbert space. Nevertheless all \emph{local} properties of
SG($\beta$,2) coincide with the corresponding quantities in the
sine-Gordon model. 
In order to impose boundary conditions we again go over to transfer matrix
eigenstates, which we denote by 
\be
\ket{\theta_1,\dots,\theta_N}_{\tt a}^{s}\ ,\qquad {\tt a}={\rm R,NS}.
\ee
Periodic boundary conditions imply that
\begin{equation}
\label{REA:28}
e^{\ri \ell \sinh \theta_i}\ket{\theta_1,\dots,\theta_N}_{\tt a}^{s}
=\sigma_{\tt a}\T^{-1}(\theta_i|\{\theta_k\})
\ket{\theta_1,\dots,\theta_N}_{\tt a}^{s}\ ,\quad i=1,\dots,N\ ,
\end{equation}  
where $\sigma_{\rm R}=1=-\sigma_{\rm NS}$. Eqs. (\ref{REA:28}) lead to finite-volume quantisation conditions of the form
\be
\label{REA:33}
e^{\ri \ell \sinh \theta_i}\Lambda^{s}(\theta_i|\{\theta_k\})=\sigma_{\tt
  a},\quad i=1,\dots,N\ ,\quad s=1,\dots,2^{N},\quad
{\tt a}={\rm R,NS}.
\ee
Taking the logarithm we obtain
\bea
\label{REA:34}
Q_{i}^{s}(\theta_1,\dots,\theta_N)=\ell\sinh{\theta_i}-\ri\log{\Lambda^{s}(\theta_i|\{\theta_k\})}=2\pi\left(I_i^{s}+\kappa_{\tt
    a}\right),\qquad\kappa_{\tt a}\equiv\frac{1-\sigma_{\tt a}}{4}.
\eea
Using again the (half-odd) integers $\{I_k\}$ to label the states we denote our basis of transfer matrix eigenstates in a large, finite volume $L$ by
\be
\{\ket{\{I_1,\dots,I_N\}}_{\tt a}^{s}, \;N\;\text{even}, \;s=1,\dots,2^N,\ {\tt a}={\rm R,NS}\}.
\label{finiteVbasis}
\ee
Here distinct sets of $I_j$ give rise to different basis states, and a
natural choice would therefore be $I_1<I_2<\ldots<I_N$. However,
in the following we will be interested in the case where our solutions consist of pairs,
i.e. $\{-I_1,I_1,-I_2,I_2,\ldots\}$. Having this situation in mind, we
choose the following odd-looking convention for labelling our basis states
\be 
I_{N-1}<I_{N-3}<\ldots<I_3<I_1<I_{2}<I_4\ldots<I_{N-2}<I_N\,.
\ee

Before concluding this subsection we note that the operator $\oO(t,x)=\exp{(\ri\beta\Phi(t,x)/2)}$ has non-vanishing matrix elements only between different sectors. This follows because the operator is odd under the symmetry $T$  
\begin{equation}
\label{REA:36}
T\oO(t,x)T^{\dag}=-\oO(t,x),
\end{equation}
while the states are either even or odd under $T$. 

\subsection{Matrix elements of local operators in the finite volume}
A general method for determining matrix elements of local operators in
a large, finite volume was developed in Refs.~[\onlinecite{fehtak11}]
and [\onlinecite{pozsgaytakacs}]. The leading corrections to the form
factors in the infinite-volume limit arises from the quantisation of
the rapidities. Given infinite-volume form factors
${f^{\oO}}^{a_{M}\cdots  a_1}_{b_1\cdots
  b_N}({\theta}_M,\dots,{\theta}_1|\tilde{\theta}_1,\dots,\tilde{\theta}_N)$
of an operator ${\cal O}(x)$, one first forms appropriate linear
combinations 
\begin{equation}
{f^{\oO}}({\theta}_M,\dots,{\theta}_1|\tilde{\theta}_1,\dots,\tilde{\theta}_N)^{s}_{\tilde{s}}=\sum\limits_{\substack{{a_1\dots
      a_{M}} \\ {b_1\dots b_{N}}}}\Psi^{\tilde{s}}_{b_1\cdots
  b_{N}}(\{\tilde{\theta}_k\})\Psi^{s}_{a_1\cdots
  a_{M}}(\{\theta_k\})^*\,
{f^{\oO}}^{a_M\cdots a_{1}}_{b_1\cdots b_{N}}
({\theta}_M,\dots,{\theta}_1|\tilde{\theta}_1,\dots,\tilde{\theta}_N),
\label{IV:10}
\end{equation}
where the amplitudes $\Psi^s_{a_1\dots a_{M}}(\{\theta_k\})$ are
defined in \fr{REA:3a}. Matrix elements in the basis \fr{finiteVbasis}
for operators ${\cal O}$ odd under the transformation $T$ are then
given by  
\begin{align}
&\tensor*[^{s}_{\rm R}]{\braket{ \{ {I}_M,\dots,{I}_1 \} |
\oO(0,0)| \{ \tilde{I}_1,\dots,\tilde{I}_N \} } }{^{\tilde{s}}_{\rm NS}}=
\frac{{f^{\oO}}({\theta}_M,\dots,{\theta}_1|\tilde{\theta}_1,\dots,\tilde{\theta}_N)^{s}_{\tilde{s}}}{\sqrt{{\rho}^s_{{M}}(\theta_1,\dots,\theta_M){\rho}^{\tilde{s}}_{N}(\tilde{\theta}_1,\dots,\tilde{\theta}_N)}}+\oO(e^{-\mu'L}).
\label{fehtak}
\end{align}
Here $N$ and $M$ are both even in our case.

\subsection{Finite-volume regularisation of integrable boundary states}
The next step is to obtain a realisation of our initial state $\ket{\Psi}_0$ in a large, finite volume $L$. Given that in the thermodynamic limit we must reproduce the spontaneous breaking of the symmetry \fr{symmetry}, the appropriate linear combination is given by (cf. Eq. \fr{REA:21})
\begin{equation}
\ket{\Psi_0}_L=\frac{1}{\sqrt{2}}\bigl[\ket{\Psi_0}_{\rm R}+\ket{\Psi_0}_{\rm NS}\bigr].
\label{II:10}
\end{equation}
We now wish to express $|\Psi_0\rangle_{\tt a}$ in terms of transfer-matrix eigenstates. This can be done by generalising the results of Ref.~[\onlinecite{kormpoz10}], which considered finite-volume realisations of integrable boundary states for diagonal scattering theories, for which both the scattering matrix $S^{ab}_{cd}(\theta)$ and the $K^{ab}(\theta)$ are scalars. This leads to an expression of the form
\begin{equation}
\ket{\Psi_0}_{\tt a}=\sum_{N=0}^{\infty}\sum_{s=1}^{2^{2N}}\sum_{0<I_1<\dots< I_N}\N^s_{2N}(\theta_1,\dots,\theta_N)\K^s_{2N}(\theta_1,\dots,\theta_N)\ket{\{-I_1,I_1,\dots,-I_N,I_N\}}^s_{{\tt a}}\,,\quad {\tt a}=\textrm{R},\textrm{NS}\,.
\label{II:1}
\end{equation}
We emphasize that the rapidities appearing in this formula are the parity symmetric solutions $\left\{-\theta_1,\theta_1,\dots,-\theta_N,\theta_N\right\}$ of the Bethe-Yang equations for every fixed $N$, $s$ and ${\tt a}$, with $0<\theta_1<\dots<\theta_N$. The functions ${\cal N}^s_{2N}$ and ${\cal K}^s_{2N}$ in \fr{II:1} are defined as
\begin{align}
&\K^s_{2N}(\theta_1,\dots,\theta_N)\equiv K^{a_1b_2}(\theta_1)\cdots K^{a_N b_N}(\theta_N)\Psi^{s}_{a_1b_1\cdots a_N b_N}(\{\theta_k\})^{*}\,,
\label{II:2}\\
&\N^s_{2N}(\theta_1,\dots,\theta_N)\equiv\frac{\sqrt{{\rho}^s_{2N}(-\theta_1,\theta_1,\dots,-\theta_N,\theta_N)}}{\bar{\rho}^s_N(\theta_1,\dots,\theta_N)}\,,
\label{II:3}
\end{align}
while $\bar{\rho}^s_N(\theta_1,\dots,\theta_N)$ is the constrained
$N$-particle density of states with polarisation $s$
\begin{equation}
\bar{\rho}^s_N(\theta_1,\dots,\theta_{N})=\det\bar{\J}^s_{N}(\theta_1,\dots,\theta_{N})\,,\qquad\bar{\J}^s_{N}(\theta_1,\dots,\theta_{N})_{ij}=\partial_{\theta_j} \bar{Q}^s_i(\theta_1,\dots,\theta_{N})\,.
\label{II:5}
\end{equation}
Here $\bar{Q}^s_i(\theta_1,\dots,\theta_{N})$ is the function $Q_i^s$ in the sector with $2N$-particles and polarisation $s$ evaluated for a symmetric rapidity distribution (so it refers to a particular subset of solutions of the Bethe-Yang equations for $2N$ particles)
\begin{equation}
\bar{Q}^s_i(\theta_1,\dots,\theta_{N})=Q_i^s(-\theta_1,\theta_1,\dots,-\theta_{N},\theta_N)\,,\quad i=1,\dots,N.
\label{II:6}
\end{equation}
Expression \fr{II:1} is obtained, following Ref.~[\onlinecite{kormpoz10}], by imposing that the expectation value of an arbitrary local operator $\oO(x)$ (even under $T$) in the state $\ket{\Psi_0}_{\tt a}$ must reproduce the infinite-volume result up to exponentially small corrections in system size. Inserting resolutions of the identity in terms of the basis $\{\ket{I_1,\dots,I_N}^s_{\textrm{a}}\}$, and then using the analogue of relations (\ref{fehtak}) to compute the matrix elements, we obtain (\ref{II:1}). A complication that arises compared to the case of diagonal scattering is that not every sector $s$ allows parity symmetric rapidity distributions of the kind used in \fr{II:1}. Indeed, such solutions exist only if the transfer matrix eigenvalues fulfil the relation
\begin{equation}
{\Lambda^{s}(-\theta_i|\{-\theta_k,\theta_k\})}={\Lambda^{s}(\theta_i|\{-\theta_k,\theta_k\})}^{-1}\,.
\label{II:7}
\end{equation}
How about sectors where \fr{II:7} does not hold? It turns out that they do not contribute to the expansion of the boundary state. To see this, let us go back to the infinite volume, where we have [cf. \fr{REA:3b}]
\be
|\Psi_0\rangle=\sum_{N=0}^{\infty}\sum_{s=1}^{2^{2N}}\int \frac{d\theta_1}{2\pi}\int_{\theta_1}^\infty\frac{d\theta_2}{2\pi}\ldots \int_{\theta_{N-1}}^\infty\frac{d\theta_N}{2\pi}
\K^s_{2N}(\theta_1,\dots,\theta_N)
\ket{ -\theta_1,\theta_1,\dots,-\theta_N,\theta_N }^s\ .
\ee
In Appendix~\ref{sec:AppC} we demonstrate that 
\begin{equation}
\K^s_{2N}(\theta_1,\dots,\theta_N)\Lambda^{s}(-\theta_i|\{-\theta_k,\theta_k\})=\K^s_{2N}(\theta_1,\dots,\theta_N){{\Lambda^{s}(\theta_i|\{-\theta_k,\theta_k\})}^{-1}}\,.
\label{II:8}
\end{equation}
This implies that either \fr{II:7} holds, in which case parity-symmetric solutions exist, or the coefficients $\K^s_{2N}(\theta_1,\dots,\theta_N)$ must vanish, in which case the corresponding sector $s$ does not contribute to the expansion of the boundary state.

\subsection{Determination of the representative eigenstate}
The final step required to set up our calculation of expectation
values of local operators is to obtain an expression for the
representative state. In Ref.[\onlinecite{CE_PRL13}] a method for
constructing the representative state for a given $|\Psi_0\rangle$ via
a generalized thermodynamic Bethe Ansatz\cite{mosselcaux} was presented.
Here we follow a different route. By definition $\ket{\Phi}_L$ is a
finite-volume basis state \fr{finiteVbasis}, which fulfils the
requirements that
\bea
\lim_{L\to\infty}\frac{1}{L}\frac{{}_L\langle\Phi|I_L^{(n)}|\Phi\rangle_L}{{}_L\langle\Phi|\Phi\rangle_L}=
\lim_{L\to\infty}\frac{1}{L}\frac{{}_L\langle\Psi_0|I_L^{(n)}|\Psi_0\rangle_L}{{}_L\langle\Psi_0|\Psi_0\rangle_L}\,,\quad n=1,2,\ldots\,.
\label{intofmotionV}
\eea
Here $\{I^{(n)}_L\}$ are suitable finite-volume regularisations of
the conservation laws $\{I^{(n)}\}$. As already pointed out in
Ref.~[\onlinecite{CE_PRL13}], one may use \fr{intofmotionV} to determine
the root density $\rho_{\Phi}(\theta)$ specifying the representative state.
Here the root density corresponding to a solution
$\{\theta_1,\ldots,\theta_N\}$ of the Bethe-Yang equations such that
$\theta_{j+1}-\theta_j={\cal O}(L^{-1})$ for large $L$ is defined as 
\be
\rho(\theta_j)=\lim_{L\to\infty}\frac{1}{L(\theta_{j+1}-\theta_j)}\,.
\ee
Given the density $\rho_{\Phi}(\theta)$, a particular representative microstate can be constructed. It takes the form
\be
\ket{\Phi}_L=\ket{\{-\tilde{I}_1,\tilde{I}_1,\dots,-\tilde{I}_N,\tilde{I}_N\}}^{\tilde{s}}_{\textrm{NS}}\equiv\repS \ ,
\label{RE:4}
\ee
where
\begin{itemize}
\item{} the number of rapidities $N$ is given by $2N=\lceil{\delta L}\rceil$, where $\delta=\int_{\RR} d\theta\rho_{\Phi}(\theta)$ is the total density of particles;
\item{} we have chosen the state to occur in the NS sector;
\item{} the set of integers $\{\tilde{I}_k\}$ is such that the corresponding root density is equal to $\rho_{\Phi}(\theta)$; 
\item{} the sector $\tilde{s}$ is chosen such that \fr{II:7} holds, i.e. symmetric solutions of the Bethe-Yang solutions exist. Furthermore we require the topological charge to fulfil $Q(\tilde{s})=2m=\lceil q L \rceil+k$ where $k\,\in\,\ZZ$ fixed and $q$ is the density of the expectation value of the topological charge in the initial state ($\abs{q}\leq\delta$).
\end{itemize}
It turns out that in the regime we are working in (low densities) we do not require an explicit expression of $q$ in terms of $K^{ab}(\theta)$ (our ``initial'' data), because at late times the leading contribution to the expectation value \fr{repstate} does not depend on the value of $q$.

\subsubsection{Calculation of the root density}\label{calcroot}
We start by rewriting the conditions \fr{intofmotionV} in terms of the root density $\rho_{\Phi}(\theta)$
\begin{equation}
\lim_{L\rightarrow\infty}\frac{1}{L}\frac{{}_L\braket{\Psi_0|I^{(n)}_L|\Psi_0}_L}{{}_L\braket{\Psi_0|\Psi_0}_L}=\int_{-\infty}^{\infty} d\theta\,\rho_{\Phi}(\theta)i^{(n)}(\theta)\,,\quad n=1,2,\ldots\,,
\label{III:1}
\end{equation}
where the functions $i^{(n)}(\theta)$ parameterize the eigenvalues of
the conservation laws $I^{(n)}_L$  
\be
I^{(n)}_L\ket{\{I_1,\dots,I_{N}\}}^{s}_{\tt a}=\left\{ \sum_{i=1}^{N}
i^{(n)}(\theta_i)\right\}\ket{\{I_1,\dots,I_{N}\}}^{s}_{\tt a},
\ee
Here $\{\theta_1,\ldots,\theta_N\}$ is the solution of the Bethe-Yang
equations in sector $s$ corresponding to the integers
$I_1,\ldots,I_N$. A set of local conservation laws for the sine-Gordon
model is in principle known\cite{Destri}. However, obtaining the root
density from an explicit calculation of the expectation values of
these conserved charges is a very challenging problem (assuming that
the local conservation laws are sufficient). Here we proceed
in a different way. The idea is to use \fr{III:1} in reverse:
if the functions $i^{(n)}$ form a complete set, we may determine
$\rho_\Phi(\theta)$ from a known set of expectation values of
conservation laws $I^{(n)}$. Assuming this to be the case, the
requirements \fr{III:1} are equivalent to
\begin{equation}
\lim_{L\rightarrow\infty}\frac{1}{L}\frac{{}_L\braket{\Psi_0|\hat{N}_{\zeta}|\Psi_0}_L}{{}_L\braket{\Psi_0|\Psi_0}_L}=\int_{-\infty}^{\infty} d\theta\,\rho_{\Phi}(\theta)\zeta(\theta)\,,
\label{III:2}
\end{equation}
where 
\be
\hat{N}_{\zeta}\ket{\{I_1,\dots,I_{N}\}}^{s}_{\tt a}=\left\{ \sum_{i=1}^{N}\zeta(\theta_i)\right\}\ket{\{I_1,\dots,I_{N}\}}^{s}_{\tt a}\,,
\ee
and $\zeta(x)$ is an \emph{arbitrary} function in the space spanned by
the $i^{(n)}(\theta)$. Our procedure essentially amounts to starting
with the conserved ``mode-occupation numbers'' in the thermodynamic limit
$I(\theta)=\sum_aZ^\dagger_a(\theta)Z_a(\theta)$.
Given that the $I(\theta)$ are conserved, it follows that
\be
\hat{N}_\zeta=\int d\theta\ \zeta(\theta)\ \sum_aZ^\dagger_a(\theta)Z_a(\theta)
\ee
are conserved as well. The idea is to find an appropriate finite
volume regularization of these operators, which together with the
arbitrariness of $\zeta(\theta)$ can be used to determine $\rho_\Phi$
from conditions \fr{III:2}.

It is convenient to express $\hat{N}_{\zeta}$ as  
\begin{equation}
\hat{N}_{\zeta}\equiv \sum_{I \in \ZZ}\hat{n}_{\zeta}(I)\,,
\label{III:2b}
\end{equation}
where $\hat{n}_{\zeta}(I)$ acts on the basis \fr{finiteVbasis} as follows
\begin{equation}
\hat{n}_{\zeta}(I)\ket{\{I_1,\dots,I_{N}\}}^{s}_{\tt a}=\left\{
\sum_{i=1}^{N}\delta_{I_i,I+\kappa_{\tt a}}
\zeta(\theta_i)\right\}\ket{\{I_1,\dots,I_{N}\}}^{s}_{\tt a} \,. 
\label{III:3}
\end{equation}
Here $\kappa_{\tt a}$ have been defined in \fr{REA:34}. 
Using the form \fr{II:10} for the initial state in the finite volume 
and fixing $I>0$ we obtain
\begin{equation}
{}_L\braket{\Psi_0|\hat{n}_{\zeta}(I)|\Psi_0}_L=\frac{1}{2}\sum_{{\tt a}}\sum_{N'=1}^{\infty}\sum_{s=1}^{2^{N'}}\sum_{m=1}^{N'}\sum_{\{I_1,\dots, I_{N'}\}^m_I}\N^s_{2N'}(\theta_1,\dots,\theta_{N'})^2\abs{\K^s_{2N'}(\theta_1,\dots,\theta_{N'})}^2\zeta(\theta_m)\,,
\label{III:7}
\end{equation}
where $\{I_1,\dots, I_{N'}\}^m_I$ denotes the set $\{I_1,
\dots,I_{N'}\}$ with ${I}_m$ removed and the
integers are ordered as 
\begin{equation}
\label{order}
0<\underbrace{I_1<\cdots<I+\kappa_{\tt a}}_{m}<\cdots< I_{N'}\,.
\end{equation} 
The case $I<0$ is dealt with analogously.
Our strategy is now to convert the sum
\begin{equation}
\frac{1}{L}\sum_{I\in\ZZ}\frac{{}_L\braket{\Psi_0|\hat{n}_{\zeta}(I)|\Psi_0}_L}{ {}_L\braket{\Psi_0|\Psi_0}_L}\,,
\label{III:10}
\end{equation}
into an integral in the thermodynamic limit, and then to use the arbitrariness of the function $\zeta(\theta)$ to determine $\rho_{\Phi}(\theta)$. In order for this to be possible we have to compute  
\be
\frac{{}_L\braket{\Psi_0|\hat{n}_{\zeta}(I)|\Psi_0}_L}{ {}_L\braket{\Psi_0|\Psi_0}_L}
\label{expn}
\ee
and show that is a function of $I/L$. The problem of calculating
\fr{expn} exactly still presents a formidable task. We therefore
restrict our attention to the case of ``small'' quenches in the sense
of Refs.~[\onlinecite{CEF}],[\onlinecite{SE:2012}], namely to cases
where the densities of excitations of the post-quench Hamiltonian in
the initial state are small. Then we may use $\abs{K^{ab}(\theta)}$ as
formal expansion parameters, and determine \fr{expn} by means of a
linked-cluster expansion first introduced for the finite-temperature
case in Refs.~[\onlinecite{finiteT}]. We start by expanding the
denominator in \fr{expn} as 
\bea
\label{III:8}
{}_L\braket{\Psi_0|\Psi_0}_L&=&1+\sum_{n\geq 1}\Upsilon_{2N}\,,\nn
\Upsilon_{2N}&\equiv&\frac{1}{2}\sum_{\tt
  a}\sum_{s=1}^{2^{2N}}\sum_{0<I_1<\dots<
  I_N}\N^s_{2N}(\theta_1,\dots,\theta_{N})^2\abs{\K^s_{2N}(\theta_1,\dots,\theta_{N})}^2\, .
\eea
The rapidities appearing in the expression for $\Upsilon_{2N}$ are solutions of the Bethe-Yang equations in the sector determined by $N$, $s$ and $\tt a$. Denoting the contribution at order $2M$ in $\abs{K^{ab}(\theta)}$ in \fr{III:7} by $\C_{\zeta,I}^{2M}$, i.e.
\be
{}_L\braket{\Psi_0|\hat{n}_{\zeta}(I)_{\zeta}|\Psi_0}_L=
\sum_{M\geq 1}\C_{\zeta,I}^{2M},
\ee
we define \emph{linked clusters} $\D_{\zeta,I}^{2M}$ recursively by
\begin{equation}
\C_{\zeta,I}^{2M}=\D_{\zeta,I}^{2M}+\sum_{N=1}^{M-1}\Upsilon_{2N}\D_{\zeta,I}^{2(M-N)}
\,,\qquad M=1,2,\dots
\label{III:11}
\end{equation}
In terms of the linked clusters our quantity of interest reads
\begin{equation}
\frac{{}_L\braket{\Psi_0|\hat{n}_{\zeta}(I)|\Psi_0}_L}{
  {}_L\braket{\Psi_0|\Psi_0}_L}=\sum_{N\geq 1}\D_{\zeta,I}^{2N}\,.
\label{III:12}
\end{equation}
Under the assumption that for small quenches \fr{III:12} has a
well-defined low-density expansion, the leading contribution is simply
given by the first term in the series 
\begin{equation}
\D_{\zeta,I}^{2}=\frac{1}{2}\sum_{\tt
  a}\sum_{s=1}^{2^2}\N^{s}_2\bigl(\theta^s_{\tt a}(I)\bigr)\,
\abs{\K^s_2\bigl(\theta^s_{\tt a}(I)\bigr)}^2\,
  \zeta\bigl(\theta^s_{\tt a}(I)\bigr)\, .
\label{III:13}
\end{equation}
Here $\theta^s_{\tt a}(I)$ is the solution of the Bethe-Yang equations
\begin{equation}
\bar{Q}^s(\theta)=2\pi (I+\kappa_{\tt a}).
\label{III:14}
\end{equation}
Up to finite-size corrections, the Bethe-Yang equations can be easily solved
\be
\theta^s_{\tt a}(I)=\theta(I)+\oO(1/\ell)\ ,\quad
\theta(I)=\textrm{arcsinh}\left({2 \pi I/\ell}\right).
\ee
Noting that $\N^{s}_2(\theta_{\tt a}^s(I))=1+\oO(1/\ell)$ $\forall\,s\,,\tt a$, we have 
\begin{equation}
\D_{\zeta,I}^{2}=\zeta\bigl(\theta(I)\bigr)\,G\bigl(\theta(I)\bigr)+\oO(1/\ell)\,,
\label{III:15}
\end{equation}
where we introduced 
\begin{equation}
G(\theta)\equiv\sum_{a,b=\pm}\abs{K^{ab}(\theta)}^2\,.
\label{III:16}
\end{equation}
The result for the expectation value (\ref{III:10}) is then
\begin{equation}
\frac{{}_L\braket{\Psi_0|\hat{n}_{\zeta}(I)|\Psi_0}_L}{ {}_L\braket{\Psi_0|\Psi_0}_L}={\zeta\bigl(\theta(I)\bigr)} G\bigl(\theta(I)\bigr)+\oO(K^{4})\,.
\label{III:17}
\end{equation}
The sum of this expression over all integers can be expressed as an integral using contour integration methods
\begin{equation}
\lim_{L\rightarrow\infty}\frac{1}{L}\frac{{}_L\braket{\Psi_0|\hat{N}_{\zeta}|\Psi_0}_L}{ {}_L\braket{\Psi_0|\Psi_0}_L}\simeq\lim_{L\rightarrow\infty}\sum_{I}\oint\limits_{\cC_{\theta(I)}}\frac{d\eta}{2\pi}  \frac{\Delta}{v}\cosh\eta \frac{G(\eta){\zeta}{(\eta)}}{e^{\ri \ell \sinh\eta}-1}\,,
\label{III:18}
\end{equation}
where  the contour $\cC_{\theta(I)}$ encircles $\theta(I)$. Since the integrand is analytic except for the residues at $\{\theta(I)\}$, we may join the paths $\cC_{\theta(I)}$ and obtain a single contour $\cC$ encircling the real axis 
\begin{equation}
\lim_{L\rightarrow\infty}\frac{1}{L}\frac{{}_L\braket{\Psi_0|\hat{N}_{\zeta}|\Psi_0}_L}{
  {}_L\braket{\Psi_0|\Psi_0}_L}\simeq\lim_{L\rightarrow\infty}\oint\limits_{\cC}\frac{d\theta}{2\pi}
\frac{\Delta}{v}\cosh\theta
\frac{G(\theta){\zeta}{(\theta)}}{e^{\ri\ell\sinh\theta}-1}
=\frac{\Delta}{v}\int_{-\infty}^{+\infty}\frac{d\theta}{2\pi}{G(\theta)\,{\zeta}{(\theta)}}\,\cosh\theta.
\label{III:19}
\end{equation}
In the last step we have used that the contribution from the path below the real axis is exponentially suppressed in $\ell$, and will not contribute in the thermodynamic limit. Finally, equating \fr{III:19} and \fr{III:2} and then using the arbitrariness of $\zeta(\theta)$, we are able to find the root density $\rho_\Phi(\theta)$
\begin{equation}
\rho_{\Phi}(\theta)=\frac{\Delta}{2\pi v} G(\theta)\,\cosh\theta +\oO(K^{4}).
\label{III:21}
\end{equation}

\subsection{Time evolution of the expectation value}
\label{Cal}
With an expression for the representative eigenstate in hand, we are
now in a position to write the explicit Lehmann representation for the
expectation value \fr{repstate} and work out its time evolution. We
focus on the first term on the right hand side of Eq.~\fr{repstate},
as the other can be obtained by taking complex conjugation combined
with the change $\beta\rightarrow-\beta$. The Lehmann representation
is given by  
\begin{align}
\frac{{}_{\rm R}\langle\Psi_0|e^{\ri\beta\Phi(t,x)/2}\repS}
{{}_{\rm NS}\langle\Psi_0\repS}=\sum_{M\geq  1}\sum_{s=1}^{2^{2M}}
&\sum_{0<I_1<
  \dots <
  I_M}\frac{\N^s_{2M}(\theta_1,\dots,\theta_M)\K^s_{2M}(\theta_1,\dots,\theta_M)}{\N^{\tilde{s}}_{2N}(\tilde{\theta}_1,\dots,\tilde{\theta}_N)\K^{\tilde{s}}_{2N}(\tilde{\theta}_1,\dots,\tilde{\theta}_N)}\exp\left[{2
    \ri t \Delta
    \Bigl(\sum_{i=1}^{M}\cosh\theta_i-\sum_{i=1}^{N}\cosh\tilde{\theta}_i}\Bigr)\right]\notag\\ 
&\times\tensor*[^{s}_{\textrm{R}}]{\braket{\{{I}_M,-{I}_M,\dots,{I}_1,-{I}_1\}|e^{{\ri}{\beta \Phi(0,0)}/2}|\{-\tilde{I}_1,\tilde{I}_1,\dots,-\tilde{I}_N,\tilde{I}_N\}}}{^{\tilde{s}}_{\textrm{NS}}}\,,
\label{IV:1}
\end{align}
where $\{I_k\}$ and $\{\tilde{I}_k\}$ are respectively integers and half-odd integers, while the rapidities $\{\theta_k\}_{k=1,\dots,M}$ and  $\{\tilde{\theta}_k\}_{k=1,\dots,N}$ are the corresponding solutions of the Bethe-Yang equations
\begin{align}
&\bar{Q}_i^{s}(\theta_1,\dots,\theta_M)=2\pi I_i\,, &i=1,\dots,M\,, \label{IV:2}\\
&\bar{Q}_j^{\tilde{s}}(\tilde{\theta}_1,\dots,\tilde{\theta}_N)=2\pi \tilde{I}_j\,, &j=1,\dots,N\,.\label{IV:3}
\end{align}
As discussed before, both sectors $(2M,s,{\textrm{R}})$ and
$(2N,\tilde{s},{\textrm{NS}})$ permit parity symmetric solutions. The
matrix elements are related to infinite-volume form factors by
\fr{fehtak}, which in the case of interest reads 
\begin{align}
&\tensor*[^{s}_{\textrm{R}}]{\braket{\{{I}_M,\dots,-{I}_1\}|e^{{\ri}{\beta \Phi(0,0)}/2}|\{-\tilde{I}_1,\dots,\tilde{I}_N\}}}{^{\tilde{s}}_{\textrm{NS}}}=
\frac{{f^{\beta/2}}({\theta}_M,\dots,-{\theta}_1|-\tilde{\theta}_1,\dots,\tilde{\theta}_N)^{s}_{\tilde{s}}}{\sqrt{{\rho}^s_{2M}(-\theta_1,\dots,\theta_M){\rho}^{\tilde{s}}_{2N}(-\tilde{\theta}_1,\dots,\tilde{\theta}_N)}}+\oO(e^{-\mu'L})\,.
\label{fehtak1}
\end{align}
Using the crossing symmetry in the infinite-volume from factors gives an expression of the form
\begin{align}
{f^{\beta/2}}({\theta}_M,\dots,-{\theta}_1|-\tilde{\theta}_1,\dots,\tilde{\theta}_N)^{s}_{\tilde{s}}&=\sum\limits_{\substack{{a_1\dots a_{2M}} \\ {b_1\dots b_{2N}}}}\Psi^{\tilde{s}}_{b_1\cdots b_{2N}}(\{\tilde{\theta}_k\})\Psi^{s}_{a_1\cdots a_{2M}}(\{\theta_k\})^*\notag\\*
&\qquad\qquad\times f^{\beta/2}_{ \bar{a}_{2M}\cdots\bar{a}_1, b_1\cdots b_{2N}}({\theta}_M+\ri\pi,\cdots,-{\theta}_1+\ri\pi,-\tilde{\theta}_1,\dots,\tilde{\theta}_N)\notag\\*
&=  f^{\beta/2}({\theta}_M+\ri\pi,\dots,-{\theta}_1+\ri\pi,-\tilde{\theta}_1,\dots,\tilde{\theta}_N)_{s,\tilde{s}}\,.
\label{IV:11}
\end{align}
In the following we will take into account only the contribution to (\ref{IV:1}) that arises from states with\cite{CE_PRL13} $N=M$, as this represents the leading one. It is shown in Appendix~\ref{app:M>N} that the contribution from states with $N=M$ dominates over those from states with $M>N$ at late times. The same conclusion holds also in the case of contributions from states with $M<N$ and can be proven analogously.

\subsubsection{Contributions from states with $M=N$}
\label{ssec:M=N}
Retaining only states with $M=N$ we have 
\begin{align}
\frac{{}_{\rm R}\langle\Psi_0|e^{\ri\beta\Phi(t,x)/2}\repS}
{{}_{\rm NS}\langle\Psi_0\repS}
\approx \sum_{s=1}^{2^{2N}}\sum_{0<I_1< \dots <
  I_N}&\frac{\bar{\rho}^{\tilde{s}}_{N}(\tilde{\theta}_1,\dots,\tilde{\theta}_N)}{\bar{\rho}^{{s}}_{N}({\theta}_1,\dots,{\theta}_N){\rho}^{\tilde{s}}_{2N}(-\tilde{\theta}_1,\dots,\tilde{\theta}_N)}\frac{\K^s_{2N}(\theta_1,\dots,\theta_N)}{\K^{\tilde{s}}_{2N}(\tilde{\theta}_1,\dots,\tilde{\theta}_N)}\notag\\ 
&\times e^{2\ri \Delta t
  {\sum_{i=1}^{N}[\cosh\theta_{i}-\cosh\tilde{\theta}_{i}]}} 
f^{\beta/2}({\theta}_N+\ri\pi,\dots,\tilde{\theta}_N)_{s,\tilde{s}}\,.
\label{IV:13}
\end{align}
Following the method introduced by Pozsgay and Takacs in~[\onlinecite{finiteT2}], we rewrite each term in \fr{IV:13} as an integral on an appropriate multi-contour ${\cal C}$
in $\CC^N$ by means of the multidimensional residue calculus 
\begin{align}
\frac{{}_{\rm R}\langle\Psi_0|e^{\ri\beta\Phi(t,x)/2}\repS}
{{}_{\rm NS}\langle\Psi_0\repS}
\approx\sum_{s=1}^{2^{2N}}\sum_{0<I_1< \dots < I_N}&\oint\limits_{\C_{\{\theta_\ell\}}}\prod_{k=1}^{N}\frac{d\eta_k}{2\pi}\frac{\bar{\rho}^{\tilde{s}}_{N}(\tilde{\theta}_1,\dots,\tilde{\theta}_N)}{{\rho}^{\tilde{s}}_{2N}(-\tilde{\theta}_1,\dots,\tilde{\theta}_N)}\frac{\K^s_{2N}(\eta_1,\dots,\eta_N)}{\K^{\tilde{s}}_{2N}(\tilde{\theta}_1,\dots,\tilde{\theta}_N)}\notag\\
&\times e^{2\ri \Delta t
  {\sum_{i=1}^{N}[\cosh\eta_{i}-\cosh\tilde{\theta}_{i}]}}
\frac{ f^{\beta/2}({\eta}_N+\ri\pi,\dots,\tilde{\theta}_N)_{s,\tilde{s}}}
{\prod_{k=1}^{N}\left\{e^{\ri
    \bar{Q}^s_k(\eta_1,\dots,\eta_N)}-1\right\}}\,.
\label{IV:14}
\end{align}
Here
$\C_{\{\theta_\ell\}}=\C_{\theta_1}\otimes\cdots\otimes\C_{\theta_N}$,
where the contour $\C_{\theta_i}$ encircles the rapidity $\theta_i$,
i.e. the $i$-th rapidity in the solution $\{\theta_k\}$ of the
Bethe-Yang equations in the sector $(2N,\,s,\,\text{R})$. We now wish
to express the multiple summations in \fr{IV:14} in terms of
appropriately defined contour integrals. In the case of a single
variable this is straightforward, see Fig.~\ref{fig:I} for a simple
example. Let $f(\eta)$ be a meromorphic function with simple poles at
the points $\eta=\theta_j$, and $C_{\theta_j}$ a very small contour
encircling $\theta_j$. Residue calculus then gives 
\begin{equation}
\sum_{j}\oint\limits_{\C_{\theta_j}}\frac{d\eta}{2\pi}f(\eta)=\oint\limits_{\C_{\text{tot}}}\frac{d\eta}{2\pi}f(\eta)-\sum_{i=1}^{N_{\R}}\oint\limits_{\,\C_{\tilde{\theta}_i}}\frac{d\eta}{2\pi}f(\eta)
\label{IV:14a}
\end{equation}
where $\C_{\text{tot}}$ is a contour encircling the region ${\cal R}$, which contains all points $\theta_j$ as well as simple poles of $f(\eta)$ at positions $\tilde{\theta}_1,\ldots,\tilde{\theta}_{N_{\cal R}}$. 
\tikzset{->-/.style={decoration={
  markings,
  mark=at position .5 with {\arrow{>}}},postaction={decorate}}}
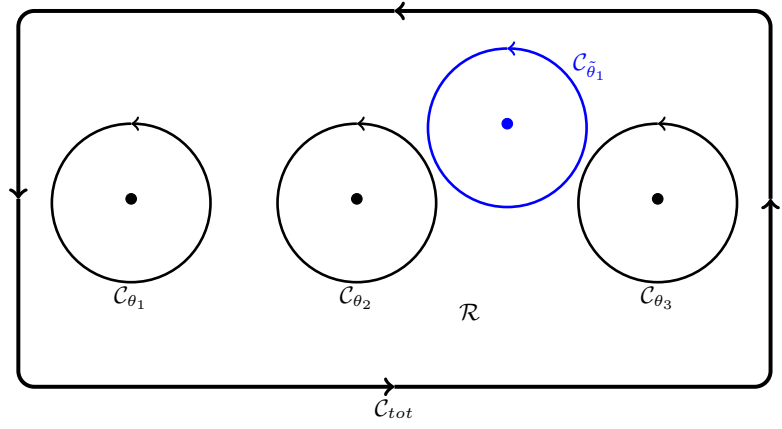
\begin{figure}
\begin{tikzpicture}
\draw[->,line width =1.5pt,rounded corners =6pt] (0,2.5) -- (0,0) --(5,0);
\draw[->,line width =1.5pt,rounded corners =6pt] (5,0) -- (10,0)--(10,2.5);
\draw[->,line width =1.5pt,rounded corners =6pt] (10,2.5) -- (10,5) --(5,5);
\draw[->,line width =1.5pt,rounded corners =6pt] (5,5) -- (0,5)--(0,2.5);
\draw (5,0) node[below=1.5pt] {$\displaystyle \C_{tot}$};
\draw[->,line width =1pt] (8.5,3.5) arc (90:450:30pt);
\draw[->,line width =1pt] (4.5,3.5) arc (90:450:30pt);
\draw[->,line width =1pt] (1.5,3.5) arc (90:450:30pt);
\draw[->,line width =1pt, blue] (6.5,4.5) arc (90:450:30pt);
\filldraw [black] (8.5,2.5) circle (2pt)                   (4.5,2.5) circle (2pt)                   (1.5,2.5) circle (2pt);
\filldraw [blue] (6.5,3.5) circle (2pt);
\draw (8.5,1.5) node[below=1.5pt] {$\displaystyle \C_{\theta_3}$};
\draw (4.5,1.5) node[below=1.5pt] {$\displaystyle \C_{\theta_2}$};
\draw (1.5,1.5) node[below=1.5pt] {$\displaystyle \C_{\theta_1}$};
\draw[blue] (7.6,4.6) node[below=1.5pt] {$\displaystyle \C_{\tilde{\theta}_1}$};
\draw (6,1) node {$\R$};
\end{tikzpicture}
\caption{Example of the contours in the one-variable case.}
\label{fig:I}
\end{figure}

The case of $N$ integration variables $\eta_1,\ldots,\eta_N$ is more
involved. Our function of interest is
\be
G^{(s,\tilde{s})}(\eta_1,\dots, \eta_N)\equiv \frac{\bar{\rho}^{\tilde{s}}_{N}(\tilde{\theta}_1,\dots,\tilde{\theta}_N)}{{\rho}^{\tilde{s}}_{2N}(-\tilde{\theta}_1,\dots,\tilde{\theta}_N)}\frac{\K^s_{2N}(\eta_1,\dots,\eta_N)}{\K^{\tilde{s}}_{2N}(\tilde{\theta}_1,\dots,\tilde{\theta}_N)}
\frac{  f^{\beta/2}({\eta}_N+\ri\pi,\dots,\tilde{\theta}_N)_{s,\tilde{s}}}{\prod_{k=1}^{N}\left\{e^{\ri \bar{Q}^s_k(\eta_1,\dots,\eta_N)}-1\right\}}e^{2\ri \Delta t {\sum_{i=1}^{N}[\cosh\eta_{i}-\cosh\tilde{\theta}_{i}]}}\,.
\label{G}
\ee
We first consider a contour $\C_{\rm tot}$, which is the $N$-fold tensor product
$\C(\RR^+)\otimes\ldots\otimes\C(\RR^+)$, where $\C(\RR^+)$ is a 
contour encircling $\RR^+$. Introducing $\C_{\tilde{\theta}_{i}}$ as
the contour encircling  $\tilde{\theta}_{i}$, we can express $\C_{\rm
  tot}$ in the form
\be
\C_{\rm tot}=\cG\otimes\ldots\otimes\cG-\sum_{m=1}^{N}(-1)^{m}\sum_{1\leq j_1<\cdots<j_m\leq N}\sum_{i_1,\cdots,i_m=1}^{N}\C^{j_1,\cdots,j_m}_{i_1,\cdots,i_m}\,,
\label{contour}
\ee
where $\cG$ is defined as 
\be
\cG\equiv \C(\RR^+)-\sum_{i=1}^{N}\C_{\tilde{\theta}_{i}}\,,
\ee
and $\C^{j_1,\cdots,j_m}_{i_1,\cdots,i_m}$ is defined as the multi-contour obtained from $\C_{\rm tot}$ by replacing the components $j_1,\cdots,j_m$ in the tensor product by $\C_{\tilde{\theta}_{i_1}},\cdots,\C_{\tilde{\theta}_{i_m}}$, i.e.
\be
\C^{j_1,\cdots,j_m}_{i_1,\cdots,i_m}=\otimes_{j=1}^{j_1-1}\C(\RR^+)\otimes
\C_{\tilde{\theta}_{i_1}}\otimes_{j=j_1+1}^{j_2-1}\C(\RR^+)\otimes
\C_{\tilde{\theta}_{i_2}}\otimes_{j=j_2+1}^{j_3-1}\C(\RR^+)\otimes\dots
\otimes_{j=j_{m-1}+1}^{j_m-1}\C(\RR^+)
\otimes\C_{\tilde{\theta}_{i_m}}
\otimes_{j=j_m+1}^{N}\C(\RR^+).
\ee
We may use \fr{contour} to rewrite the integral of $G^{(s,\tilde{s})}$
over the multi-contour $\C_{\rm tot}$ as
\begin{align}
\oint\limits_{\C_{\rm
    tot}}\prod_{i=1}^{N}\frac{d\eta_i}{2\pi}G^{(s,\tilde{s})}(\eta_1,\dots,
\eta_N)=
&\oint\limits_{\cG\otimes\cdots\otimes\cG}
\prod_{i=1}^{N}\frac{d\eta_i}{2\pi}G^{(s,\tilde{s})}(\eta_1,\dots, \eta_N)\notag\\ 
&-\sum_{m=1}^{N}(-1)^{m}\sum_{1\leq j_1<\cdots<j_m\leq N}\sum_{i_1,\cdots,i_m=1}^{N}\oint\limits_{\C^{j_1,\cdots,j_m}_{i_1,\cdots,i_m}}\prod_{i=1}^{N}\frac{d\eta_i}{2\pi}G^{(s,\tilde{s})}(\eta_1,\dots, \eta_N)
\,.\label{IV:14ab}
\end{align}
Let us now focus on the first term in the right-hand side of \fr{IV:14ab}. 
As the singularities of the form factor $f^{\beta/2}$ lie outside the contour
$\cG\otimes\cdots\otimes\cG$ and all other functions are expected to be
well behaved, the only contributions to the integral arise from the regions
characterized by 
\be
e^{\ri\bar{Q}^s_k(\eta_1,\dots,\eta_N)}-1\approx 0\,,\qquad
k=1,\dots,N\ .
\label{reg1}
\ee
By choosing $\C(\RR^+)$ sufficiently close to the real axis, we can
ensure that \fr{reg1} is fulfilled only when $\{\eta_k\}$ is a real
solution of the Bethe-Yang equations in the sector
$(2N,\,s,\,\text{R})$ for some set $\{I_k\}$. We can ensure that
\emph{all} solutions to the Bethe-Yang equations in the sector
$(2N,\,s,\,\text{R})$ are enclosed in $\cG\otimes\cdots\otimes\cG$ by
choosing the contours $\C_{\tilde{\theta}_i}$ to be sufficiently close
to $\tilde{\theta}_i$, because $\{\tilde{\theta}_k\}$ is a solution of
the Bethe-Yang equations in the NS sector and therefore cannot be
arbitrarily close to any solution in the R sector. This allows us to
conclude that
\be
\oint\limits_{\cG\otimes\cdots\otimes\cG}
\prod_{i=1}^{N}\frac{d\eta_i}{2\pi}G^{(s,\tilde{s})}(\eta_1,\dots, \eta_N)=\sum_{I_1\ldots,I_N}\oint\limits_{\C_{\{\theta_\ell\}}}
\prod_{i=1}^{N}\frac{d\eta_i}{2\pi}G^{(s,\tilde{s})}(\eta_1,\dots, \eta_N)\,.
\label{BYcontour}
\ee
Let us now turn to the contributions with $m=N$ on the right-hand side of Eq.~\fr{IV:14ab}. For these the relevant multi-contours are of the form
\be
\C_{\tilde{\theta}_{\sigma(1)}}\otimes\dots\otimes\C_{\tilde{\theta}_{\sigma(N)}}\,,
\ee
where $\sigma\in\S_N$ is a permutation. Exploiting again the fact that
$\{\tilde{\theta}_k\}$ is a solution of the Bethe-Yang equations in
the NS sector, we may conclude that for contours $\C_{\tilde{\theta}_i}$
sufficiently close to the points $\tilde{\theta}_i$ we have for any
point $\eta_1,\ldots\eta_N$ inside the multi-contour
\be
e^{\ri\bar{Q}^s_k(\eta_1,\dots,\eta_N)}-1\neq0\,,\qquad\forall\,k\ .
\ee 
Hence the only contributions to the integral arise from the
annihilation poles of the form factor $f^{\beta/2}$. These
considerations will be very useful in the following.

Substituting \fr{BYcontour} into \fr{IV:14ab} we obtain the desired
$N$-variable generalisation of \fr{IV:14a} 
\begin{align}
\sum_{I_1\ldots,I_N}\oint\limits_{\C_{\{\theta_\ell\}}}
\prod_{i=1}^{N}\frac{d\eta_i}{2\pi}G^{(s,\tilde{s})}(\eta_1,\dots, \eta_N)=
&\oint\limits_{\C_{\rm
    tot}}\prod_{i=1}^{N}\frac{d\eta_i}{2\pi}G^{(s,\tilde{s})}(\eta_1,\dots,
\eta_N)\notag\\ 
&+\sum_{m=1}^{N}(-1)^{m}\sum_{1\leq j_1<\cdots<j_m\leq N}\sum_{i_1,\cdots,i_m=1}^{N}\oint\limits_{\C^{j_1,\cdots,j_m}_{i_1,\cdots,i_m}}\prod_{i=1}^{N}\frac{d\eta_i}{2\pi}G^{(s,\tilde{s})}(\eta_1,\dots, \eta_N)
\,.\label{IV:14aa}
\end{align}

In order to deal with Eq.~\fr{IV:14} we require one more step: since
in \fr{IV:14} we are considering a fixed order of integers $\{I_k\}$, the
corresponding solutions of the Bethe-Yang equations must satisfy
$0<\theta_1<\cdots<\theta_N$. In order to 
accommodate this ordering of rapidities we need to modify our
multi-contours such that the $i$-th component depends on the variable
$z_{i+1}$. For example, the integral on $\C_{\rm tot}$ changes to 
\be
\oint\limits_{\D}\frac{dz_N}{2\pi}\oint\limits_{\D(z_N)}\frac{dz_{N-1}}{2\pi}\cdots\oint\limits_{\D(z_2)}\frac{dz_1}{2\pi}f(z_1,\dots,z_N)\equiv
\oint\limits_{\bar{\C}_{\rm
    tot}}\prod_{i=1}^{N}\frac{dz_i}{2\pi}f(z_1,\dots, z_N)\,,  
\label{IV:ab}
\ee
where the contour $\D(z_N)$ encircles the interval $[0,\Re z_N)$. Reduced multi-contours $\bar{\C}^{j_1,\cdots,j_m}_{i_1,\cdots,i_m}$ are defined in a way analogous to ${\C}^{j_1,\cdots,j_m}_{i_1,\cdots,i_m}$, but include additional constraints ensuring the ordering $\Re(z_N)>\Re(z_{N-1})>\ldots>\Re(z_1)$. We may now use these considerations to turn the sums in \fr{IV:14} into integrals over appropriate multi-contours in the form
\begin{align}
\frac{{}_{\rm R}\langle\Psi_0|e^{\ri\beta\Phi(t,x)/2}\repS}
{{}_{\rm NS}\langle\Psi_0\repS}\approx
&\sum_s\oint\limits_{\bar{\C}_{\text{tot}}}\prod_{i=1}^{N}\frac{d\eta_i}{2\pi}G^{(s,\tilde{s})}(\eta_1,\dots, \eta_N)\notag\\
&+\sum_s\sum_{m=1}^{N-1}(-1)^{m}\sum_{1\leq j_1<\cdots<j_m\leq N}\sum_{i_1,\cdots,i_m=1}^{N}\oint\limits_{\bar{\C}^{j_1,\cdots,j_m}_{i_1,\cdots,i_m}}\prod_{i=1}^{N}\frac{d\eta_i}{2\pi}G^{(s,\tilde{s})}(\eta_1,\dots, \eta_n)\notag\\
&+(-1)^N\sum_s \oint\limits_{\,\C_{\tilde{\theta}_{1}}\otimes\dots\otimes\C_{\tilde{\theta}_{N}}}\prod_{i=1}^{N}\frac{d\eta_i}{2\pi}G^{(s,\tilde{s})}(\eta_1,\dots, \eta_N)\,.
\label{IV:15}
\end{align}
The crucial simplification occurring at this stage is that in the thermodynamic limit $L\to\infty$ and at late times we need to retain only the last term, as it features the most singular contribution, which occurs when~\cite{noteF}
\be
\eta_j\approx \tilde{\theta}_j\ ,\quad j=1,\ldots N.
\ee
As we have argued above, the only singularities we need to consider in
the last term of \fr{IV:15} are those arising from the form factor.
The \emph{leading} singularities of the form factors can be calculated using
the annihilation pole axiom, see Appendix \ref{app:annihilationpole},
and gives 
\begin{equation}
f^{\beta/2}({\eta}_N+\ri\pi,-{\eta}_N+\ri\pi,\dots,-\tilde{\theta}_N,\tilde{\theta}_N)_{s,\tilde{s}}\Bigg|_{\tilde{\theta}_j \approx\eta_j}=
\cG_{\beta/2}\delta_{\tilde{s},s}\ \prod_{i=1}^{N}
\frac{4}{(\eta_i-\tilde{\theta}_i)^2}\,+\text{less singular}.
\label{IV:16}
\end{equation}
An explicit expression for the vacuum expectation value
$\cG_{\beta/2}\equiv\braket{0|e^{\ri \beta \Phi/2}|0}$ is given in
Ref.~[\onlinecite{LZ}]. Substituting the expression \fr{IV:16} into
the last term on the right-hand side of Eq.~\fr{IV:15} gives the
leading contribution at late times. Carrying out the contour integrals
we obtain 
\begin{align}
\frac{{}_{\rm R}\langle\Psi_0|e^{\ri\beta\Phi(t,x)/2}\repS}
{{}_{\rm NS}\langle\Psi_0\repS}\approx
&(-4\ri)^N\cG_{\beta/2}\frac{\bar{\rho}^{\tilde{s}}_{N}(\tilde{\theta}_1,\dots,\tilde{\theta}_N)}{{\rho}^{\tilde{s}}_{2N}(-\tilde{\theta}_1,\dots,\tilde{\theta}_N)}\partial_{\eta_1}
\bigg|_{\eta_1=\tilde{\theta}_1}
\cdots\partial_{\eta_N}\bigg|_{\eta_N=\tilde{\theta}_N}
\prod_{i=1}^{N}\frac{e^{2\ri \Delta t {[\cosh\eta_{i}-\cosh\tilde{\theta}_{i}]}}}{e^{\ri
    \bar{Q}^{\tilde{s}}_i(\eta_1,\dots,\eta_N)}-1}+\ldots
\label{IV:17}
\end{align}
The derivatives can be computed in leading order in the density $\delta$ using that $\partial_j\bar{Q}^{\tilde{s}}_j={\cal O}(\ell)$, while $\partial_j\bar{Q}^{\tilde{s}}_i={\cal O}(1)$ (see Appendix  \ref{app:mixedderivatives})
\bea
\partial_{\eta_1}
\bigg|_{\eta_1=\tilde{\theta}_1}
\cdots\partial_{\eta_N}\bigg|_{\eta_N=\tilde{\theta}_N}
\prod_{i=1}^{N}\frac{e^{2\ri \Delta t {[\cosh\eta_{i}-\cosh\tilde{\theta}_{i}]}}}{e^{\ri
    \bar{Q}^{\tilde{s}}_i(\eta_1,\dots,\eta_N)}-1}&=&\left(\frac{\ri}{4}\right)^N\prod_{i=1}^{N}\left\{\partial_i\bar{Q}^{\tilde{s}}_i(\tilde{\theta}_1,\dots,\tilde{\theta}_N)-4
\Delta t\sinh\tilde{\theta}_{i}\right\}\nn
&&+\text{higher order in $\delta$}.
\eea
Here we have used that
\be
e^{\ri\bar{Q}^{\tilde{s}}_i(\tilde{\theta}_1,\dots,\tilde{\theta}_N)}=-1.
\ee
The densities of states $\bar{\rho}_N^{\tilde{s}}$ and
${\rho}_{2N}^{\tilde{s}}$ are defined in terms of partial derivatives
of the functions $\bar{Q}^{\tilde{s}}_j$ and ${Q}^{\tilde{s}}_j$, see
Eqs.~\fr{II:5} and \fr{REA:9}; their ratio can be calculated to
leading order in the density $\delta$ by using the results derived in
Appendix \ref{app:mixedderivatives} as well 
\begin{equation}
\frac{\bar{\rho}^{\tilde{s}}_{N}(\tilde{\theta}_1,\dots,\tilde{\theta}_N)}{{\rho}^{\tilde{s}}_{2N}(-\tilde{\theta}_1,\dots,\tilde{\theta}_N)}=
\prod_{i=1}^{N}\frac{1}{\partial_i\bar{Q}^{\tilde{s}}_i(\tilde{\theta}_1,\dots,\tilde{\theta}_N)}
+\text{higher order in $\delta$}.
\label{IV:21}
\end{equation}
Putting everything together, we obtain
\bea
\frac{{}_{\rm R}\langle\Psi_0|e^{\ri\beta\Phi(t,x)/2}\repS}
{{}_{\rm NS}\langle\Psi_0\repS}
&\approx&\cG_{\beta/2}\prod_{i=1}^{N}\left\{1-\frac{4\Delta
  t\sinh\tilde{\theta}_{i}}{\partial_i\bar{Q}^{\tilde{s}}_i(\tilde{\theta}_1,\dots,\tilde{\theta}_N)}\right\}+\ldots\nn*
&=&\cG_{\beta/2}\exp\left[{\sum_{i=1}^{N}\log\left(1-\frac{4 \Delta t\sinh\tilde{\theta}_{i}}{\partial_i\bar{Q}^{\tilde{s}}_i(\tilde{\theta}_1,\dots,\tilde{\theta}_N)}\right)}\right]+\ldots
\label{IV:23}
\eea
Since $\partial_i\bar{Q}^{\tilde{s}}_i(\tilde{\theta}_1,\dots,\tilde{\theta}_N)={\cal O}(\ell)\gg \Delta t$, we may expand the logarithm 
\begin{equation}
\sum_{i=1}^{N}\log\left(1-\frac{4 \Delta t\sinh\tilde{\theta}_{i}} {\partial_i\bar{Q}^{\tilde{s}}_i(\tilde{\theta}_1,\dots,\tilde{\theta}_N)}\right)=-\frac{4v t}{L}\left\{\sum_{i=1}^{N}\tanh\tilde{\theta}_{i}\right\}+\ldots,
\end{equation}
where we used 
\begin{equation}
\frac{4 \Delta t\sinh\tilde{\theta}_{i}}{\partial_i\bar{Q}^{\tilde{s}}_i(\tilde{\theta}_1,\dots,\tilde{\theta}_N)}=\frac{4\Delta t}{\ell}\tanh\tilde{\theta}_{i}+\text{higher order in $\delta$}.
\label{IV:24}
\end{equation}
Finally, we convert the sum over $\tilde{\theta}_j$ into an integral in the $L\to\infty$ limit
\bea
\lim_{L \rightarrow \infty}
\frac{{}_{\rm R}\langle\Psi_0|e^{\ri\beta\Phi(t,x)/2}\repS}
{{}_{\rm NS}\langle\Psi_0\repS}
&\approx&\lim_{L\to\infty}
\cG_{\beta/2}\exp\left[-\frac{4vt}{L}\sum_{i=1}^{N}\tanh\tilde{\theta}_i
\right]+\ldots\nn
&=&\cG_{\beta/2}\exp\left[-4vt\int_0^\infty
  d\theta\ \rho_\Phi(\theta)\
 \tanh\theta\right]+\ldots\nn
&=&\cG_{\beta/2}e^{-{t}/{\tau}}+\ldots
\label{IV:25}
\eea
The decay time $\tau$ defined in this way is given by  
\begin{equation}
\tau^{-1}\equiv \frac{2 \Delta}{\pi} \int_{0}^{\infty}d\theta\, \left[G(\theta) \sinh{\theta} +{\cal O}(K^4)\right]\,.
\label{eq:decaytime}
\end{equation}
Using $\cG_{a}=\cG_{-a}=\cG_{a}^*$, we obtain our final result for the
time evolution of the one-point function in the thermodynamic limit at
late times $\Delta t\gg 1$, and in the low-density limit $\delta\ll 1$
\begin{equation}
\lim_{L\rightarrow\infty}\braket{e^{\ri\beta \Phi(t,x)/2}}_L
=\lim_{L\rightarrow\infty}\frac{_L\bra{\Psi_t}e^{\ri\beta \Phi(x)/2}\ket{\Psi_t}_L}{_L\langle\Psi_t\ket{\Psi_t}_L}
=\cG_{\beta/2}e^{-{t}/{\tau}}+\ldots
\label{IV:28}
\end{equation}
In the next section we show that this result can also be obtained
using a complementary approach. 


\section{Linked-cluster expansion approach}\label{sec:LCE}
A characteristic feature of initial states after quantum quenches is
that they involve finite densities of excitations of the post-quench
Hamiltonian. In this sense the quench problem is similar to the one of
finite-temperature dynamics in equilibrium. Building on recent
progress in the latter problem\cite{finiteT,finiteT2},
Refs.~\onlinecite{CEF,SE:2012} developed an approach to quench
problems based on a linked-cluster expansion for initial states of the
form \fr{psi0}. Here we generalize this approach to the case of the
sine-Gordon model. 

\subsection{Formal expansion of one-point functions}
Our goal is the derivation of a formal expansion of the one-point function
\begin{equation}
\frac{\langle\Psi_t|\mathcal{O}|\Psi_t\rangle}{\langle\Psi_t|\Psi_t\rangle}=
\frac{\langle\Psi_0|e^{\ri{\cal H}_{\rm SG}t}\mathcal{O}e^{-\ri{\cal H}_{\rm SG}t}|\Psi_0\rangle}{\langle\Psi_0|\Psi_0\rangle}
\label{eq:At}
\end{equation}
in powers of the quench matrix $K^{ab}$, which physically corresponds to an expansion in the density of post-quench excitations. The operator $\mathcal{O}$ is assumed to be semi-local with respect to the fundamental field creating solitons and antisolitons, and to possess Lorentz spin $s$. This is reflected in its form factors and the axioms which they must satisfy. For completeness we state them in App.~\ref{sec:FFaxioms}. For the operator $\mathcal{O}=e^{\ri\alpha\Phi}$ one has $l_\pm^\alpha=e^{\pm 2\pi\ri\alpha/\beta}$ and Lorentz spin $s=0$. Thus for $\alpha=\beta/2$, the semi-locality factor simplifies to $l_a^{\beta/2}=-1$ for both solitons and antisolitons. The initial state $\ket{\Psi_0}$ is assumed to be of the form \eqref{psi0} with a suitable ultra-violet regularisation provided for example by introducing an extrapolation time \eqref{extrapolationtime}.

First we expand the initial states in both the numerator and denominator of \eqref{eq:At} to obtain the formal expansions
\begin{eqnarray}
  \langle\Psi_t|\mathcal{O}|\Psi_t\rangle&=&
  \sum_{M,N=0}^\infty\int_{0}^\infty
  \frac{d\theta_1'\ldots d\theta_M'}{M!(2\pi)^M}
  \frac{d\theta_1\ldots d\theta_N}{N!(2\pi)^N}
  \prod_{i=1}^M \bigl(K^{a_ib_i}(\theta_i')\bigr)^*\,\prod_{j=1}^N K^{c_jd_j}(\theta_j)\,
  e^{2\Delta \ri t\sum_i\cosh\theta_i'}\,e^{-2\Delta \ri t\sum_j\cosh\theta_j}\nonumber\\*
  & &\hspace{20mm}\times_{b_1a_1\ldots b_Ma_M}
  \bra{\theta_1',-\theta_1',\ldots,\theta_M',-\theta_M'}\mathcal{O}
  \ket{-\theta_N,\theta_N,\ldots,-\theta_1,\theta_1}_{c_Nd_N\ldots c_1d_1}\\
  &\equiv&\sum_{M,N=0}^\infty C_{2M,2N}(t)
  \label{eq:Cmn}
\end{eqnarray}
as well as 
\begin{eqnarray}  
  \bra{\Psi_0}\Psi_0\rangle&=&
  1+\sum_{N=1}^\infty\int_{0}^\infty
  \frac{d\theta_1'\ldots d\theta_N'}{N!(2\pi)^N}
  \frac{d\theta_1\ldots d\theta_N}{N!(2\pi)^N}\prod_{i=1}^N \bigl(K^{a_ib_i}(\theta_i')\bigr)^*\,K^{c_id_i}(\theta_i)\nonumber\\*
  & &\qquad\times_{b_1a_1\ldots b_Na_N}
  \bra{\theta_1',-\theta_1',\ldots,\theta_N',-\theta_N'}-\theta_N,\theta_N,\ldots,-\theta_1,\theta_1\rangle_{c_Nd_N\ldots c_1d_1}\\*
  &\equiv&\sum_{N=0}^\infty Z_{2N}.\label{eq:Z2n}
\end{eqnarray}
Note that the indices $2M$ and $2N$ correspond to the number of
particles originating from the left and right initial state
respectively. Furthermore, here and in the following we implicitly sum
over all indices $a_i,b_i,c_i,d_i=\pm$ appearing twice. The expansion
\eqref{eq:Z2n} implies 
\begin{equation}
  \frac{1}{\braket{\Psi_0 | \Psi_0 }}=1-Z_2+Z_2^2-Z_4+\mathcal{O}(K^6).
  \label{eq:1BB}
\end{equation}
We note that \eqref{eq:1BB} defines linked clusters, i.e. it identifies the parts of the numerator in \eqref{eq:At}
that diverge in the infinite volume.

The matrix elements in the terms $C_{2M,2N}(t)$ and $Z_{2N}$ possess
kinematical poles whenever $\theta_i'=\theta_j$ and therefore have to
be regularized. This is done following Smirnov\cite{Smirnov92book} by
shifting the rapidities of the, say, outgoing states into the complex
plane as outlined in detail in Appendix~\ref{sec:Smirnov}. After this
regularisation the denominator and numerator in \eqref{eq:At} still
contain divergencies due to the intertwining of particles with
rapidities $\theta_i$ and $-\theta_i$ in the initial state
\eqref{psi0}. These divergences are a consequence of working in the
infinite volume and have to be canceled against each other. In order
to exhibit these cancellations explicitly we use the
$\kappa$-regularisation scheme originally introduced in the study of
finite-temperature correlation functions\cite{finiteT} and recently
generalized to the quench dynamics of Ising
systems\cite{CEF,SE:2012}. The details of this are given in
Appendix~\ref{sec:kappascheme}. 

Combining these two regularisations the expansions \eqref{eq:Cmn} and
\eqref{eq:Z2n} are finite but may contain terms proportional to the
system size, $\propto L^k$, that diverge in the infinite-volume
limit. However, considering their ratio and expanding again formally
in powers of the quench matrix we find 
\begin{equation}
\frac{\langle\Psi_t|\mathcal{O}|\Psi_t\rangle}{\langle\Psi_t|\Psi_t\rangle}=\frac{\sum_{M,N=0}^\infty C_{2M,2N}(t)}{\sum_{N=0}^\infty Z_{2N}}\equiv
\sum_{M,N=0}^\infty D_{2M,2N}(t),
\label{eq:Dmn}
\end{equation}
where all terms $\propto L^k$ with $k\ge 1$ cancel each other and the
remaining functions $D_{2M,2N}(t)$ are finite in the infinite-volume
limit. This is analogous to a linked-cluster expansion of
finite-temperature correlators\cite{finiteT,finiteT2}. In the
following we calculate the first few terms of this expansion,
specifically the leading terms up to $\mathcal{O}(K^4)$ in the
expansion \eqref{eq:Dmn} for the operator
$\mathcal{O}=e^{\ri\beta\Phi/2}$, and the terms up to
$\mathcal{O}(K^2)$ for the more general operator
$\mathcal{O}=e^{\ri\alpha\Phi}$ in Section~\ref{sec:LCalpha}.  

\subsection{Terms in $\boldsymbol{\mathcal{O}(K^0)}$ and $\boldsymbol{\mathcal{O}(K)}$}\label{sec:K0K1}
The terms in these orders do not contain any divergencies and can be straightforwardly calculated with the results
\begin{eqnarray}
D_{00}(t)&=&C_{00}(t)=\mathcal{G}_{\beta/2},\\
D_{20}(t)&=&C_{20}(t)=\int_0^\infty \frac{d\theta}{2\pi}\,\bigl(K^{ab}(\theta)\bigr)^*\,
  f_{ab}^{-\beta/2}(-\theta,\theta)\,e^{2\Delta \ri t\cosh\theta},\\
D_{02}(t)&=&C_{02}(t)=\int_0^\infty \frac{d\theta}{2\pi}\,K^{ab}(\theta)\,f_{ab}^{\beta/2}(-\theta,\theta)\,e^{-2\Delta \ri t\cosh\theta}.
\end{eqnarray}
Here $f^{\pm\beta/2}_{ab}(\theta_1,\theta_2)=\bra{0}e^{\pm \ri\beta\Phi/2}\ket{\theta_1,\theta_2}_{ab}$ is the two-particle form factor\cite{2pFF} of $e^{\pm\ri\beta\Phi/2}$, which we state for completeness in Appendix~\ref{sec:app2pFF}. The long-time behaviour of $D_{20}(t)+D_{02}(t)$ is given by
\begin{equation}
D_{20}(t)+D_{02}(t)\propto\frac{\cos(2\Delta t)}{(\Delta t)^{3/2}}.
\end{equation}

\subsection{Terms in $\boldsymbol{\mathcal{O}(K^2)}$}
In this order there exist three terms. The first two originate from $C_{40}(t)$ and $C_{04}(t)$, which are completely regular. These two terms simply yield sub-leading corrections to $D_{20}(t)+D_{02}(t)$ stated above and will not be considered in the following. 

The third term, however, contains kinematical poles as well as infinite-volume divergencies. From the formal expansion \eqref{eq:Dmn} we deduce that
\begin{equation}
D_{22}(t)=C_{22}(t)-Z_2\,C_{00}(t)=C_{22}(t)-Z_2\,\mathcal{G}_{\beta/2}.
\label{eq:D22}
\end{equation}
Here the first non-trivial term in the expansion \eqref{eq:Z2n} reads using the infinite-volume regularisation discussed in Appendix~\ref{sec:kappascheme}
\begin{eqnarray}
Z_2&\equiv&
  \int d\kappa\,P(\kappa)\int_0^\infty\frac{d\theta'd\theta}{(2\pi)^2}\,\bigl(K^{ab}(\theta')\bigr)^*\,K^{cd}(\theta)
   \,_{ba}\!\bra{\theta',-\theta'}-\theta+\kappa,\theta+\kappa\rangle_{cd}\\
  &=&\int d\kappa\,P(\kappa)\,\delta(-2\kappa)\int_0^\infty d\theta\,\bigl(K^{ab}(\theta+\kappa)\bigr)^*\,K^{ab}(\theta)
  =\frac{L}{2}\int_0^\infty d\theta\,G(\theta),\label{eq:Z2calc2}
\end{eqnarray}
where we implicitly sum over $a,b=\pm$ and used the notation \eqref{III:16}, i.e. $G(\theta)=\big| K^{ab}(\theta)\big|^2$. The infinite-volume divergence is now clearly exhibited. 

On the other hand, we have to consider 
\begin{equation}
    C_{22}(t)=\int_{0}^\infty\frac{d\theta'd\theta}{(2\pi)^2}\,\bigl(K^{ab}(\theta')\bigr)^*\,K^{cd}(\theta)
    \,_{ba}\!\bra{\theta',-\theta'}e^{\ri\beta\Phi/2}\ket{-\theta,\theta}_{cd}\,e^{2\Delta \ri t(\cosh\theta'-\cosh\theta)}.
    \label{eq:C22}
\end{equation}
Again shifting the rapidities $\pm\theta$ by the auxiliary parameter $\kappa$ and regularising the form factor by analytically continuing in the rapidities $\pm\theta'$ as discussed in Appendix~\ref{sec:Smirnov} we obtain
\begin{eqnarray}
   \,_{ba}\!\bra{\theta',-\theta'}e^{\ri\beta\Phi/2}\ket{-\theta,\theta}_{cd}&=&
  \,_{ba}\!\bra{\theta',-\theta'}e^{\ri\beta\Phi/2}\ket{-\theta+\kappa,\theta+\kappa}_{cd}\label{eq:regC22}\\
&=&(2\pi)^2\,\mathcal{G}_{\beta/2}\,\delta_a^c\,\delta_b^d\,\delta(-2\kappa)\,
   \delta(\theta'-\theta+\kappa)\label{eq:C22reg1}\\*
   &&+2\pi\,S_{ba}^{ef}(2\theta-2\kappa)\,S_{cd}^{fh}(-2\theta)\,\delta(\theta'-\theta+\kappa)\,
   _e\!\bra{\theta'+\ri 0}e^{\ri\beta\Phi/2}\ket{\theta+\kappa}_h\label{eq:C22reg2}\\*
   &&+2\pi\,\delta_b^d\,\delta(\theta'-\theta-\kappa)\,_a\!\bra{-\theta'+\ri 0}e^{\ri\beta\Phi/2}\ket{-\theta+\kappa}_c\label{eq:C22reg3}\\*
   &&+\,_{ba}\!\bra{\theta'+\ri 0,-\theta'+\ri 0}e^{\ri\beta\Phi/2}\ket{-\theta+\kappa,\theta+\kappa}_{cd},\label{eq:C22reg4}
\end{eqnarray}
where we have used \eqref{eq:Sunitarity} and \eqref{eq:Srelations}. Furthermore, we have dropped two contributions which result in expressions like $\delta(-\theta'-\theta-\kappa)\delta(\theta'+\theta-\kappa)$, since they yield only terms $\propto\kappa$ vanishing after multiplying with $P(\kappa)$ and integrating over $\kappa$ (see Ref.~\onlinecite{SE:2012} for a detailed discussion of similar terms for the Ising field theory). 

Plugging \eqref{eq:regC22} into \eqref{eq:C22} we obtain three different terms which can be thougth of as being disconnected, semi-connected and fully connected. Specifically we have
\begin{equation}
C_{22}(t)=C_{22}^0(t)+C_{22}^1(t)+C_{22}^2(t).
\end{equation}
In the following we will analyse each term separately.

\subsubsection{Disconnected contribution: $C_{22}^0$}
The fully disconnected contribution follows from \eqref{eq:C22reg1}. Explicitly we find
\begin{equation}
    C_{22}^0(t)=\mathcal{G}_{\beta/2}\,\delta(-2\kappa)\int_0^\infty d\theta\,\big|K^{ab}(\theta)\big|^2
    =\mathcal{G}_{\beta/2}\,\delta(-2\kappa)\int_0^\infty d\theta\,G(\theta).
\end{equation}
Now multiplying with $P(\kappa)$ and integrating over $\kappa$ we obtain
\begin{equation}
    C_{22}^0(t)=\frac{L}{2}\,\mathcal{G}_{\beta/2}\int_0^\infty d\theta\,G(\theta)
    =Z_2\,\mathcal{G}_{\beta/2},
\end{equation}
i.e. this cancels the second term in \eqref{eq:D22}. The remainder of $D_{22}(t)$ does not contain terms $\propto\delta(\kappa)$.

\subsubsection{Semi-connected contribution: $C_{22}^1$}
Next consider the terms \eqref{eq:C22reg2} and \eqref{eq:C22reg3}. We find
\begin{eqnarray}
	C_{22}^1(t)&=&
	\int_0^\infty\frac{d\theta}{2\pi}\,\bigl(K^{ab}(\theta)\bigr)^*\,K^{cd}(\theta+\kappa)\,
	S_{ba}^{ef}(2\theta)\,S_{cd}^{fh}(-2\theta-2\kappa)\,
	\,_e\!\bra{\theta+\ri 0}e^{\ri\beta\Phi/2}\ket{\theta+2\kappa}_h\,e^{2\Delta \ri t[\cosh\theta-\cosh(\theta+\kappa)]}\qquad\nonumber\\*
	&&
	+\int_0^\infty\frac{d\theta}{2\pi}\,\bigl(K^{ab}(\theta)\bigr)^*\,K^{cb}(\theta-\kappa)\,
	\,_a\!\bra{-\theta+\ri 0}e^{\ri\beta\Phi/2}\ket{-\theta+2\kappa}_c\,e^{2\Delta \ri t[\cosh\theta-\cosh(\theta-\kappa)]}\\
	&=&-\,f^{\beta/2}_{\bar{e}h}(\ri\pi+\ri 0,2\kappa)
	\int_0^\infty\frac{d\theta}{2\pi}\,\bigl(K^{ab}(\theta)\bigr)^*\,K^{cd}(\theta+\kappa)\,
	S_{ba}^{ef}(2\theta)\,S_{cd}^{fh}(-2\theta-2\kappa)\,
	e^{2\Delta \ri t[\cosh\theta-\cosh(\theta+\kappa)]}\nonumber\\*
	&&-\,f^{\beta/2}_{\bar{a}c}(\ri\pi+\ri 0,2\kappa)
	\int_0^\infty\frac{d\theta}{2\pi}\,\bigl(K^{ab}(\theta)\bigr)^*\,K^{cb}(\theta-\kappa)\,
	e^{2\Delta \ri t[\cosh\theta-\cosh(\theta-\kappa)]}.
\end{eqnarray}
Now we use the annihilation pole axiom (see Appendix~\ref{sec:FFaxioms})
\begin{equation}
f^{\beta/2}_{ab}(\ri\pi+\ri 0,2\kappa)=-f^{\beta/2}_{ba}(2\kappa+2\ri\pi,\ri\pi+\ri 0)=
-2\ri \frac{C_{ab}\,\mathcal{G}_{\beta/2}}{2\kappa-\ri 0}-F^{\beta/2}_{ba}(\kappa),
\label{eq:poleFF2}
\end{equation}
where $F^{\beta/2}_{ba}(\kappa)$ is analytic for $\kappa\to 0$. Finally we expand $C_{22}^1$ in $\kappa$ and keep only the terms $\propto 1/\kappa$ and $\propto t$, since all terms $\propto\kappa^n$ with $n\ge 1$ vanish in the infinite-volume limit. We obtain
\begin{eqnarray}
	C_{22}^1(t)&=&\frac{2\ri\,\mathcal{G}_{\beta/2}}{2\kappa-\ri 0}
	\int_0^\infty\frac{d\theta}{2\pi}\,\bigl(K^{ab}(\theta)\bigr)^*\,K^{cd}(\theta)\,
	S_{ab}^{fe}(2\theta)\,S^{cd}_{fe}(-2\theta)\,
	\bigl[1-2\Delta \ri t\kappa\sinh\theta\bigr]\nonumber\\*
	&&+\frac{2\ri\,\mathcal{G}_{\beta/2}}{2\kappa-\ri 0}
	\int_0^\infty\frac{d\theta}{2\pi}\,\bigl|K^{ab}(\theta)\bigr|^2\,
	\bigl[1+2\Delta \ri t\kappa\sinh\theta\bigr]+\ldots\label{eq:C221}
\end{eqnarray}
Obviously this contains static terms $\propto 1/\kappa$, i.e. there are still remnants of infinite-volume divergencies. As we are going to show in the next subsection, these terms will cancel against contributions from the fully connected piece $C_{22}^2$. 

\subsubsection{Fully connected contribution: $C_{22}^2$}
Finally let us consider the term \eqref{eq:C22reg4}. Plugging it into \eqref{eq:C22} yields 
\begin{equation}
    C_{22}^2(t)=\int_{0}^\infty\frac{d\theta'd\theta}{(2\pi)^2}\,\bigl(K^{ab}(\theta')\bigr)^*\,K^{cd}(\theta)
    f^{\beta/2}_{\bar{b}\bar{a}cd}(\theta'+\ri\pi+\ri 0,-\theta'+\ri\pi+\ri 0,-\theta+\kappa,\theta+\kappa)
    \,e^{2\Delta \ri t(\cosh\theta'-\cosh\theta)}.
\end{equation}
The form factor possesses annihilation poles at the positions $\theta=\pm\theta'-\kappa+\ri0$ and $\theta=\mp\theta'+\kappa-\ri0$. We deal with these poles by shifting the contour of integration for $\theta$ to the lower half plane. Doing this we pick up a pole at $\theta=\theta'+\kappa-\ri 0$ and obtain
\begin{equation}
C_{22}(t)=C_{22}'(t)+C_{22}^p(t),
\end{equation}
where
\begin{equation}
C_{22}'(t)=\int_{0}^\infty\frac{d\theta'}{2\pi}\int_{\gamma_-}\frac{d\theta}{2\pi}\,\bigl(K^{ab}(\theta')\bigr)^*\,K^{cd}(\theta)
    f^{\beta/2}_{\bar{b}\bar{a}cd}(\theta'+\ri\pi+\ri 0,-\theta'+\ri\pi+\ri 0,-\theta+\kappa,\theta+\kappa)
    \,e^{2\Delta \ri t(\cosh\theta'-\cosh\theta)}
    \label{eq:C22prime}
\end{equation}
with the contour of integration $\gamma_-$ parametrized by $(0<\phi_0\le\pi/4$)
\begin{equation}
\gamma_-(s)=\left\{\begin{array}{ll}-\ri s,& 0\le s\le \phi_0,\\
(s-\phi_0)-\ri\phi_0,& \phi_0\le s<\infty.
\end{array}\right.
\label{eq:contour}
\end{equation}  
and
\begin{eqnarray}
C_{22}^p(t)&=&
	-\ri\int_{0}^\infty\frac{d\theta'}{2\pi}\,\bigl(K^{ab}(\theta')\bigr)^*\,K^{cd}(\theta'+\kappa)
   	\,e^{2\Delta \ri t[\cosh\theta'-\cosh(\theta'+\kappa)]}\nonumber\\*
   & &\qquad\qquad\qquad\times
   \text{Res}\bigl[f^{\beta/2}_{\bar{b}\bar{a}cd}(\theta'+\ri\pi+\ri 0,-\theta'+\ri\pi+\ri 0,-\theta+\kappa,\theta+\kappa),
   \theta=\theta'+\kappa-\ri 0\bigr].
\end{eqnarray}
The leading long-time behaviour comes from the pole contribution
$C_{22}^p(t)$, while $C_{22}'(t)$ yields a correction and will not be
considered further. The corresponding term for the time evolution of
the order parameter field in the Ising field theory gives rise to a
correction $C_{22}'(t)\sim 1/t$ at late times~\cite{SE:2012}. In
order to evaluate $C_{22}^p(t)$ we first have to determine the
appearing residue. This is done using the annihilation pole axiom
stated in Appendix~\ref{sec:FFaxioms} as well as
$\text{Res}\bigl[f(z),z=z_0\bigr]=-\text{Res}\bigl[f(-z),z=-z_0\bigr]$,
resulting in 
\begin{eqnarray}
   &&\text{Res}\bigl[f^{\beta/2}_{\bar{b}\bar{a}cd}(\theta'+\ri\pi+\ri 0,-\theta'+\ri\pi+\ri 0,-\theta+\kappa,\theta+\kappa),
   \theta=\theta'+\kappa-\ri 0\bigr]\nonumber\\
   &&\qquad=
   \text{Res}\bigl[S_{\bar{b}\bar{a}}^{ef}(2\theta')\,S_{cd}^{gh}(-2\theta)\,
   f^{\beta/2}_{fehg}(-\theta'+\ri\pi+\ri 0,\theta'+\ri\pi+\ri 0,\theta+\kappa,-\theta+\kappa),
   \theta=\theta'+\kappa-\ri 0\bigr]\nonumber\\
   &&\qquad=-S_{\bar{b}\bar{a}}^{ef}(2\theta')\,S_{cd}^{gh}(-2\theta'-2\kappa)\,
   \text{Res}\bigl[
   f^{\beta/2}_{gfeh}(-\theta+\kappa+2\ri\pi,-\theta'+\ri\pi+\ri 0,\theta'+\ri\pi+\ri 0,\theta+\kappa),
   \theta=\theta'+\kappa-\ri 0\bigr]\nonumber\\
   &&\qquad=\ri C_{gk}\,S_{\bar{b}\bar{a}}^{ef}(2\theta')\,S_{cd}^{gh}(-2\theta'-2\kappa)\,
   f^{\beta/2}_{ij}(\ri\pi+\ri 0,2\kappa)\,
   \bigl[\delta_e^i\delta_h^j\delta_f^k+S_{fe}^{li}(-2\theta')S_{lh}^{kj}(-2\theta'-2\kappa+\ri\pi)\bigr].
\end{eqnarray}
Hence we get
\begin{eqnarray}
C_{22}^p(t)&=&C_{gk}\,f^{\beta/2}_{ij}(\ri\pi+\ri 0,2\kappa)
	\int_{0}^\infty\frac{d\theta}{2\pi}\,\bigl(K^{ab}(\theta)\bigr)^*\,K^{cd}(\theta+\kappa)\,
	S_{\bar{b}\bar{a}}^{ef}(2\theta)\,S_{cd}^{gh}(-2\theta-2\kappa)\nonumber\\*
   &&\qquad\qquad\qquad\times
   \bigl[\delta_e^i\delta_h^j\delta_f^k+S_{fe}^{li}(-2\theta)S_{lh}^{kj}(-2\theta-2\kappa+\ri\pi)\bigr]\,
	e^{2\Delta \ri t[\cosh\theta-\cosh(\theta+\kappa)]},
\end{eqnarray}
which is further simplified by expanding in $\kappa$ up to terms $\propto 1/\kappa$ and $\propto t$
and using \eqref{eq:poleFF2}. This yields
\begin{eqnarray}
C_{22}^p(t)&=&
	-\ri\frac{2\mathcal{G}_{\beta/2}}{2\kappa-\ri 0}
	\int_{0}^\infty\frac{d\theta}{2\pi}\,\bigl(K^{ab}(\theta)\bigr)^*\,K^{cd}(\theta)\,
	S_{\bar{b}\bar{a}}^{ef}(2\theta)\,S_{cd}^{\bar{f}\bar{e}}(-2\theta)
	\bigl[1-2\Delta \ri t\kappa\sinh\theta\bigr]\nonumber\\*
   &&-\ri\frac{2\mathcal{G}_{\beta/2}}{2\kappa-\ri 0} 
	\int_{0}^\infty\frac{d\theta}{2\pi}\,\bigl(K^{ab}(\theta)\bigr)^*\,K^{cd}(\theta)
	\bigl[1-2\Delta \ri t\kappa\sinh\theta\bigr]\nonumber\\*
	&&\hspace{50mm}\times S_{\bar{b}\bar{a}}^{ef}(2\theta)\,S_{cd}^{gh}(-2\theta)\,
	S_{fe}^{l\bar{j}}(-2\theta)S_{lh}^{\bar{g}j}(-2\theta+\ri\pi)+\ldots
\end{eqnarray}
Now using the crossing relation \eqref{eq:Scrossing} as well as \eqref{eq:Srelations} we obtain
\begin{eqnarray}
C_{22}^p(t)&=&
	-\ri\frac{2\mathcal{G}_{\beta/2}}{2\kappa-\ri 0} 
	\int_{0}^\infty\frac{d\theta}{2\pi}\,\bigl(K^{ab}(\theta)\bigr)^*\,K^{cd}(\theta)\,
	S_{ab}^{ef}(2\theta)\,S^{cd}_{ef}(-2\theta)\,
	\bigl[1-2\Delta \ri t\kappa\sinh\theta\bigr]\nonumber\\*
   &&-\ri\frac{2\mathcal{G}_{\beta/2}}{2\kappa-\ri 0} 
	\int_{0}^\infty\frac{d\theta}{2\pi}\,\bigl|K^{ab}(\theta)\bigr|^2\bigl[1-2\Delta \ri t\kappa\sinh\theta\bigr]+\ldots\label{eq:C22p}
\end{eqnarray}

\subsubsection{Final result for $D_{22}$}
We are now in a position to write down the final result for \eqref{eq:D22}. As already discussed the terms $\propto Z_2$ cancel and we are left with
\begin{equation}
D_{22}(t)=C_{22}^1(t)+C_{22}^p(t)+C_{22}'(t).
\end{equation}
The leading long-time behaviour is given by the first and second term which we will analyse in the following. In contrast, the last term stated in \eqref{eq:C22prime} constitutes a sub-leading correction and will not be considered further. 

For the leading terms \eqref{eq:C221} and \eqref{eq:C22p} we find that the ill-defined terms $\propto 1/\kappa$ as well as the terms containing scattering matrices cancel each other. We are thus left with
\begin{equation}
C_{22}^1(t)+C_{22}^p(t)=-\frac{2\kappa}{2\kappa-i 0}\,4\mathcal{G}_{\beta/2}\Delta t
	\int_{0}^\infty\frac{d\theta}{2\pi}\,G(\theta)\sinh\theta.
\end{equation}
Now employing the $\kappa$-regularisation scheme using
\begin{equation}
\int d\kappa\,P(\kappa)\,\frac{2\kappa}{2\kappa-\ri 0}=1
\end{equation}
we obtain the final result
\begin{equation}
D_{22}(t)=-\mathcal{G}_{\beta/2}\frac{t}{\tau}+C_{22}'(t),
\label{eq:D22finalresult}
\end{equation}
with the decay rate in $\mathcal{O}(K^2)$ given by
\begin{equation}
\tau^{-1}=\frac{2 \Delta}{\pi} \int_{0}^{\infty}d\theta\,G(\theta) \sinh{\theta},
\label{eq:taurate}
\end{equation}
which was already obtained in the representative state approach \eqref{eq:decaytime} with $G(\theta)=\sum_{ab}\bigl|K^{ab}(\theta)\big|^2$. 

\subsection{Leading long-time behaviour in $\boldsymbol{\mathcal{O}(K^4)}$ and final result}\label{sec:LCK4}
As we have derived above, the contributions in $\mathcal{O}(K^2)$ grow linear in time, see Eq.~\eqref{eq:D22finalresult}. The same behaviour was observed in the linked-cluster expansion performed for Ising systems~\cite{CEF,SE:2012}, where it was also shown that the leading behaviour in $\mathcal{O}(K^{2n})$ is given by terms growing as $t^n$ at late times. While an analysis of the leading long-time behaviour at arbitrary order in the quench matrix is out of reach in the sine-Gordon model, we have extracted the long-time asymptotics at $\mathcal{O}(K^4)$. As we show in Appendix~\ref{sec:appK4} it is given by
\begin{equation}
D_{44}(t)\sim C_{44}(t)=\frac{2\Delta^2t^2}{\pi^2}\mathcal{G}_{\beta/2}
\int_0^\infty d\theta_1d\theta_2\,G(\theta_1)\,G(\theta_2)\,\sinh\theta_1\,\sinh\theta_2+\ldots
=\frac{\mathcal{G}_{\beta/2}}{2}\left(\frac{t}{\tau}\right)^2+\ldots
\label{eq:D44asymptotics}
\end{equation}
In the derivation of \eqref{eq:D44asymptotics} we have assumed the
absence of diagonal terms in the quench matrix,
i.e. $K^{aa}(\theta)=0$ (this is the case for fixed boundary
conditions). Furthermore, the derivation of Appendix~\ref{sec:appK4}
is restricted to terms quadratic in $t$, i.e. we are not able to
extract corrections to the decay time $\tau$ in $\mathcal{O}(K^4)$.  

Together with the results derived above we thus obtain for the leading behaviour up to $\mathcal{O}(K^4)$ 
\begin{eqnarray}
\frac{\langle\Psi_t|e^{\ri\beta\Phi/2}|\Psi_t\rangle}{\langle\Psi_t|\Psi_t\rangle}&=&
\mathcal{G}_{\beta/2}\left[1-\frac{t}{\tau}+\frac{1}{2}\left(\frac{t}{\tau}\right)^2+\ldots\right]=\mathcal{G}_{\beta/2}e^{-t/\tau}(1+\ldots),\\*
\tau^{-1}&=&\frac{2 \Delta}{\pi} \int_{0}^{\infty}d\theta\,G(\theta) \sinh{\theta}+\mathcal{O}(K^4).
\label{eq:decaytimeLCE}
\end{eqnarray}
As we clearly see the leading terms up to the order considered here are consistent with an exponential decay of the one-point function also deduced from the representative-state approach in Eq.~\eqref{IV:28}. The dots represent corrections which are, up to $\mathcal{O}(K^2)$ given by $D_{20}(t)+D_{02}(t)\propto\cos(2\Delta t)/(\Delta t)^{3/2}$, $D_{40}(t)+D_{04}(t)$ and $C_{22}'(t)$.

\subsection{Time evolution for $\boldsymbol{\alpha\neq\beta/2}$}\label{sec:LCalpha}
The calculation of the time evolution of $e^{\ri\alpha\Phi}$ for $\alpha\neq\beta/2$ is significantly more involved since the semi-locality factor $l_a^\alpha$ now depends on the index $a$. Up to $\mathcal{O}(K)$ this complications do not show up and the results are simply given by the ones of Section~\ref{sec:K0K1} with the normalisation\cite{LZ} $\mathcal{G}_\alpha$ and two-particle form factors\cite{Lukyanov97} $f^\alpha_{ab}(\theta_1,\theta_2)$. 

In $\mathcal{O}(K^2)$, however, there appear more pronounced
changes. Starting with \eqref{eq:C22} the form factor can still be
regularized using \eqref{eq:C22reg1}--\eqref{eq:C22reg4} yielding a
disconnected, semi-connected and fully connected contribution. The
first is essentially unchanged,
$C_{22}^0(t)=Z_2\,\mathcal{G}_\alpha$. Differences appear in
$C_{22}^1$ due to the non-trivial semi-locality factors in the
crossing relation \eqref{eq:crossing} and the annihilation pole
axiom. We find after using  
\begin{equation}
f^\alpha_{ab}(\ri\pi+\ri 0,2\kappa)=l_{\bar{b}}^\alpha f^\alpha_{ba}(2\kappa+2\ri\pi,\ri\pi+\ri 0)=
\ri(l_{\bar{b}}^\alpha-1) \frac{C_{ab}\,\mathcal{G}_\alpha}{2\kappa-\ri 0}+l_{\bar{b}}^\alpha {F_{ba}(\kappa),}
\label{eq:poleFF2alpha}
\end{equation}
where $F_{ba}(\kappa)$ is again analytic for $\kappa\to 0$, and expanding in $\kappa$ keeping only the terms $\propto 1/\kappa$ and $\propto t$
\begin{eqnarray}
	C_{22}^1(t)&=&\ri(1-l_e^\alpha)\frac{\mathcal{G}_\alpha}{2\kappa-\ri 0}
	\int_0^\infty\frac{d\theta}{2\pi}\,\bigl(K^{ab}(\theta)\bigr)^*\,K^{cd}(\theta)\,
	S_{ab}^{fe}(2\theta)\,S^{cd}_{fe}(-2\theta)\,
	\bigl[1-2\Delta \ri t\kappa\sinh\theta\bigr]\nonumber\\*
	&&+\ri(1-l_a^\alpha)\frac{\mathcal{G}_\alpha}{2\kappa-\ri 0}
	\int_0^\infty\frac{d\theta}{2\pi}\,\bigl|K^{ab}(\theta)\bigr|^2\,
	\bigl[1+2\Delta \ri t\kappa\sinh\theta\bigr]+\ldots\label{eq:C221alpha}
\end{eqnarray}
For the fully connected contribution we again have to determine the residue, which is now given by
\begin{eqnarray}
   &&\text{Res}\bigl[f^\alpha_{\bar{b}\bar{a}cd}(\theta'+\ri\pi+\ri 0,-\theta'+\ri\pi+\ri 0,-\theta+\kappa,\theta+\kappa),
   \theta=\theta'+\kappa-\ri 0\bigr]\nonumber\\*
   &&\qquad=-\ri C_{gk}l_{\bar{g}}^\alpha\,S_{\bar{b}\bar{a}}^{ef}(2\theta')\,S_{cd}^{gh}(-2\theta'-2\kappa)\,
   f^\alpha_{ij}(\ri\pi+\ri 0,2\kappa)\,
   \bigl[\delta_e^i\delta_h^j\delta_f^k-
   l_g^\alpha S_{fe}^{li}(-2\theta')S_{lh}^{kj}(-2\theta'-2\kappa+\ri\pi)\bigr].
\end{eqnarray}
Now performing the same steps as previously we obtain for the expansion in $\kappa$
\begin{eqnarray}
C_{22}^p(t)&=&
	-\ri l_a^\alpha l_b^\alpha l_{\bar{e}}^\alpha (l_{\bar{f}}^\alpha -1)\frac{\mathcal{G}_\alpha}{2\kappa-\ri 0} 
	\int_{0}^\infty\frac{d\theta}{2\pi}\,\bigl(K^{ab}(\theta)\bigr)^*\,K^{cd}(\theta)\,
	S_{ab}^{ef}(2\theta)\,S^{cd}_{ef}(-2\theta)
	\bigl[1-2\Delta \ri t\kappa\sinh\theta\bigr]\\
   &&+\ri l_a^\alpha l_b^\alpha (l_{\bar{b}}^\alpha-1)\frac{\mathcal{G}_\alpha}{2\kappa-\ri 0}
	\int_{0}^\infty\frac{d\theta}{2\pi}\,\bigl|K^{ab}(\theta)\bigr|^2\,
	\bigl[1-2\Delta \ri t\kappa\sinh\theta\bigr]+\ldots
\end{eqnarray}
Summing these two contributions we see that the terms $\propto 1/\kappa$ cancel, i.e. the result in this order is well-defined in the infinite-volume limit, and we obtain
\begin{equation}
D_{22}(t)=-\mathcal{G}_\alpha \frac{t}{\tau_\alpha},\quad
\tau_\alpha^{-1}=\Delta\int_0^\infty\frac{d\theta}{2\pi}
\Bigl[\bigl|K^{aa}(\theta)\bigr|^2\,(1-l_a^\alpha)^2+2\bigl|K^{a\bar{a}}(\theta)\bigr|^2\,(1-l_a^\alpha)\Bigr]
\sinh\theta.
\label{D22t}
\end{equation}
For $\alpha=\beta/2$ this simplifies to our previous result
\eqref{eq:taurate}. However, for the local operator
$e^{\ri\beta\Phi}$ studied by Gritsev et al.~\cite{gritsev08b} the
contribution \fr{D22t} vanishes, and our results do not
provide information on the time evolution of $F_{\Psi_0}^\beta(t)$.  

For initial states fulfilling \eqref{eq:DirichletK} (the state
corresponding to Dirichlet boundary conditions being an example)
this can be simplified further
to  
\be
\tau_\alpha^{-1}=\frac{4\Delta}{\pi}\,\sin^2\left(\frac{\pi\alpha}{\beta}\right)\int_0^\infty d\theta\,\bigl|K(\theta)\bigr|^2\,\sinh\theta,
\ee
where $K(\theta)=K^{a\bar{a}}(\theta)=K^{\bar{a}a}(\theta)$ is independent of $a$.
 
\section{Summary and Conclusions}\label{sec:conclusions}
We have considered the time evolution of the semi-local operator $e^{\ri\beta\Phi/2}$ after an ``integrable'' quench to the sine-Gordon model. The system was assumed to be initialised in an integrable boundary state of the form
\be
|\Psi_0\rangle=\exp\left(\int_0^\infty \frac{d\theta}{2\pi} 
K^{ab}(\theta)Z^\dagger_a(-\theta)Z^\dagger_b(\theta)\right)|0\rangle,
\label{BSsumm}
\ee
and we focussed on the behaviour of the expectation value
$\langle\Psi_0|e^{\text{i}\beta\Phi(t)/2}|\Psi_0\rangle/\langle\Psi_0|\Psi_0\rangle$
at late times after the quench. We developed two novel methods to
evaluate this expectation value for the case of a ``small'' quench,
defined as involving small densities of excitations of the post-quench
Hamiltonian. The first approach is based on the concept of a
\emph{representative state}\cite{CE_PRL13}, while the second entails
the generalisation of linked cluster expansions first developed for
studying finite-temperature dynamics in integrable
models\cite{finiteT,finiteT2} to integrable models with non-trivial
scattering out of equilibrium. Our main result is that in the
thermodynamic limit at late times $\Delta t\gg 1$, and for low
densities 
$\delta\ll 1$ 
\begin{equation}
\frac{\langle\Psi_0|e^{\ri\beta \Phi(x,t)/2}|\Psi_0\rangle}
{\langle\Psi_0|\Psi_0\rangle}=\cG_{\beta/2}e^{-{t}/{\tau}}+\ldots,
\label{eq:finalresult}
\end{equation}
where the decay time $\tau$ is given in Eq.~\fr{eq:decaytime}
or~\eqref{eq:decaytimeLCE}.
We note that exponential decay has also been observed in numerical
simulations\cite{Barmettler} of the staggered magnetisation after
interaction quenches in the XXZ Heisenberg chain.

To the best of our knowledge we provide here the first successful
analytic calculation of the time dependence of a local observable for
a quench to an \emph{interacting} integrable model. In order to be
able to carry out our calculations we required several simplifying
assumptions: 
\begin{enumerate}
\item{} the initial state was taken to be of the form \fr{BSsumm};
\item{} we required the functions $K^{ab}(\theta)$ to be uniformly small, which corresponds to the limit of low densities of excitations after the quench;
\item{} we focussed on the simplest semi-local operator $e^{\ri\beta\Phi/2}$;
\item{} we considered the repulsive regime of the sine-Gordon model;
\item{} the time $t$ was taken to be large.
\end{enumerate}
It clearly would be interesting to go beyond these restrictions. The
treatment of more general vertex operators of the form
$e^{\ri\alpha\Phi}$ appears possible, although the results of
Section~\ref{sec:LCalpha} show that for the local operator
$e^{\ri\beta\Phi}$ the leading term in the linked-cluster expansion
vanishes and thus an analysis of higher orders is necessary.
Similarly we believe that (a combination of) our methods can be
applied with relatively minor extensions to the attractive regime of
the sine-Gordon model. To dispense with our other assumptions appears
significantly more difficult. Going beyond the leading term in the
low-density approximation has proved possible but complicated for the
much simpler case of the transverse field Ising
chain\cite{CE_PRL13,CEF,SE:2012}. To do the same for the
sine-Gordon case appears to be a daunting task. For example, in the
linked-cluster expansion one is faced the problem of late-time
singularities at higher orders in the expansion that may no
longer simply exponentiate. In the representative state approach,
detailed properties of the solutions of the Bethe-Yang equation
entering the Lehmann representation \fr{IV:1} will now play a
role. Finally, initial states of the form \fr{BSsumm} are clearly
rather special \cite{spyros}. It is not generally known how to express
the initial state in terms of Hamiltonian eigenstates after quenching
a system parameter. Given the difficulty of this problem, a more
fruitful avenue of research would be to follow
Ref.~[\onlinecite{FCEC}] and consider special initial states
characterized by low entanglement. 

\acknowledgements
We thank Pasquale Calabrese and Gabor Takacs for helpful
discussions. DS thanks the Institute for Theory of Statistical
Physics, RWTH Aachen University, where parts of this work have been
performed. This work is part of the D-ITP consortium, a program of the
Netherlands Organisation for Scientific Research (NWO) that is funded
by the Dutch Ministry of Education, Culture and Science (OCW). This
work was supported by the EPSRC under grants EP/I032487/1 (FHLE and
BB) and EP/J014885/1 (FHLE and BB). DS was partially supported by the
German Research Foundation (DFG) through the Emmy-Noether Program
under SCHU 2333/2-1.

\appendix
\section{Form-factor axioms}\label{sec:FFaxioms}
For completeness we state here the used form-factor axioms. We follow Delfino\cite{Delfino04}. The $n$-particle form factor of an arbitrary operator $\mathcal{O}$ is defined by 
\begin{equation}
  f^\mathcal{O}_{a_1\ldots a_n}(\theta_1,\ldots,\theta_n)=
  \bra{0}\mathcal{O}\ket{\theta_1,\ldots,\theta_N}_{a_1\ldots a_N}=
  \bra{0}\mathcal{O}\,Z^\dagger_{a_1}(\theta_1)\ldots Z^\dagger_{a_N}(\theta_N)\ket{0}.
  \label{eq:formfactordefinition}
\end{equation}
The creation and annihilation operators for solitons and antisolitons
are $Z_\pm^\dagger(\theta)$ and $Z_\pm(\theta)$ introduced in Section
\ref{sec:integrFT}. The form-factor axioms read:  
\begin{enumerate}
\item The form factors $f^\mathcal{O}_{a_1\ldots a_N}(\theta_1,\ldots,\theta_{N})$ are meromorphic functions of $\theta_N$ in the physical strip $0\le\mathfrak{Im}\,\theta_N\le 2\pi$. There exist only simple poles in this   strip.
\item Scattering axiom:
\begin{equation*}
\begin{split}
&f^\mathcal{O}_{a_1\ldots a_{i}a_{i+1}\ldots a_N}
(\theta_1,\ldots,\theta_{i},\theta_{i+1},\ldots,\theta_N)\\
&\hspace{10mm}
=S_{a_{i}a_{i+1}}^{b_ib_{i+1}}(\theta_{i}-\theta_{i+1})\,
f^\mathcal{O}_{a_1\ldots b_{i+1}b_{i}\ldots a_N}
(\theta_1,\ldots,\theta_{i+1},\theta_{i},\ldots,\theta_N),
\end{split}
\end{equation*}
where $S^{b_{i}b_{i+1}}_{a_ia_{i+1}}(\theta_{i}-\theta_{i+1})$ is the scattering matrix introduced in Section \ref{sec:integrFT}.
\item Periodicity axiom:
\begin{equation*}
f^\mathcal{O}_{a_1\ldots a_N}(\theta_1+2\pi i,\theta_2,\ldots,\theta_N)=
l_{a_1}(\mathcal{O})\,
f^\mathcal{O}_{a_2\ldots a_Na_1}(\theta_2,\ldots,\theta_{N},\theta_1),
\label{periodicity}
\end{equation*}
where $l_{\pm}(\mathcal{O})$ is the mutual semi-locality factor between the operator $\mathcal{O}$ and the fundamental fields $\mathcal{O}_0^{\pm}$. In particular, we have for $\mathcal{O}=e^{\ri\alpha\Phi}$ the factor $l_\pm^\alpha=e^{\pm \ri 2\pi\alpha/\beta}$.
\item Lorentz transformations:
\begin{equation*}
f^\mathcal{O}_{a_1\ldots a_N}(\theta_1+\Lambda,\ldots,\theta_N+\Lambda)=
e^{s(\mathcal{O})\Lambda}\,f^\mathcal{O}_{a_1\ldots a_N}(\theta_1,\ldots,\theta_N),
\end{equation*}
where $s(\mathcal{O})$ denotes the Lorentz spin of $\mathcal{O}$. Here we have $s(e^{\ri\alpha\Phi})=0$, i.e. $\mathcal{O}=e^{\ri\alpha\Phi}$ is a Lorentz scalar.
\item Annihilation pole axiom:
\begin{equation*}
\begin{split}
&\mathrm{Res}\bigl[
f^\mathcal{O}_{aba_1\ldots a_N}(\theta',\theta,\theta_1,\ldots,\theta_N),
\theta'=\theta+\ri\pi\bigr]\\
&\hspace{5mm}=\ri\,C_{ac}\,
f^\mathcal{O}_{b_1\ldots b_N}(\theta_1,\ldots,\theta_N)
\left[\delta_{a_1}^{b_1}\ldots\delta_{a_N}^{b_N}\delta_{b}^{c}-
l_a(\mathcal{O}) S_{b_{\phantom{1}}a_1}^{c_1b_1}(\theta-\theta_1)
S_{c_1a_2}^{c_2b_2}(\theta-\theta_2)
\ldots
S_{c_{N-1}a_N}^{c_{\phantom{N-1}}b_N}(\theta-\theta_N)\right],
\label{apoleaxiom}
\end{split}
\end{equation*}
with the charge conjugation matrix of the sine-Gordon model given by $C_{ab}=\delta_{a+b,0}$.  If there do not exist bound states in the model, i.e. for $\beta^2\ge 1/2$, these are the only poles of the form factors.
\end{enumerate}
We note in passing that the stated annihilation pole axiom corrects a
misprint in Ref.~\onlinecite{SEJF11} (which had no bearing on the
results in said reference).

\section{Hamiltonian eigenstates in the finite volume and the quantum inverse scattering method}
\label{app:QISM}
In this appendix we summarize some elements of the quantum inverse scattering method\cite{korepinbook}, which are used to construct Hamiltonian eigenstates in a large, finite volume\cite{fehtak11}. We define a \emph{monodromy matrix} by
\begin{equation}
\M_{ab}(\lambda|\{\theta_k\})_{i_1\dots i_N}^{ j_1\dots
  j_N}=S_{a\,\,i_1}^{c_1j_1}(\lambda-\theta_1)\cdots S_{c_{N-1}
  i_N}^{b\,\,j_N}(\lambda-\theta_N)\,. 
\label{AA:1}
\end{equation}
It can be written as a $2\times 2$ matrix
\begin{equation}
\M(\lambda|\{\theta_k\})=
\begin{pmatrix}
\A(\lambda|\{\theta_k\}) & \B(\lambda|\{\theta_k\})\\
\C(\lambda|\{\theta_k\}) & \D(\lambda|\{\theta_k\})
\end{pmatrix}\,,
\label{AA:2}
\end{equation}
where $\A(\lambda|\{\theta_k\})$, $\B(\lambda|\{\theta_k\})$, $\C(\lambda|\{\theta_k\})$, $\D(\lambda|\{\theta_k\})$ are operators acting on the $2^N$ dimensional space of scattering states with $N$ particles and fixed rapidities 
\be
\M_{ab}(\lambda|\{\theta_k\})\ket{\theta_1,\cdots,\theta_N}_{i_1\dots
  i_N}=\M_{ab}(\lambda|\{\theta_k\})_{i_1\dots i_N}^{j_1\dots
  j_N}\ket{\theta_1,\cdots,\theta_N}_{j_1\dots j_N}.
\ee
The rapidities $\theta_j$ play the roles of inhomogeneities\cite{korepinbook} in the monodromy matrix. The \emph{transfer matrix} is the trace of the monodromy matrix
\begin{equation}
\T(\lambda|\{\theta_k\})=\sum_{a}\M_{aa}(\lambda|\{\theta_k\})=\A(\lambda|\{\theta_k\})+\D(\lambda|\{\theta_k\}),
\label{AA:3}
\end{equation}
As a consequence of the S-matrix being a solution of the Yang-Baxter
equation, the monodromy matrices fulfil the relation 
\begin{equation}
S_{c_1c_2}^{b_1b_2}(\lambda-\mu)\M_{a_1b_1}(\lambda|\{\theta_k\})
\M_{a_2b_2}(\mu|\{\theta_k\})=\M_{d_1c_1}(\lambda|\{\theta_k\})
\M_{d_2c_2}(\mu|\{\theta_k\})S_{a_1a_2}^{d_1d_2}(\lambda-\mu),
\label{AA:5}
\end{equation}
which implies that the transfer matrices form a commuting family
\begin{equation}
\bigl[\T(\lambda|\{\theta_k\}),\T(\mu|\{\theta_k\})\bigr]=0.
\label{AA:6}
\end{equation}
Simultaneous eigenstates of all transfer matrices can be constructed by algebraic Bethe Ansatz. Starting point is the ``reference state''
\be
\Omega(\{\theta_k\})=\ket{\theta_1,\dots,\theta_N}_{+\dots+}.
\ee
Using the definition of the monodromy matrix, we have
\begin{equation}
\M(\lambda|\{\theta_k\})\Omega(\{\theta_k\})=
\begin{pmatrix}
a(\lambda|\{\theta_k\}) & *\\
0 & d(\lambda|\{\theta_k\})
\end{pmatrix}\Omega(\{\theta_k\})\,,
\label{AA:12}
\end{equation}
where $*$ denotes a non-zero entry not needed in the following, and 
\begin{equation}
a(\lambda|\{\theta_k\})=\prod_{k=1}^{N}S_0(\lambda-\theta_k)\,,\qquad d(\lambda|\{\theta_k\})=\prod_{k=1}^{N}S_T(\lambda-\theta_k)\prod_{k=1}^{N}S_0(\lambda-\theta_k)\,.
\label{AA:13}
\end{equation}
Here $S_T(\theta)$ is defined in \fr{Smatrix}. Taking the trace in \fr{AA:12} we see that $\Omega(\{\theta_k\})$ is in fact a transfer matrix eigenstate
\be
\T(\lambda|\{\theta_k\})\Omega(\{\theta_k\})=
\Big[a(\lambda|\{\theta_k\})+d(\lambda|\{\theta_k\})\Big]
\Omega(\{\theta_k\}).
\ee
The other eigenvectors of the transfer matrix can be constructed by
the Ansatz 
\begin{equation}
\Psi(\{\lambda_k\}|\{\theta_k\})=\left[\prod_{j=1}^r
\B(\lambda_j+\tfrac{\ri\pi}{2}|\{\theta_k\})\right]\Omega(\{\theta_k\})\ ,\quad
0\leq r\leq \frac{N}{2}.
\label{AA:15}
\end{equation} 
This gives simultaneous eigenvectors of the transfer matrix and the
solitonic charge operator $Q$ with a positive eigenvalue of $Q$;
cf. Subsection \ref{AppB2}. States with negative eigenvalues of $Q$
can be obtained acting with the charge conjugation operator $C$ on 
\fr{AA:15}. 
Using the commutation relations for $\A(\lambda|\{\theta_k\}),\B(\lambda|\{\theta_k\}),\D(\lambda|\{\theta_k\})$ that follow from (\ref{AA:5}), one observes that (\ref{AA:13}) are eigenvectors of the transfer matrix with eigenvalues
\begin{equation}
\Lambda(\lambda,\{{\lambda}_k\}|\{\theta_i\})=\left\{\prod_{k=1}^{r}S_T({\lambda}_k-\lambda+\tfrac{\ri\pi}{2})^{-1}+\prod_{k=1}^{N}S_T(\lambda-\theta_k)\prod_{k=1}^{r}S_T(\lambda-\lambda_k-\tfrac{\ri\pi}{2})^{-1}\right\}\,\prod_{k=1}^{N}S_0(\lambda-\theta_k),
\label{AA:16}
\end{equation}
if the parameters $\{\lambda_k\}$ satisfy the following set of non-linear algebraic equations
\begin{equation}
\prod_{k=1}^{N}\frac{\sinh\left(\frac{\ri\pi}{2\xi}-\frac{\lambda_j-\theta_k}{\xi}\right)}{\sinh\left(\frac{\ri\pi}{2\xi}+\frac{\lambda_j-\theta_k}{\xi}\right)}=\prod\limits_{\substack{ \mathllap{k} = \mathrlap{1} \\ \mathllap{k} \neq \mathrlap{j}}}^{r}-\frac{\sinh\left(\frac{\ri\pi}{\xi}-\frac{\lambda_j-\lambda_k}{\xi}\right)}{\sinh\left(\frac{\ri\pi}{\xi}-\frac{\lambda_k-\lambda_j}{\xi}\right)}\,.
\label{AA:17}
\end{equation}
Since the equations are $\ri\pi\xi$-periodic we impose the restriction $0\leq\Im\,{\lambda}_k\leq\pi\xi$ $\forall$ $k$. Taking the logarithm of \fr{AA:17} results in equations of the form
\begin{equation}
\M_j(\theta_1,\dots,\theta_N|\lambda_1,\dots,\lambda_r)=2\pi I_j
\,,\qquad I_j\in\ZZ\,,\qquad j=1,\dots,r\,, 
\label{AA:18}
\end{equation}
where we defined
\begin{equation}
\M_j(\theta_1,\dots,\theta_N|\lambda_1,\dots,\lambda_r)\equiv
{-\ri}\sum_{k=1}^{N}\log\frac{\sinh\left(\frac{\ri\pi}{2\xi}-\frac{\lambda_j-\theta_k}{\xi}\right)}{\sinh\left(\frac{\ri\pi}{2\xi}+\frac{\lambda_j-\theta_k}{\xi}\right)}+{\ri}\sum\limits_{\substack{ \mathllap{k} = \mathrlap{1} \\ \mathllap{k} \neq \mathrlap{j}}}^{r}\log-\frac{\sinh\left(\frac{\ri\pi}{\xi}-\frac{\lambda_j-\lambda_k}{\xi}\right)}{\sinh\left(\frac{\ri\pi}{\xi}-\frac{\lambda_k-\lambda_j}{\xi}\right)}\,.
\label{AA:19}
\end{equation} 

\subsection{Periodic boundary conditions}
Imposing periodic boundary conditions \fr{REA:8} on the transfer matrix eigenstates gives rise to the following set of ``nested'' Bethe Ansatz equations 
\bea
&&e^{\ri \ell \sinh \theta_i}=\prod_{m=1}^NS_0(\theta_i-\theta_m)
\prod_{k=1}^{r}\frac{\sinh\Big(\frac{i\pi}{2\xi}+
\frac{\lambda_k-\theta_i}{\xi}\Big)}
{\sinh\Big(\frac{i\pi}{2\xi}-\frac{\lambda_k-\theta_i}{\xi}\Big)}
\ ,\qquad i=1,\dots,N,\nn
&&\prod_{k=1}^{N}\frac{\sinh\left(\frac{\ri\pi}{2\xi}-\frac{\lambda_j-\theta_k}{\xi}\right)}{\sinh\left(\frac{\ri\pi}{2\xi}+\frac{\lambda_j-\theta_k}{\xi}\right)}=\prod\limits_{\substack{ 
    \mathllap{k} = \mathrlap{1} \\ \mathllap{k} \neq
    \mathrlap{j}}}^{r}\frac{
\sinh\left(\frac{\lambda_j-\lambda_k-i\pi}{\xi}\right)}{\sinh\left(\frac{\lambda_j-\lambda_k+i\pi}{\xi}\right)}\ ,\quad
j=1,\ldots, r.
\label{nestedBAE}
\eea
We recall that $\ell=L\Delta/v$. It is customary to express the
logarithmic form of the equations in terms of \emph{counting
  functions} defined as 
\bea
\ell z(\theta)&=&\ell\sinh\theta+\ri\sum_{m=1}^N\ln\bigl[S_0(\theta-\theta_m)\bigr]
+\ri\sum_{k=1}^r\ln\left[
\frac{\sinh\Big(\frac{\theta-\lambda_k-\frac{\ri\pi}{2}}{\xi}\Big)}
{\sinh\Big(\frac{\theta-\lambda_k+\frac{\ri\pi}{2}}{\xi}\Big)}\right],\nn
\ell y(\lambda)&=&\ri
\sum_{k=1}^{N}\ln\left[
\frac{\sinh\left(\frac{\lambda-\theta_k}{\xi}-\frac{\ri\pi}{2\xi}\right)}
{\sinh\left(\frac{\lambda-\theta_k}{\xi}+\frac{\ri\pi}{2\xi}\right)}
\right]-\ri\sum_{k=1}^r
\ln\left[\frac{\sinh\left(\frac{\lambda-\lambda_k}{\xi}-\frac{\ri\pi}{\xi}\right)}
{\sinh\left(\frac{\lambda-\lambda_k}{\xi}+\frac{\ri\pi}{\xi}\right)}\right],
\label{countingfns}
\eea
where the branch cuts of the logarithms need to be chosen appropriately. The Bethe-Yang equations then read
\bea
z(\theta_j)&=&\frac{2\pi I_j}{\ell}\ ,\quad i=1,\ldots, N,\nn
y(\lambda_k)&=&\frac{2\pi J_k}{\ell}\ ,\quad k=1,\ldots, r,
\eea
where $I_j$ and $J_k$ are integer or half-odd integer numbers.

\subsection{Topological charge and charge conjugation operators}
\label{AppB2}
It is useful to know how the \emph{solitonic charge operator} $Q$ and \emph{charge conjugation operator} $C$ act of the transfer matrix eigenstates. Their respective actions on scattering states \fr{scatteringstates} are
\begin{align}
&Q\ket{\theta_1,\dots,\theta_N}_{a_1\dots a_N}=\left\{\sum_{k=1}^{N}a_k\right\}\ket{\theta_1,\dots,\theta_N}_{a_1\dots a_N}, \label{AA:20}\\
&C\ket{\theta_1,\dots,\theta_N}_{a_1\dots a_N}=\ket{\theta_1,\dots,\theta_N}_{\bar{a}_1\dots \bar{a}_N}. \label{AA:21}
\end{align}
It is easily checked that
\begin{align}
&Q\M_{ab}(\lambda|\{\theta_k\})-\M_{ab}(\lambda|\{\theta_k\})Q=(b-a)\M_{ab}(\lambda|\{\theta_k\}), \label{AA:22}\\
&C\M_{ab}(\lambda|\{\theta_k\})C^{-1}=C\M_{ab}(\lambda|\{\theta_k\})C=
\M_{\bar{a}\bar{b}}(\lambda|\{\theta_k\}). \label{AA:23}
\end{align}
This implies that $Q$ and $C$ commute with the transfer matrix, while
\begin{align}
&\bigl[Q,\B(\lambda|\{\theta_k\})\bigr]=-2\B(\lambda|\{\theta_k\})\,, \label{AA:24}\\
&\bigl[Q,\C(\lambda|\{\theta_k\})\bigr]=2\C(\lambda|\{\theta_k\})\,, \label{AA:25}\\
&C\B(\lambda|\{\theta_k\})C=\C(\lambda|\{\theta_k\})\,, \label{AA:26}\\
&C\C(\lambda|\{\theta_k\})C=\B(\lambda|\{\theta_k\})\ . \label{AA:27}
\end{align}
The action of $Q$ on the reference state is
\begin{equation}
Q\Omega(\{\theta_k\})=N\Omega(\{\theta_k\})\,,
\label{AA:29}
\end{equation}
and hence (\ref{AA:15}) are eigenstates of $Q$ with eigenvalue $N-2r$. 

\section{Proof of property \fr{II:8}}\label{sec:AppC}
In this appendix we show that the eigenvalues $\Lambda^s(\lambda|\{\theta_k\})$ of the transfer matrix (\ref{AA:6}) satisfy the relations 
\begin{equation}
\K^s_{2N}(\theta_1,\dots,\theta_N)\Lambda^s(\theta_i|\{\theta_k\})=\K^s_{2N}(\theta_1,\dots,\theta_N){\Lambda^s(-\theta_i|\{\theta_k\})^{-1}}\ ,
\label{KLKL}
\end{equation}
where $\{\theta_k\}=\{-\theta_1,\theta_1,\dots,-\theta_N,\theta_N\}$. The proof of \fr{KLKL} is divided into three main steps.

\subsection{Step I}
We first establish the following identity
\begin{align}
&K^{a_1 b_1}(\theta_1)\cdots K^{a_N b_N}(\theta_N)\T(\sigma\theta_i|{\{-\theta_1,\theta_1,\dots,-\theta_N,\theta_N\}})_{a_1b_1\dots a_N b_{N}}^{a'_1b'_1\dots a'_N b'_{N}}\notag\\
&\quad=K^{a_1 b_1}(\theta_1)\cdots K^{a_N b_N}(\theta_N)\T(\sigma\theta_i|{\{-\theta_N,\theta_N,\dots,-\theta_1,\theta_1\}})_{b_Na_N\dots b_1 a_{1}}^{b'_Na'_N\dots b'_1 a'_{1}}\,,\label{AB:16}
\end{align}
where $\sigma=\pm 1$ and $\T$ is the transfer matrix 
\begin{equation}
\T(\lambda|\{-\theta_1,\theta_1,\dots,-\theta_N,\theta_N\})_{a_1\dots
  b_{N}}^{a'_1\dots
  b'_{N}}=S_{c_{2N}a_1}^{c_1a'_1}(\lambda+\theta_1)S_{c_1b_1}^{c_2b_1'}(\lambda-\theta_1)\cdots S_{c_{2N-1}
  b_{N}}^{c_{2N}b'_{N}}(\lambda-\theta_N)\ .
\label{AB:15}
\end{equation}
We focus on the case $i=1$ in \fr{AB:16}, all other cases can be proved analogously. Setting $i=1$ and $\sigma=-1$, substituting the expression of $\T$ in terms of scattering matrices \fr{AB:15}, and finally using that
\be
S_{ab}^{cd}(0)=-\delta_{a}^{d}\delta_{b}^{c},
\label{S0}
\ee
we find that \fr{AB:16} is reduced to
\begin{align}
&K^{a_1 b_1}(\theta_1)\cdots K^{a_N b_N}(\theta_N)S_{a_{1}b_1}^{c_1 b'_1}(-2\theta_1)S_{c_{1} a_2}^{c_{2} a'_2}(-\theta_1+\theta_2)\cdots S_{c_{2N-1} b_{N}}^{a'_{1}b'_{N}}(-\theta_1-\theta_N) =\notag\\
&\qquad K^{a_1 b_1}(\theta_1)\cdots K^{a_N b_N}(\theta_N)S_{b_1 a_1}^{c_{1} a'_1 }(-2\theta_1)S_{c_{1}b_N}^{c_{2}b'_N}(-\theta_1+\theta_N)\cdots S_{c_{2N-1} a_{2}}^{b'_1 a'_{2}}(-\theta_1-\theta_{2})  \,.
\label{AB:17}
\end{align}  
Employing the boundary unitarity property \fr{BUE} of the $K$-matrix, the left-hand side of \fr{AB:17} can be written in the form
\begin{align}
&K^{a_1 b_1}(\theta_1)\cdots K^{a_N b_N}(\theta_N)S_{a_{1}b_1}^{c_1 b'_1}(-2\theta_1)S_{c_{1} a_2}^{c_{2} a'_2}(-\theta_1+\theta_2)\cdots S_{c_{2N-1} b_{N}}^{a'_{1}b'_{N}}(-\theta_1-\theta_N) =\notag\\
&\qquad K^{b'_1 c_1}(-\theta_1)\cdots K^{a_N b_N}(\theta_N)S_{c_{1} a_2}^{c_{2} a'_2}(-\theta_1+\theta_2)\cdots S_{c_{2N-1} b_{N}}^{a'_{1}b'_{N}}(-\theta_1-\theta_N)\,.
\label{AB:18}
\end{align} 
Now we use the boundary Yang-Baxter equation \fr{BYBE} to rewrite the terms involving $\theta_1$ and $\theta_2$
\begin{align}
&K^{b'_1 c_1}(-\theta_1)\cdots K^{a_N b_N}(\theta_N)S_{c_{1} a_2}^{c_{2} a'_2}(-\theta_1+\theta_2)\cdots S_{c_{2N-1} b_{N}}^{a'_{1}b'_{N}}(-\theta_1-\theta_N)=\label{AB:23}\\
&\qquad K^{c_1 c_3}(-\theta_1)\cdots K^{a_N b_N}(\theta_N)S_{c_{3} a_3}^{c_{4} a'_3}(-\theta_1+\theta_3)\cdots S_{c_{2N-1} b_{N}}^{a'_{1}b'_{N}}(-\theta_1-\theta_N)S_{c_{1} b_{2}}^{c_{2} b'_{2}}(-\theta_1+\theta_{2})S_{c_{2} a_{2}}^{b'_1 a'_{2}}(-\theta_1-\theta_{2})\,.\notag
\end{align}
The index structure on the right-hand side of \fr{AB:23} is such that we may use the boundary Yang-Baxter equation to rewrite the terms involving $\theta_1$ and $\theta_3$, then $\theta_1$ and $\theta_4$ and so on. This brings the left-hand side of \fr{AB:17} to the form
\begin{align}
&K^{a_1 b_1}(\theta_1)\cdots K^{a_N b_N}(\theta_N)S_{a_{1}b_1}^{c_1 b'_1}(-2\theta_1)S_{c_{1} a_2}^{c_{2} a'_2}(-\theta_1+\theta_2)\cdots S_{c_{2N-1} b_{N}}^{a'_{1}b'_{N}}(-\theta_1-\theta_N)=\\
&\qquad K^{c_1 a'_1}(-\theta_1)\cdots K^{a_N b_N}(\theta_N)S_{c_1 b_N}^{c_{2} b'_N}(-\theta_1+\theta_N)\cdots S_{c_{2N-1} a_{2}}^{b'_1 a'_{2}}(-\theta_1-\theta_{2})\,.\notag
\end{align} 
Finally we use the reflection equations \fr{BUE} and parity invariance of the S-matrix
\be
S_{ab}^{cd}(\theta)=S_{ba}^{dc}(\theta)\,,
\label{cex}
\ee
to arrive at
\begin{align}
&K^{a_1 b_1}(\theta_1)\cdots K^{a_N b_N}(\theta_N)S_{a_{1}b_1}^{c_1 b'_1}(-2\theta_1)S_{c_{1} a_2}^{c_{2} a'_2}(-\theta_1+\theta_2)\cdots S_{c_{2N-1} b_{N}}^{a'_{1}b'_{N}}(-\theta_1-\theta_N)=\label{AB:24}\\
&\qquad K^{a_1 b_1}(\theta_1)\cdots K^{a_N b_N}(\theta_N)S_{b_1 a_1}^{c_1 a'_1}(-2\theta_1)S_{c_1 b_N}^{c_{2} b'_N}(-\theta_1+\theta_N)\cdots S_{c_{2N-1} a_{2}}^{b'_1 a'_{2}}(-\theta_1-\theta_{2})\,.
\end{align}
This establishes \fr{AB:17} for $i=1$ and $\sigma=-1$. The case $\sigma=+1$ is proved in the same way.

\subsection{Step II}
Next we establish the following identity
\begin{equation}
\left[\T(\theta_i|{\{-\theta_N,\theta_N,\dots,-\theta_1,\theta_1\}})\right]_{b_Na_N\dots b_1 a_{1}}^{b'_Na'_N\dots b'_1 a'_{1}}=\left[\T(-\theta_i|\{-\theta_1,\theta_1,\dots,-\theta_{N},\theta_{N}\})^{-1}\right]_{a_1b_1\dots a_N b_{N}}^{a'_1b'_1\dots a'_N b'_{N}}\,,
\label{AB:32}
\end{equation}
which is equivalent to
\be
\sum_{a'_1\dots b'_N} \left[\T(\theta_i|{\{-\theta_N, \dots,
    \theta_1\}})\right]_{b_N\dots a_{1}}^{b'_N\dots
  a'_{1}}\left[\T(-\theta_i|{\{-\theta_1,\dots,\theta_{N}\}})\right]_{a'_1\dots
  b'_{N}}^{a''_1\dots
  b''_{N}}=\prod_{i=1}^{N}\delta_{a_i}^{a''_i}\delta_{b_i}^{b''_i}\,. 
\label{prod}
\ee
Using the explicit expression of the two transfer matrices
\bea
\left[\T(\theta_i|{\{-\theta_N, \dots, \theta_1\}})\right]_{b_N\dots a_{1}}^{b'_N\dots a'_{1}}&=&S_{b_{i} a_{i-1}}^{c_{1} a'_{i-1}}(\theta_i+\theta_{i-1})\cdots S_{c_{N-1} a_{i+1}}^{c_N a'_{i+1}}(\theta_i-\theta_{i+1})S_{c_{N} b_{i}}^{a'_{i} b'_{i}}(2\theta_{i})\,,\nn
\left[\T(-\theta_i|{\{-\theta_1,\dots,\theta_{N}\}})\right]_{a_1\dots b_{N}}^{a'_1\dots b'_{N}}&=&S^{c_1 b'_{i}}_{a_i b_{i}}(-2\theta_i)S^{c_{2} a'_{i+1}}_{c_1 a_{i+1}}(-\theta_i-\theta_{i+1})\cdots S_{c_{N} b_{i-1}}^{a'_i b'_{i-1}}(-\theta_i-\theta_{i-1})\,,
\label{AB:34}
\eea
the left-hand side of Eq. \fr{prod} can be expressed in the form
\be
\sum_{a'_1\dots b'_N}S_{b_{i} a_{i-1}}^{c_{1} a'_{i-1}}(\theta_i+\theta_{i-1})\cdots S_{c_{N} b_{i}}^{a'_{i} b'_{i}}(2\theta_{i}) S^{c_1 b''_{i}}_{a'_i b'_{i}}(-2\theta_i)\cdots S_{c_{N} b'_{i-1}}^{a''_i b''_{i-1}}(-\theta_i-\theta_{i-1})\,.
\label{AB:34b}
\ee
The product of the two S-matrices involving $\theta_i$ can be simplified using the unitarity condition
\be
S_{ab}^{cd}(\theta)S_{cd}^{ef}(-\theta)=\delta_{a}^{e}\delta_{b}^{f}\ .
\label{US}
\ee 
Eq. \fr{AB:34b} then becomes
\be
\delta_{b_i}^{b''_i} \sum_{a'_1\dots b'_N}S_{b_{i} a_{i-1}}^{c_{1} a'_{i-1}}(\theta_i+\theta_{i-1})\cdots S_{c_{N-1} a_{i+1}}^{c_1 a'_{i+1}}(\theta_i-\theta_{i+1}) S^{c_{2} a''_{i+1}}_{c_1 a'_{i+1}}(-\theta_i-\theta_{i+1}) \cdots S_{c_{N} b'_{i-1}}^{a''_i b''_{i-1}}(-\theta_i-\theta_{i-1})\,.
\ee 
We observe that now the product of the two S-matrices involving $\theta_{i+1}$ can be simplified using \fr{US}. Repeating this procedure for $\theta_{i+2},\dots,\theta_{N},\theta_1,\dots,\theta_{i-1}$ then establishes \fr{prod}.

\subsection{Step III}
We now substitute (\ref{AB:32}) into (\ref{AB:16}) to obtain
\begin{align}
&K^{a_1 b_1}(\theta_1)\cdots K^{a_N b_N}(\theta_N)\left[\T(-\theta_i|{\{-\theta_1,\theta_1,\dots,-\theta_N,\theta_N\}})\right]_{a_1b_1\dots a_N b_{N}}^{a'_1b'_1\dots a'_N b'_{N}}=\notag\\
&=K^{a_1 b_1}(\theta_1)\cdots K^{a_N
    b_N}(\theta_N)\left[{\T(\theta_i|{\{-\theta_1,\theta_1,\dots,-\theta_N,\theta_N\}})^{-1}}\right]_{a_1b_1\dots
    a_N b_{N}}^{a'_1b'_1\dots a'_N b'_{N}}\, .
\label{AB:35}
\end{align}
Combining the eigenvalue equation \fr{REA:4} with the completeness relation \fr{REA:6} we have
\be
\Psi^s_{a_1'\ldots b_N'}(\{\theta_k\})^*
\left[\T(-\theta_i|{\{-\theta_1,\theta_1,\dots,-\theta_N,\theta_N\}})\right]_{a_1b_1\dots
  a_N b_{N}}^{a'_1b'_1\dots a'_N b'_{N}}=\Lambda^s(-\theta_i|
\{-\theta_1,\dots,\theta_N\})
\Psi^s_{a_1\ldots b_N}(\{\theta_k\})^*.
\label{AB:36}
\ee
Finally we contract both sides of \fr{AB:35} with $\Psi^s_{a_1'\ldots b_N'}(\{\theta_k\})^*$ and then use \fr{AB:36} to obtain the desired result 
\begin{equation}
\K^s_{2N}(\theta_1,\dots,\theta_N)\Lambda^s(-\theta_i|\{-\theta_1,\dots,\theta_N\})=\K^s_{2N}(\theta_1,\dots,\theta_N){\Lambda^s(\theta_i|\{-\theta_1,\dots,\theta_N\})^{-1}}\, .
\label{AB:38}
\end{equation} 

\section{Derivatives of the functions $\boldsymbol{\bar Q^s_i(\theta_1,\ldots,\theta_N)}$}
\label{app:mixedderivatives}
The functions $\bar Q^s_i(\theta_1,\ldots,\theta_N)$ are defined by  \fr{II:6} and can be written in the form [cf. \fr{countingfns}]
\bea
\bar Q_i^{s}(\theta_1,\dots,\theta_N)&=&\ell\sinh\theta_i+\ri\sum_{m=1}^N\ln\bigl[S_0(\theta_i-\theta_m)S_0(\theta_i+\theta_m)\bigr]
+\ri\sum_{k=1}^r\ln\left[
\frac{\sinh\Big(\frac{\theta_i-\lambda_k-\frac{\ri\pi}{2}}{\xi}\Big)}
{\sinh\Big(\frac{\theta_i-\lambda_k+\frac{\ri\pi}{2}}{\xi}\Big)}\right],
\eea
where the parameters $\lambda_k$ are obtained by solving the set of equations
\bea
2\pi J^s_m&=&\ri
\sum_{k=1}^{N}\ln\left[
\frac{\sinh\left(\frac{\lambda_m-\theta_k}{\xi}-\frac{\ri\pi}{2\xi}\right)\sinh\left(\frac{\lambda_m+\theta_k}{\xi}-\frac{\ri\pi}{2\xi}\right)}
{\sinh\left(\frac{\lambda_m-\theta_k}{\xi}+\frac{\ri\pi}{2\xi}\right)\sinh\left(\frac{\lambda_m+\theta_k}{\xi}+\frac{\ri\pi}{2\xi}\right)}
\right]-\ri\sum_{k=1}^r
\ln\left[\frac{\sinh\left(\frac{\lambda_m-\lambda_k}{\xi}-\frac{\ri\pi}{\xi}\right)}
{\sinh\left(\frac{\lambda_m-\lambda_k}{\xi}+\frac{\ri\pi}{\xi}\right)}\right],\quad
m=1,\ldots r,
\label{nesting}
\eea
for a given set of (half-odd) integers $\{J^s_m\}$ specifying the polarisation $s$. We are interested in the situation where
\be
\theta_j\approx\begin{cases}
-\tilde{\theta}_{N+1-j}  & j\leq N\\
\quad\tilde{\theta}_{j-N} & j>N
\end{cases}
\,,\quad j=1,\ldots,2N,
\ee
where $\{\tilde{\theta}_j\}$ is the solution of the Bethe-Yang equations corresponding to our representative state, i.e.
\be
\bar{Q}_j^{\tilde{s}}(\tilde\theta_1,\dots,\tilde\theta_N)=2\pi \tilde{I}_j\ ,\quad j=1,\ldots,N.
\ee
Crucially, for this solution we have
\be
\tilde\theta_{j+1}-\tilde\theta_j=\frac{1}{\ell\rho_\Phi(\theta_j)}+{\cal O}(\ell^{-2}),
\ee
where $\rho_\Phi(\theta)$ is the root density \fr{III:21} describing the representative state in the thermodynamic limit. We may use this fact to recast \fr{nesting} in the form
\bea
\frac{2\pi J^s_m}{\ell}&=&\ri\int_{-\infty}^\infty d\theta\ \rho_\Phi(\theta)
\ln\left[
\frac{\sinh\left(\frac{\lambda_m-\theta}{\xi}-\frac{\ri\pi}{2\xi}\right)}
{\sinh\left(\frac{\lambda_m-\theta}{\xi}+\frac{\ri\pi}{2\xi}\right)}
\right]-\frac{\ri}{\ell}\sum_{k=1}^r
\ln\left[\frac{\sinh\left(\frac{\lambda_m-\lambda_k}{\xi}-\frac{\ri\pi}{\xi}\right)}
  {\sinh\left(\frac{\lambda_m-\lambda_k}{\xi}+\frac{\ri\pi}{\xi}\right)}\right]
+{\cal O}(\ell^{-1}),\quad  m=1,\ldots r.
\label{nesting2}
\eea
Denoting the solution to Eqs. \fr{nesting2} when the ${\cal O}(\ell^{-1})$ correction terms are dropped by $\big\{\lambda_m^{(\infty)}\big\}$, we conclude that 
\be
\lambda_m=\lambda_m^{(\infty)}+{\cal O}(\ell^{-1}).
\ee
Importantly, this conclusion holds for any set $\{\theta_j\}$ that is described by the root density $\rho_\Phi(\theta)$ in the thermodynamic limit. For any such set we then have
\bea
\bar{Q}_i^{\tilde{s}}(\theta_1,\dots,\theta_N)&=&\ell\sinh\theta_i+\ri\sum_{m=1}^N\ln\bigl[S_0(\theta_i-\theta_m)S_0(\theta_i+\theta_m)\bigr]
+\ri\sum_{k=1}^r\left(\ln\left[
\frac{\sinh\Big(\frac{\theta_i-\lambda^{(\infty)}_k-\frac{\ri\pi}{2}}{\xi}\Big)}
{\sinh\Big(\frac{\theta_i-\lambda^{(\infty)}_k+\frac{\ri\pi}{2}}{\xi}\Big)}\right]+{\cal  O}(\ell^{-1})\right).
\eea
We conclude that
\bea
\frac{\partial
  \bar{Q}_i^{\tilde{s}}(\theta_1,\dots,\theta_N)}{\partial\theta_j}=
\begin{cases}
{\cal O}(\ell) & i=j\ ,\\
{\cal O}(1) & i\neq j.
\end{cases}
\eea

\section{Contributions from states with $\boldsymbol{M>N}$}
\label{app:M>N}
In this appendix we use the ideas developed in the Subsection \ref{Cal} to argue that terms with $M>N$ only lead to sub-leading contributions in the Lehmann representation \fr{IV:1}. Repeating the reasoning employed to arrive at formula \fr{IV:15}, we obtain
\begin{align}
\frac{{}_{\rm R}\langle\Psi_0|e^{\ri\beta\Phi(t,x)/2}\repS}
{{}_{\rm NS}\langle\Psi_0\repS}\bigg|_{M>N}=&\sum_{s=1}^{2^{2M}}\oint\limits_{\bar{\C}_{\text{tot}}}\prod_{i=1}^{M}\frac{d\eta_i}{2\pi}E(\eta_1,\dots, \eta_M)_{\tilde{s},s}\notag\\*
&+\sum_{s=1}^{2^{2M}}\sum_{m=1}^{N}(-1)^{m}\sum_{1\leq
  j_1<\cdots<j_m\leq
  M}\sum_{i_1,\cdots,i_m=1}^{N}\oint\limits_{\bar{\C}^{j_1,\cdots,j_m}_{i_1,\cdots,i_m}}\prod_{i=1}^{M}\frac{d\eta_i}{2\pi}
E(\eta_1,\dots, \eta_M)_{\tilde{s},s}\ .
\label{APD:1}
\end{align}
Here the multi-contours $\bar{\C}_{\text{tot}}$ and $\bar{\C}^{j_1,\cdots,j_m}_{i_1,\cdots,i_m}$ are defined analogously to Subsection \ref{ssec:M=N}, while 
\be
E(\eta_1,\dots, \eta_M)_{\tilde{s},s}\equiv 
\frac{\bar{\rho}^{\tilde{s}}_{N}(\tilde{\theta}_1,\dots,\tilde{\theta}_N)}{{\rho}^{\tilde{s}}_{2N}(-\tilde{\theta}_1,\dots,\tilde{\theta}_N)}\frac{\K^s_{2M}(\eta_1,\dots,\eta_M)}{\K^{\tilde{s}}_{2N}(\tilde{\theta}_1,\dots,\tilde{\theta}_N)}
\frac{f^{\beta/2}({\eta}_M+\ri\pi,\dots,\tilde{\theta}_N)_{s,\tilde{s}}}{\prod_{k=1}^{M}\left\{e^{\ri \bar{Q}^s_k(\eta_1,\dots,\eta_M)}-1\right\}}e^{2\ri \Delta t \sum_{i=1}^{M}\cosh\eta_{i}-\sum_{i=1}^{N}\cosh\tilde{\theta}_{i}}\,.
\label{APD:1a}
\ee
As we are dealing with a local operator, we expect that significant contributions in the large-$N$ limit can only arise from states with
\be
M-N={\cal O}(1).
\ee
The leading contribution to \fr{APD:1} arises from the $M\choose N$ regions where $N$ of the $\eta_j$ are integrated around the singularities at $\tilde{\theta}_1,\dots,\tilde{\theta}_N$. In order to obtain an estimate for these contributions we consider the case where $\eta_j\approx\tht_j$, $j=1,\ldots, N$ in more detail. The leading singularity of the form factor is given by
\begin{equation}
 f^{\beta/2}({\eta}_M+\ri\pi,\dots,\tilde{\theta}_N)_{s,\tilde{s}}\approx
 f(\eta_1, \dots, \eta_M)_{\tilde{s},s}
 \prod_{j=1}^{N}\frac{1}{(\eta_j-\tht_j)^2}\ ,
\label{singM>N}
\end{equation}
where $f(\eta_1, \dots, \eta_M)_{s,\tilde{s}}$ is a regular function. Substituting \fr{singM>N} back into \fr{APD:1a} and then carrying out the integrals over $\eta_1,\ldots,\eta_N$ gives a leading contribution of the form
\begin{align}
\sim\oint\limits_{{\D}^{1,\cdots,N}_{1,\cdots,N}}\prod_{i=N+1}^{M}
&\frac{d\eta_i}{2\pi}g({\eta}_{N+1},\dots, {\eta}_M)_{\tilde{s},s}
\prod_{j=1}^N\frac{\partial}{\partial_{\eta_j}}
\bigg|_{\eta_j=\tilde{\theta}_j}
\left[\prod_{i=1}^{M}\frac{e^{2\ri \Delta t
      \cosh\eta_{i}}}{e^{\ri
      \bar{Q}^s_i(\eta_1,\dots,\eta_M)}-1}\right]
e^{-2\ri\Delta t\sum_{m=1}^N \cosh\tht_m}
\,.\label{APD:3}
\end{align}
Here $g({\eta}_{N+1},\dots,{\eta}_M)_{\tilde{s},s}$ is a regular function scaling as $L^{-N}$ and 
${\D}^{1,\cdots,N}_{1,\cdots,N}$ is a multi-contour in $\CC^{M-N}$ obtained by removing the first $N$ components from $\bar{\C}^{1,\cdots,N}_{1,\cdots,N}$. We may now proceed as in Appendix~\ref{app:mixedderivatives}. In particular, the nested Bethe-Yang equations for solutions
$\{\theta_j\}$ such that
\be
\theta_k\approx\tht_k\ ,\quad k=1,\ldots,N,
\ee
can still be cast in the form \fr{nesting2}. Hence we again have
\be
\lambda_m=\lambda_m^{(\infty)}+{\cal O}(\ell^{-1}),
\ee
which in turn implies that
\bea
\bar{Q}_i^{s}(\theta_1,\dots,\theta_M)&=&\ell\sinh\theta_i+\ri\sum_{m=1}^M\ln\bigl[S_0(\theta_i-\theta_m)S_0(\theta_i-\theta_m)\bigr]
+\ri\sum_{k=1}^r\ln\left[
\frac{\sinh\Big(\frac{\theta_i-\lambda_k^{(\infty)}-\frac{\ri\pi}{2}}{\xi}\Big)}
{\sinh\Big(\frac{\theta_i-\lambda_k^{(\infty)}+\frac{\ri\pi}{2}}{\xi}\Big)}\right]+{\cal O}(1).
\eea
Following the same steps as in Subsection~\ref{ssec:M=N} we then find
\bea
\prod_{j=1}^N\frac{\partial}{\partial_{\eta_j}}
\bigg|_{\eta_j=\tilde{\theta}_j}
\left[\prod_{i=1}^{M}\frac{e^{2\ri \Delta t\cosh\eta_{i}}}{e^{\ri
      \bar{Q}^{s}_i(\eta_1,\dots,\eta_M)}-1}\right]e^{-2\ri\Delta t\sum_{m=1}^N \cosh\tht_m}
&\approx&\left(\frac{\ri}{4}\right)^N\prod_{i=1}^{N}\left\{\partial_i\bar{Q}^{\tilde{s}}_i(\tilde{\theta}_1,\dots,\tilde{\theta}_N)-4
\Delta t\sinh\tilde{\theta}_{i}\right\}\nn
&&\times\prod_{j=N+1}^M\frac{e^{2\ri \Delta t\cosh\eta_j}}
{e^{\ri\bar{Q}^s_j(\tht_1,\ldots,\tht_N,\eta_{N+1},\ldots,\eta_M)}-1}.
\eea
The main difference as compared to the $N=M$ case is the presence of additional integrals over $\eta_{N+1},\ldots,\eta_M$. Using that
\be
\frac{\partial
 \bar Q_i^{\tilde{s}}(\theta_1,\dots,\theta_M)}{\partial\theta_j}=\oO(1)\,,\qquad i \neq j\,,
\ee
we see that for small but fixed $\{\epsilon_i\}$ 
\begin{equation}
\bar{Q}_i^s(\tilde{\theta}_1, \dots\tilde{\theta}_N, \theta_{N+1}+\ri\epsilon_{N+1} ,\dots \theta_M+\ri\epsilon_M)=\bar{Q}_i^s(\tilde{\theta}_1,\dots, \theta_M)+\ri\epsilon_i\ell \bigl[\cosh\theta_i +\oO(1/\ell)\bigr],\quad i=N+1,\ldots,M.
\label{APD:4}
\end{equation}
Formula \fr{APD:4} implies that for $L\to\infty$ the parts of the paths composing $\hat{\C}_{\text{tot}}$ below the real axis will give a vanishing contribution. On the remaining parts of the paths we have $\Im\,\eta_j>0$, and assuming that we can deform the integration contours a finite distance up into the upper half plane without encountering singularities, we conclude that the resulting contributions are exponentially suppressed in time. 

\section{Most singular parts of the form factors}
\label{app:annihilationpole}
In this appendix we determine the most singular contribution to matrix elements of the form
\be
{}_{b_N \cdots
    b_1}\braket{\theta_N,\ldots,\theta_1 |
  e^{\ri\beta\Phi(0)/2} |
  \tilde{\theta}_1,\ldots,\tilde{\theta}_N}_{{a}_1\ldots{a}_N}
=f^{\beta/2}_{b_N\dots b_1a_1\dots a_N}(\theta_N+\ri\pi,\dots,\theta_1+\ri\pi,\tht_1,\dots,\tht_N).
\ee
We will show by induction that
\be
f^{\beta/2}_{b_{N}\dots b_1a_1\dots
  a_{N}}(\theta_{N}+\ri\pi,\dots,\theta_1+\ri\pi,\tht_1,\dots,\tht_{N})
\Bigg|_{\genfrac{}{}{0pt}{}{\theta_j\approx\tht_j}{j=1,\ldots,{N}}}={\cal
  G}_{\beta/2}\prod_{k=1}^{{N}}
\frac{2\ri}{\theta_k-\tht_k}\delta_{a_k,\bar{b}_k}+\text{less singular}.
\label{maxsing}
\ee
The case ${N}=1$ is an immediate consequence of the annihilation pole axiom (see Appendix~\ref{sec:FFaxioms}) for the semi-local operator $e^{\ri\beta\Phi/2}$. We will now assume that \fr{maxsing} holds and consider
\be
f^{\beta/2}_{b_{{N}+1}\dots b_1a_1\dots
  a_{{N}+1}}(\theta_{{N}+1}+\ri\pi,\dots,
\theta_1+\ri\pi,\tht_1,\dots,\tht_{{N}+1})
\Bigg|_{\genfrac{}{}{0pt}{}{\theta_j\approx\tht_j}{j=1,\ldots,{N}+1}}.
\ee
Using the periodicity axiom ${N}$ times, this can be rewritten as
\be
(-1)^{N}\ f^{\beta/2}_{b_1a_1\dots a_{{N}+1}b_{{N}+1}\ldots
  b_2}(\theta_1+\ri\pi,\tht_1,\dots,\tht_{{N}+1},
\theta_{{N}+1}-\ri\pi,\ldots, \theta_{2}-\ri\pi)
\Bigg|_{\genfrac{}{}{0pt}{}{\theta_j\approx\tht_j}{j=1,\ldots,{N}+1}}.
\ee
The annihilation pole axiom allows us to extract the leading singularity of this expression for $\theta_1\approx\tht_1$
\bea
&&(-1)^{N}\frac{\ri\delta_{c,\bar{b}_1}}{\theta_1-\tht_1}
f^{\beta/2}_{a'_2\dots a'_{{N}+1}b'_{{N}+1}\ldots  b'_2}(\tht_2,\dots,\tht_{{N}+1},
\theta_{{N}+1}-\ri\pi,\ldots, \theta_{2}-\ri\pi)
\Bigg|_{\genfrac{}{}{0pt}{}{\theta_j\approx\tht_j}{j=2,\ldots,{N}+1}}\nn
&&\times\ 
\left[\delta_{a_1}^c\prod_{j=2}^{{N}+1}\delta_{a_j}^{a_j'}\delta_{b_j}^{b_j'}+
S_{a_1a_2}^{c_2a_2'}(\tht_1-\tht_2)\ldots
S_{c_{{N}}a_{{N}+1}}^{c_{{N}+1}a'_{{N}+1}}(\tht_1-\tht_{{N}+1})
S_{c_{{N}+1}b_{{N}+1}}^{d_{{N}}b'_{{N}+1}}(\tht_1-\theta_{{N}+1}-\ri\pi)\ldots
S^{cb'_2}_{d_2b_2}(\tht_1-\theta_2-\ri\pi)
\right].
\eea
We now use the periodicity axiom ${N}$ times to bring the form factor into a form where we can use the induction assumption \fr{maxsing}. The most singular contribution is then given by
\bea
&&\frac{\ri\delta_{c,\bar{b}_1}}{\theta_1-\tht_1}
\prod_{k=2}^{{N}+1}\frac{2\ri}{\theta_k-\tht_k}\delta_{a'_k,\bar{b}'_k}\nn
&&\times\ 
\left[\delta_{a_1}^c\prod_{j=2}^{{N}+1}\delta_{a_j}^{a_j'}\delta_{b_j}^{b_j'}+
S_{a_1a_2}^{c_2a_2'}(\tht_1-\tht_2)\ldots
S_{c_{{N}-1}a_{N}}^{c_{N}a'_{N}}(\tht_1-\tht_{{N}+1})
S_{c_{{N}}b_{N}}^{d_{{N}-1}b'_{N}}(\tht_1-\tht_{{N}+1}-\ri\pi)\ldots
S^{cb'_2}_{d_2b_2}(\tht_1-\tht_2-\ri\pi)
\right].
\eea
The product of S-matrices can be simplified by repeatedly using the identity (starting with the two S-matrices involving $\tht_{{N}+1}$ and then moving outwards in the product)
\be
S_{a_1a_2}^{c_1c_2}(\theta)
S_{c_1\bar{b}_2}^{b_1\bar{c}_2}(\theta+\ri\pi)
=\delta_{a_1}^{b_1}\delta_{a_2}^{b_2}.
\ee
This completes the induction step.

\subsection{Singularities of form factors with respect to transfer
  matrix eigenstates}
As the most singular contribution to the form factors \fr{maxsing} is \emph{diagonal} in the indices $a_j$ and $b_j$, we conclude that the corresponding contribution to form factors involving transfer matrix eigenstates is
\be
{}^s\braket{\theta_{N},\ldots,\theta_1|
  e^{\ri\beta\Phi(0)/2}
  |\tilde{\theta}_1,\ldots,\tilde{\theta}_{N}}^{s'}
\Bigg|_{\genfrac{}{}{0pt}{}{\theta_j\approx\tht_j}{j=1,\ldots,{N}}}
=\delta_{s,s'}{\cal G}_{\beta/2}\prod_{k=1}^{{N}}
\frac{2\ri}{\theta_k-\tht_k}+\text{less singular}.
\label{maxsing2}
\ee

\section{Regularisation of annihilation poles}\label{sec:Smirnov}
The correlation functions to be calculated contain matrix elements with incoming and outgoing particles,
\begin{equation}
_{a_1\ldots a_N}\!\!\bra{\theta_1',\ldots,\theta_N'}
{\cal O}\ket{\theta_M,\ldots,\theta_1}_{b_M\ldots b_1},
\label{eq:generalmatrixelement}
\end{equation}
which possess kinematical poles whenever $\theta_i'=\theta_j$ and $a_i=b_j$.  These matrix elements can be decomposed into ``connected'' and ``disconnected'' contributions. The latter are characterized by the appearance of terms like $\delta(\theta_i'-\theta_j)$, signalling that some of the particles do not encounter the operator $\mathcal{O}$ in the process described by the matrix element. We follow Smirnov\cite{Smirnov92book} to analytically continue form factors. Let $\overrightarrow{A}=\{\theta_1',\ldots,\theta_N'\}$ with $\theta_1'<\theta_2'<\ldots<\theta_N'$ and $\overleftarrow{B}=\{\theta_M,\ldots,\theta_1\}$ with $\theta_M>\theta_{M-1}>\ldots>\theta_1$ denote two sets of ordered rapidities and introduce the notations
\begin{eqnarray}
Z[\overrightarrow{A}]_{a_1\ldots
  a_N}&\equiv& Z_{a_1}(\theta_1')Z_{a_2}(\theta_2')\ldots Z_{a_N}(\theta_N'),\\*
Z^\dagger[\overleftarrow{B}]_{b_M\ldots b_1}
&\equiv&Z^\dagger_{b_M}(\theta_M) Z^\dagger_{b_{M-1}}(\theta_{M-1})\ldots
Z^\dagger_{b_1}(\theta_1).  
\end{eqnarray}
Now let $A_1$ and $A_2$ be a partition of $A$, i.e. $A=A_1\cup A_2$, where $A_1$ contains $n(A_1)=N-k$ rapidities.  As a consequence of the Faddeev--Zamolodchikov algebra we have
\begin{equation}
Z[\overrightarrow{A}]_{a_1\ldots a_N}=S(\overrightarrow{A}|\overrightarrow{A_1})^{c_1\ldots c_N}_{a_1\ldots a_N}\,Z[\overrightarrow{A_2}]_{c_1\ldots c_k}\,Z[\overrightarrow{A_1}]_{c_{k+1}\ldots c_N},
\end{equation}
where $S(\overrightarrow{A}|\overrightarrow{A_1})$ is the product of two-particle scattering matrices needed to rearrange the order of Faddeev--Zamolodchikov operators in $Z[\overrightarrow{A}]$ to arrive at $Z[\overrightarrow{A_2}]Z[\overrightarrow{A_1}]$.  For example, if $\overrightarrow{A}=\{\theta_1',\ldots,\theta_4'\}$ and $\overrightarrow{A_1}=\{\theta_2',\theta_3'\}$ it is given by
\begin{equation}
S(\overrightarrow{A}|\overrightarrow{A_1})^{c_1\ldots c_4}_{a_1\ldots a_4}=\delta_{a_1}^{c_4}\,S_{a_2b}^{c_2c_4}(\theta_2'-\theta_4')\,S_{a_3a_4}^{c_3b}(\theta_3'-\theta_4').
\end{equation}
Similarly we have
\begin{equation}
Z^\dagger[\overleftarrow{B}]_{b_M\ldots b_1}=
Z^\dagger[\overleftarrow{B_1}]_{d_M\ldots d_{l+1}}
Z^\dagger[\overleftarrow{B}_2]_{d_l\ldots d_1}
\ S(\overleftarrow{B_1}|\overleftarrow{B})_{b_M\ldots b_1}^{d_M\ldots d_1}.
\end{equation}
Finally we define
\begin{equation}
\delta[\overrightarrow{A},\overleftarrow{B}]_{a_1\ldots a_N
\atop b_M\ldots b_1}
=\delta_{NM}\prod_{j=1}^N
2\pi\delta_{a_jb_j}\delta(\theta_j'-\theta_j).
\end{equation}
We are now in a position to analytically continue matrix elements as
\begin{eqnarray}
\langle 0|Z[\overrightarrow{A}]_{a_1\ldots a_N}\,\mathcal{O}\,
Z^\dagger[\overleftarrow{B}]_{b_M\ldots b_1}|0\rangle&=&
\sum_{{A=A_1\cup A_2}\atop{B=B_1\cup B_2}}
S(\overrightarrow{A}|\overrightarrow{A_1})^{c_1\ldots c_N}_{a_1\ldots a_N}
\,S(\overleftarrow{B_1}|\overleftarrow{B})_{b_M\ldots b_1}^{d_M\ldots d_1}
\,\delta[\overrightarrow{A_2},\overleftarrow{B_2}]
_{c_1\ldots c_k\atop d_l\ldots d_1}\nonumber\\*
&&\qquad\qquad\times
\langle 0|Z[\overrightarrow{A}_1+\ri 0]_{c_{k+1}\ldots c_N}\,\mathcal{O}\,
Z^\dagger[\overleftarrow{B}_1]_{d_M\ldots d_{l+1}}|0\rangle.
\label{eq:reg1}
\end{eqnarray}
Here the sum is over all possible ways to break the sets $A$ and $B$ into subsets and $\overrightarrow{A}_1+\ri 0$ means that all rapidities in $A_1$ are slightly moved into the upper half-plane.  Similarly, we could choose to analytically continue to the lower half-plane
\begin{eqnarray}
\langle 0|Z[\overrightarrow{A}]_{a_1\ldots a_N}\,\mathcal{O}\,
Z^\dagger[\overleftarrow{B}]_{b_M\ldots b_1}|0\rangle&=&
\sum_{{A=A_1\cup A_2}\atop{B=B_1\cup B_2}}d_{A_2}(\mathcal{O})\,
S(\overrightarrow{A}|\overrightarrow{A_2})^{c_1\ldots c_N}_{a_1\ldots a_N}
\,S(\overleftarrow{B_2}|\overleftarrow{B})_{b_M\ldots b_1}^{d_M\ldots d_1}
\,\delta[\overrightarrow{A_2},\overleftarrow{B_2}]
_{c_1\ldots c_k\atop d_l\ldots d_1}\nonumber\\*
&&\qquad\qquad\times
\langle 0|Z[\overrightarrow{A}_1-\ri 0]_{c_{k+1}\ldots c_N}\,\mathcal{O}\,
Z^\dagger[\overleftarrow{B}_1]_{d_M\ldots d_{l+1}}|0\rangle.
\label{eq:reg2}
\end{eqnarray}
The factor $d_{A_2}(\mathcal{O})$ is due to a possible semi-locality of the operator $\mathcal{O}$ with respect to the fundamental fields creating the excitations. For the operator $\mathcal{O}=e^{\ri\alpha\Phi}$ it is given by
\begin{equation}
d_A(e^{\ri\alpha\Phi})=\prod_{i=1}^{n(A)} l_{a_i}^\alpha.
\end{equation}
The remaining matrix elements in \eqref{eq:reg1} and \eqref{eq:reg2} can be evaluated using crossing
\begin{eqnarray}
&&\langle 0|Z[\overrightarrow{A}_1\pm\ri 0]_{c_{k+1}\ldots c_N}\,\mathcal{O}\,
Z^\dagger[\overleftarrow{B}_1]_{d_M\ldots d_{l+1}}|0\rangle\,=\,
_{c_{k+1}\ldots c_{N}}\!\bra{\theta_{i_{k+1}}'\!\pm\!\ri 0,\ldots,
\theta_{i_N}'\!\pm\!\ri 0}\mathcal{O}
\ket{\theta_{j_M},\ldots,\theta_{j_{l+1}}}_{d_{M}\ldots d_{l+1}}\nonumber\\*[2mm]
&&\qquad=d_{A_1}(\mathcal{O})\,C_{c_{k+1}e_{k+1}}\ldots C_{c_Ne_N}\,
f^\mathcal{O}_{e_{k+1}\ldots e_{N}d_{M} \ldots d_{l+1}}
(\theta_{i_{k+1}}'\!+\!\ri\pi\!\pm\!\ri\eta_{i_{k+1}},\ldots,
\theta_{i_N}'\!+\!\ri\pi\!\pm\!\ri\eta_{i_N},
\theta_{j_M},\ldots,\theta_{j_{l+1}}),\qquad\quad
\label{eq:crossing}
\end{eqnarray}
where $C_{ab}=\delta_{a+b,0}$ is the charge conjugation matrix and $\eta_i\rightarrow 0^+$. The analytic continuation of general matrix elements \eqref{eq:generalmatrixelement} with arbitrary orders of the rapidities can be obtained using the scattering axiom. This regularisation procedure was used previously to study finite-temperature correlations\cite{finiteT}, models with boundaries\cite{SEJF11,boundary} and quenches in Ising systems\cite{CEF,SE:2012}. 

\section{Infinite-volume regularisation}\label{sec:kappascheme}
In this appendix we review the infinite-volume regularisation originally introduced in the study of finite-temperature correlation functions\cite{finiteT} and recently generalized to the quench problem in the Ising field theory\cite{SE:2012}. The aim is to exhibit divergences originating in the intertwining of particles with rapidities $\theta_i$ and $-\theta_i$ in the initial state. These divergences are a consequence of working in the infinite volume and have to cancel against each other in observables. These cancellations can be made explicit as follows: For each pair of rapidities $\{-\theta_i,\theta_i\}$ in the, say, ket states we introduce an auxiliary real parameter $\kappa_i$ to shift the rapidities away from the singularities. The resulting expressions have to be understood as generalized functions of the auxiliary variables $\kappa_i$. Next we introduce a smooth function $P(\kappa)$ which is strongly peaked around $\kappa=0$ and satisfies 
\begin{equation}
P(0)=L,\quad \int d\kappa\,P(\kappa)=1.
\end{equation}
Here $L$ can be thought of as the length of the system in a finite-volume regularisation; a possible choice is $P(\kappa)=L\Delta v/[v^2+(\pi L\Delta\kappa)^2]$. 

\section{Two-particle form factors of $\boldsymbol{\mathcal{O}=e^{\ri\beta\Phi/2}}$}\label{sec:app2pFF}
The two-particle form factors of $e^{\ri\beta\Phi/2}$ have been obtained in Refs.~\onlinecite{2pFF,Smirnov92book,Lukyanov97}. For the form-factor axioms stated in Appendix~\ref{sec:FFaxioms} they take the form
\begin{equation}
f^{\beta/2}_{\pm\mp}(\theta_1,\theta_2)=\frac{2\ri\mathcal{G}_{\beta/2}}{\mathcal{C}_1\xi}\frac{\mathcal{G}(\theta_{12})}{\sinh\frac{\theta_{12}-\ri\pi}{\xi}}\,e^{\pm(\theta_{12}-\ri\pi)/(2\xi)},
\end{equation}
where $\theta_{12}=\theta_1-\theta_2$, $\mathcal{C}_1=\mathcal{G}(\ri\pi)>0$ and
\begin{equation}
\mathcal{G}(\theta)=-\ri\,\mathcal{C}_1\,\sinh\frac{\theta}{2}\,\exp\left(\int_0^\infty\frac{dt}{t}\frac{\sinh^2\bigl(t(1+\ri\theta/\pi)\bigr)\,\sinh\bigl(t(\xi-1)\bigr)}{\sinh(2t)\,\cosh t\,\sinh(\xi t)}\right)
\end{equation}
satisfies $\mathcal{G}(\theta+2\pi \ri)=\mathcal{G}(-\theta)$ and $\mathcal{G}(\theta)=S_0(\theta)\mathcal{G}(-\theta)$. An integral representation for the normalisation $\mathcal{G}_{\beta/2}=\bra{0}e^{\ri\beta\Phi/2}\ket{0}$ was obtained in Ref.~\onlinecite{LZ}.

\section{Leading long-time behaviour in $\boldsymbol{\mathcal{O}(K^4)}$}\label{sec:appK4}
The aim of this appendix is the extraction of the leading long-time behaviour in $\mathcal{O}(K^4)$, i.e. the contribution growing as $\propto t^2$ at late times. To simplify matters we restrict ourselves to Dirichlet-type initial states, $K^{aa}(\xi)=0$, and do not calculate any terms $\propto\delta(\kappa_i)$, $\propto 1/\kappa_i$ or $\propto t$. The latter implies that we are not able to determine the $\mathcal{O}(K^4)$ corrections to the decay rate \eqref{eq:decaytime}.

We start with 
\begin{equation}
\begin{split}
C_{44}=&\frac{1}{4}\int_0^\infty\frac{d\theta_1'd\theta_2'd\theta_1d\theta_2}{(2\pi)^4}\,
\bigl(K^{a\bar{a}}(\theta_1')\bigr)^*\,\bigl(K^{b\bar{b}}(\theta_2')\bigr)^*\,
K^{c\bar{c}}(\theta_1)\,K^{d\bar{d}}(\theta_2)\,e^{2\Delta \ri t\sum_i(\cosh\theta_i'-\cosh\theta_i)}\\
&\qquad\times\;_{\bar{a}a\bar{b}b}\bra{\theta_1',-\theta_1',\theta_2',-\theta_2'}e^{i\beta\Phi/2}
\ket{-\theta_2+\kappa_2,\theta_2+\kappa_2,-\theta_1+\kappa_1,\theta_1+\kappa_1}_{d\bar{d}c\bar{c}}.
\end{split}
\end{equation}
Since there are two pairs of rapidities in the ket state we have introduced two auxiliary variables $\kappa_{1,2}$. The resulting expressions have to be considered as generalized functions of $\kappa_1$ and $\kappa_2$ with the final results obtained by multiplication with $P(\kappa_1)P(\kappa_2)$ and integration over both. The regularisation outlined in App.~\ref{sec:Smirnov} yields
\begin{eqnarray}
& &\hspace{-10mm}\;_{\bar{a}a\bar{b}b}\bra{\theta_1',-\theta_1',\theta_2',-\theta_2'}e^{i\beta\Phi/2}
\ket{-\theta_2+\kappa_2,\theta_2+\kappa_2,-\theta_1+\kappa_1,\theta_1+\kappa_1}_{d\bar{d}c\bar{c}}\label{eq:C44reg0}\\
&=&f_{a\bar{a}b\bar{b}d\bar{d}c\bar{c}}^{\beta/2}
(\theta_1'+\ri\pi+\ri 0,-\theta_1'+\ri\pi+\ri 0,\theta_2'+\ri\pi+\ri 0,-\theta_2'+\ri\pi+\ri 0,
-\theta_2+\kappa_2,\theta_2+\kappa_2,-\theta_1+\kappa_1,\theta_1+\kappa_1)\label{eq:C44reg1}\\
&&-2\pi\delta_a^c\delta(\theta_1'-\theta_1-\kappa_1)f_{\bar{a}b\bar{b}d\bar{d}c}^{\beta/2}
(-\theta_1'+\ri\pi+\ri 0,\theta_2'+\ri\pi+\ri 0,-\theta_2'+\ri\pi+\ri 0,-\theta_2+\kappa_2,\theta_2+\kappa_2,
-\theta_1+\kappa_1)\label{eq:C44reg2}\\
&&-2\pi\delta_{\bar{a}}^g\delta(\theta_1'-\theta_2-\kappa_2)S_{\bar{d}c}^{ef}(\theta_1+\theta_2)S_{e\bar{c}}^{gh}(\theta_2-\theta_1)\nn*
&&\qquad\times f_{\bar{a}b\bar{b}dfh}^{\beta/2}
(-\theta_1'+\ri\pi+\ri 0,\theta_2'+\ri\pi+\ri 0,-\theta_2'+\ri\pi+\ri 0,-\theta_2+\kappa_2,-\theta_1+\kappa_1,
-\theta_1+\kappa_1)\label{eq:C44reg3}\\
&&-2\pi\delta_{h}^{\bar{c}}\delta(\theta_2'-\theta_1-\kappa_1)S_{a\bar{b}}^{ef}(-\theta_1'-\theta_2')S_{\bar{a}f}^{gh}(\theta_1'-\theta_2')\nn*
&&\qquad\times f_{\bar{g}\bar{e}\bar{b}d\bar{d}c}^{\beta/2}
(\theta_1'+\ri\pi+\ri 0,-\theta_1'+\ri\pi+\ri 0,-\theta_2'+\ri\pi+\ri 0,-\theta_2+\kappa_2, \theta_2+\kappa_2,
-\theta_1+\kappa_1)\label{eq:C44reg4}\\
&&-2\pi\delta_{h}^m\delta(\theta_2'-\theta_2-\kappa_2)S_{a\bar{b}}^{ef}(-\theta_1'-\theta_2')S_{\bar{a}f}^{gh}(\theta_1'-\theta_2')S_{\bar{d}c}^{ij}(\theta_1+\theta_2)S_{i\bar{c}}^{mn}(\theta_2-\theta_1)\nn*
&&\qquad\times f_{\bar{g}\bar{e}\bar{b}djn}^{\beta/2}
(\theta_1'+\ri\pi+\ri 0,-\theta_1'+\ri\pi+\ri 0,-\theta_2'+\ri\pi+\ri 0,-\theta_2+\kappa_2,-\theta_1+\kappa_1,
\theta_1+\kappa_1)\label{eq:C44reg5}\\
&&-2\pi\delta_e^f\delta(-\theta_1'+\theta_1-\kappa_1)S_{\bar{a}a}^{\bar{e}e}(2\theta_1')S_{c\bar{c}}^{f\bar{f}}(-2\theta_1)\nn*
&&\qquad\times f_{eb\bar{b}d\bar{d}\bar{f}}^{\beta/2}
(\theta_1'+\ri\pi+\ri 0,\theta_2'+\ri\pi+\ri 0,-\theta_2'+\ri\pi+\ri 0,-\theta_2+\kappa_2,\theta_2+\kappa_2,
\theta_1+\kappa_1)\label{eq:C44reg6}\\
&&-2\pi\delta_{i}^e\delta(-\theta_1'+\theta_2-\kappa_2)S_{\bar{a}a}^{\bar{e}e}(2\theta_1')S_{d\bar{d}}^{f\bar{f}}(-2\theta_2)S_{fc}^{gh}(\theta_1-\theta_2)S_{g\bar{c}}^{ij}(-\theta_1-\theta_2)\nn*
&&\qquad\times f_{eb\bar{b}\bar{f}hj}^{\beta/2}
(\theta_1'+\ri\pi+\ri 0,\theta_2'+\ri\pi+\ri 0,-\theta_2'+\ri\pi+\ri 0,\theta_2+\kappa_2,-\theta_1+\kappa_1,
\theta_1+\kappa_1)\label{eq:C44reg7}\\
&&-2\pi\delta_{j}^f\delta(-\theta_2'+\theta_1-\kappa_1)S_{\bar{b}b}^{\bar{e}e}(2\theta_2')S_{ae}^{gh}(\theta_2'-\theta_1')S_{\bar{a}h}^{ij}(\theta_1'+\theta_2')S_{c\bar{c}}^{f\bar{f}}(-2\theta_1)\nn*
&&\qquad\times f_{\bar{i}\bar{g}ed\bar{d}\bar{f}}^{\beta/2}
(\theta_1'+\ri\pi+\ri 0,-\theta_1'+\ri\pi+\ri 0,\theta_2'+\ri\pi+\ri 0,-\theta_2+\kappa_2,\theta_2+\kappa_2,
\theta_1+\kappa_1)\label{eq:C44reg8}\\
&&-2\pi\delta_{j}^p\delta(-\theta_2'+\theta_2-\kappa_2)S_{\bar{b}b}^{\bar{e}e}(2\theta_2')S_{ae}^{gh}(\theta_2'-\theta_1')S_{\bar{a}h}^{ij}(\theta_1'+\theta_2')S_{d\bar{d}}^{f\bar{f}}(-2\theta_2)S_{fc}^{mn}(\theta_1-\theta_2)S_{m\bar{c}}^{pq}(-\theta_1-\theta_2)\nn*
&&\qquad\times f_{\bar{i}\bar{g}e\bar{f}nq}^{\beta/2}
(\theta_1'+\ri\pi+\ri 0,-\theta_1'+\ri\pi+\ri 0,\theta_2'+\ri\pi+\ri 0,\theta_2+\kappa_2,-\theta_1+\kappa_1,
\theta_1+\kappa_1)\label{eq:C44reg9}\\
&&+(2\pi)^2S_{a\bar{b}}^{ef}(-\theta_1'-\theta_2')S_{\bar{d}c}^{gh}(\theta_1+\theta_2)f^{\beta/2}_{\bar{e}\bar{b}dh}(-\theta_1'+\ri\pi+\ri 0,-\theta_2'+\ri\pi+\ri 0,-\theta_2+\kappa_2,-\theta_1+\kappa_1)\nn*
&&\qquad\times\Bigl[\delta_a^c\delta_f^g\delta(\theta_1'-\theta_1-\kappa_1)\delta(\theta_2'-\theta_2-\kappa_2)+\delta_a^{\bar{j}}\delta_c^{\bar{i}}\delta(\theta_1'-\theta_2-\kappa_2)\delta(\theta_2'-\theta_1-\kappa_1)S_{gi}^{jf}(\theta_2-\theta_2')\Bigr]\label{eq:C44reg10}\\
&&+(2\pi)^2S_{\bar{a}a}^{\bar{e}e}(2\theta_1')S_{\bar{b}b}^{\bar{f}f}(2\theta_2')S_{\bar{e}f}^{ij}(\theta_1'+\theta_2')S_{c\bar{c}}^{g\bar{g}}(-2\theta_1)S_{d\bar{d}}^{h\bar{h}}(-2\theta_2)S_{h\bar{g}}^{mn}(-\theta_1-\theta_2)\nn*
&&\qquad\times f^{\beta/2}_{\bar{i}f\bar{h}n}(\theta_1'+\ri\pi+\ri 0,\theta_2'+\ri\pi+\ri 0,\theta_2+\kappa_2,\theta_1+\kappa_1)\nn*
&&\qquad\times\Bigl[\delta_e^g\delta_j^m\delta(-\theta_1'+\theta_1-\kappa_1)\delta(-\theta_2'+\theta_2-\kappa_2)+\delta_e^q\delta_g^p\delta(-\theta_1'+\theta_2-\kappa_2)\delta(-\theta_2'+\theta_1-\kappa_1)S_{mp}^{qj}(\theta_2'-\theta_2)\Bigr]\label{eq:C44reg11}\\
&&+(2\pi)^2\delta_a^c\delta_h^i\delta(\theta_1'-\theta_1-\kappa_1)\delta(-\theta_2'+\theta_2-\kappa_2)S_{\bar{b}b}^{\bar{e}e}(2\theta_2')S_{ae}^{gh}(\theta_2'-\theta_1')S_{d\bar{d}}^{f\bar{f}}(-2\theta_2)S_{fc}^{ij}(\theta_1-\theta_2)\nn*
&&\qquad\times f^{\beta/2}_{\bar{g}e\bar{f}j}(-\theta_1'+\ri\pi+\ri 0,\theta_2'+\ri\pi+\ri 0,\theta_2+\kappa_2,-\theta_1+\kappa_1)\label{eq:C44reg12}\\
&&+(2\pi)^2\delta_{\bar{a}}^m\delta_h^n\delta(\theta_1'-\theta_2-\kappa_2)\delta(-\theta_2'+\theta_1-\kappa_1)S_{\bar{b}b}^{\bar{e}e}(2\theta_2')S_{ae}^{gh}(\theta_2'-\theta_1')S_{c\bar{c}}^{f\bar{f}}(-2\theta_1)\nn*
&&\qquad\times S_{\bar{d}\bar{f}}^{ij}(\theta_2-\theta_1)S_{if}^{mn}(\theta_1+\theta_2)\,f^{\beta/2}_{\bar{g}edj}(-\theta_1'+\ri\pi+\ri 0,\theta_2'+\ri\pi+\ri 0,-\theta_2+\kappa_2,\theta_1+\kappa_1)\label{eq:C44reg13}\\
&&+(2\pi)^2\delta_j^{\bar{c}}\delta_i^m\delta(\theta_2'-\theta_1-\kappa_1)\delta(-\theta_1'+\theta_2-\kappa_2)S_{\bar{a}a}^{\bar{e}e}(2\theta_1')S_{\bar{e}\bar{b}}^{gh}(\theta_1'-\theta_2')S_{eh}^{ij}(-\theta_1'-\theta_2')S_{d\bar{d}}^{f\bar{f}}(-2\theta_2)\nn*
&&\qquad\times S_{fc}^{mn}(\theta_1-\theta_2)\,f^{\beta/2}_{\bar{g}\bar{b}\bar{f}n}(\theta_1'+\ri\pi+\ri 0,-\theta_2'+\ri\pi+\ri 0,\theta_2+\kappa_2,-\theta_1+\kappa_1)\label{eq:C44reg14}\\
&&+(2\pi)^2\delta_e^f\delta_h^i\delta(-\theta_1'+\theta_1-\kappa_1)\delta(\theta_2'-\theta_2-\kappa_2)S_{\bar{a}a}^{\bar{e}e}(2\theta_1')S_{\bar{e}\bar{b}}^{gh}(\theta_1'-\theta_2')S_{c\bar{c}}^{f\bar{f}}(-2\theta_1)S_{\bar{d}\bar{f}}^{ij}(\theta_2-\theta_1)\nn*
&&\qquad\times f^{\beta/2}_{\bar{g}\bar{b}dj}(\theta_1'+\ri\pi+\ri 0,-\theta_2'+\ri\pi+\ri 0,-\theta_2+\kappa_2,\theta_1+\kappa_1)\label{eq:C44reg15}\\
& &+\ldots,
\end{eqnarray}
where the dots represent terms that will not lead to contributions $\propto t^2$. An example for such a term is given in Appendix~\ref{sec:C2227} below. In the following we denote the contribution originating in \eqref{eq:C44reg1} by $C_{44}^4$, the one originating in \eqref{eq:C44reg2}--\eqref{eq:C44reg9} by $C_{44}^3$ and the one originating in \eqref{eq:C44reg10}--\eqref{eq:C44reg15} by $C_{44}^2$. We consider them separately.

\subsection{Leading long-time behaviour of $\boldsymbol{C_{44}^4}$}
We start with the fully connected contribution following from \eqref{eq:C44reg1}. Similar to the evaluation of  $C_{22}^2$ we shift the contours of integration for $\theta_1$ and $\theta_2$ to the lower-half plane. The leading long-time behaviour originates in the pole contributions at $\theta_1=\theta_1'+\kappa_1-\ri 0$ and $\theta_2=\theta_2'+\kappa_2-\ri 0$ or $\theta_1=\theta_2'+\kappa_1-\ri 0$ and $\theta_2=\theta_1'+\kappa_2-\ri 0$. Hence we find
\begin{eqnarray}
C_{44}^4&=&-\frac{1}{4}\int_0^\infty\frac{d\theta_1'd\theta_2'}{(2\pi)^2}\,
\bigl(K^{a\bar{a}}(\theta_1')\bigr)^*\,\bigl(K^{b\bar{b}}(\theta_2')\bigr)^*\,
e^{2\Delta \ri t(\cosh\theta_1'+\cosh\theta_2')}\nonumber\\*
&&\quad\times\left\{K^{c\bar{c}}(\theta_2')\,K^{d\bar{d}}(\theta_1')\,
e^{-2\Delta \ri t(\cosh(\theta_1'+\kappa_1)+\cosh(\theta_2'+\kappa_2))}\right.\nonumber\\*
&&\qquad\qquad\times\text{Res}\Bigl[\text{Res}\Bigl[f_{a\bar{a}b\bar{b}c\bar{c}d\bar{d}}^{\beta/2}
(\theta_1'+\ri\pi+\ri 0,-\theta_1'+\ri\pi+\ri 0,\theta_2'+\ri\pi+\ri 0,-\theta_2'+\ri\pi+\ri 0,\nonumber\\*
&&\qquad\qquad\qquad\qquad-\theta_2+\kappa_2,\theta_2+\kappa_2,-\theta_1+\kappa_1,\theta_1+\kappa_1),\theta_1=\theta_1'+\kappa_1-\ri 0\Bigr],\theta_2=\theta_2'+\kappa_2-\ri 0\Bigr]\\*
&&\quad\quad+K^{c\bar{c}}(\theta_1')\,K^{d\bar{d}}(\theta_2')\,
e^{-2\Delta \ri t(\cosh(\theta_1'+\kappa_2)+\cosh(\theta_2'+\kappa_1))}\nonumber\\*
&&\qquad\qquad\times\text{Res}\Bigl[\text{Res}\Bigl[f_{a\bar{a}b\bar{b}c\bar{c}d\bar{d}}^{\beta/2}
(\theta_1'+\ri\pi+\ri 0,-\theta_1'+\ri\pi+\ri 0,\theta_2'+\ri\pi+\ri 0,-\theta_2'+\ri\pi+\ri 0,\nonumber\\*
&&\left.\qquad\qquad\qquad\qquad-\theta_2+\kappa_2,\theta_2+\kappa_2,-\theta_1+\kappa_1,\theta_1+\kappa_1),\theta_1=\theta_2'+\kappa_1-\ri 0\Bigr],\theta_2=\theta_1'+\kappa_2-\ri 0\Bigr]\right\}\\
&&+\ldots,\nonumber
\end{eqnarray}
where the dots represent terms which grow at most as $t$. On the other hand, the leading time dependence originating from the exponentials has the form $\propto\kappa_1\kappa_2 t^2$. Since we are interested in this $t^2$-behaviour only, we thus have to extract the contributions $\propto 1/(\kappa_1\kappa_2)$ from the double residue. This is done by applying the form-factor axioms stated in Appendix~\ref{sec:FFaxioms}, resulting in
\begin{eqnarray}
C_{44}^4&=&C_{44}^{4,1}+C_{44}^{4,2}+\ldots,\label{eq:C444}\\
C_{44}^{4,1}&=&\frac{\mathcal{G}_{\beta/2}\Delta^2t^2}{\pi^2}
\int_0^\infty d\theta'_1d\theta'_2\,\big|K^{a\bar{a}}(\theta'_1)\big|^2\,
\big|K^{b\bar{b}}(\theta'_2)\big|^2\,\sinh\theta'_1\,\sinh\theta'_2,\\
C_{44}^{4,2}&=&\frac{\mathcal{G}_{\beta/2}\Delta^2t^2}{4\pi^2}\int_0^\infty d\theta'_1d\theta'_2\,
\bigl(K^{a\bar{a}}(\theta'_1)\bigr)^*\,\bigl(K^{b\bar{b}}(\theta'_2)\bigr)^*\,
K^{c\bar{c}}(\theta'_1)\,K^{d\bar{d}}(\theta'_2)\,\sinh\theta'_1\,\sinh\theta'_2\nonumber\\
&&\times\left\{S_{gf}^{\bar{d}c}(\theta'_1+\theta'_2)\,
S_{\bar{a}b}^{\bar{e}k}(-\theta'_1-\theta'_2)\,S_{be}^{\bar{g}c}(\theta'_2-\theta'_1)\,
S_{fd}^{ak}(\theta'_1-\theta'_2)\right.\\
&&+S_{a\bar{a}}^{m\bar{m}}(2\theta'_1)\,S_{ef}^{pq}(-2\theta'_1)\,
S_{g\bar{f}}^{\bar{d}c}(\theta'_1+\theta'_2)\,S_{be}^{\bar{g}c}(\theta'_2-\theta'_1)\,
S_{p\bar{d}}^{mk}(-\theta'_1-\theta'_2)\,S_{qk}^{\bar{m}\bar{b}}(\theta'_1-\theta'_2)\\
&&+S_{b\bar{b}}^{e\bar{e}}(2\theta'_2)\,S_{d\bar{d}}^{f\bar{f}}(-2\theta'_2)\,
S_{\bar{a}\bar{e}}^{gh}(\theta'_2-\theta'_1)\,S_{ah}^{i\bar{f}}(\theta'_1+\theta'_2)\,
S_{ge}^{\bar{c}k}(-\theta'_1-\theta'_2)\,S_{cf}^{ik}(\theta'_1-\theta'_2)\\
&&\left.+S_{b\bar{b}}^{e\bar{e}}(2\theta'_2)\,S_{d\bar{d}}^{f\bar{f}}(-2\theta'_2)\,
S_{ig}^{mn}(2\theta'_1)\,S_{c\bar{c}}^{k\bar{k}}(-2\theta'_1)\,
S_{\bar{a}\bar{e}}^{gh}(\theta'_2-\theta'_1)\,S_{ah}^{i\bar{f}}(\theta'_1+\theta'_2)\,
S_{k\bar{f}}^{\bar{n}p}(-\theta'_1-\theta'_2)\,S_{k\bar{p}}^{me}(\theta'_1-\theta'_2)\right\}\\
&\equiv&C_{44}^{4,2,a}+C_{44}^{4,2,b}+C_{44}^{4,2,c}+C_{44}^{4,2,d},
\end{eqnarray}
where the dots in \eqref{eq:C444} represent terms that grow at most as $t$ at large times
as well as terms $\propto\delta(\kappa_i)$ or $\propto 1/\kappa_i$.

\subsection{Leading long-time behaviour of $\boldsymbol{C_{44}^3}$}
We now turn to the disconnected pieces originating in \eqref{eq:C44reg2}--\eqref{eq:C44reg9}. If we label the resulting terms consecutively from $C_{44}^{3,1}$ to $C_{44}^{3,8}$, we find after straightforward evaluation
\begin{eqnarray}
C_{44}^{3,1}&=&C_{44}^{3,4}=-C_{44}^{3,5}=-C_{44}^{3,8}=\frac{1}{2}C_{44}^{4,1},\\
C_{44}^{3,2}&=&C_{44}^{4,2,a}+C_{44}^{4,2,b},\\
C_{44}^{3,3}&=&C_{44}^{4,2,a}+C_{44}^{4,2,c},\\
C_{44}^{3,6}&=&-C_{44}^{4,2,b}-C_{44}^{4,2,d},\\
C_{44}^{3,7}&=&-C_{44}^{4,2,c}-C_{44}^{4,2,d}.
\end{eqnarray}
Thus in total we find 
\begin{equation}
C_{44}^3=\sum_{i=1}^8C_{44}^{3,i}=2C_{44}^{4,2,a}-2C_{44}^{4,2,d}.
\label{eq:C443}
\end{equation}

\subsection{Leading long-time behaviour of $\boldsymbol{C_{44}^2}$}
Similarly we label the terms originating from \eqref{eq:C44reg10}--\eqref{eq:C44reg15} by
$C_{44}^{2,1}$ to $C_{44}^{2,6}$. Straightforward evaluation of the leading long-time behaviour then yields 
\begin{eqnarray}
C_{44}^{2,1}&=&\frac{1}{4}C_{44}^{4,1}+C_{44}^{4,2,a},\\
C_{44}^{2,2}&=&\frac{1}{4}C_{44}^{4,1}+C_{44}^{4,2,d},\\
C_{44}^{2,3}&=&C_{44}^{2,6}=-\frac{1}{4}C_{44}^{4,1},\\
C_{44}^{2,4}&=&-C_{44}^{4,2,c},\\
C_{44}^{2,5}&=&-C_{44}^{4,2,b}.
\end{eqnarray}
Hence we find in total 
\begin{equation}
C_{44}^2=\sum_{i=1}^6C_{44}^{2,i}=C_{44}^{4,2,a}-C_{44}^{4,2,b}-C_{44}^{4,2,c}+C_{44}^{4,2,d}.
\label{eq:C442}
\end{equation}

\subsection{Example for a contribution not leading to $\boldsymbol{t^2}$-behaviour}\label{sec:C2227}
As an example for a term not leading to a contribution $\propto t^2$ we consider another disconnected piece appearing in the regularisation of \eqref{eq:C44reg0}, which is given by
\begin{equation}
(2\pi)^2\delta_a^c\delta_a^f\delta(\theta_1'-\theta_1-\kappa_1)\delta(-\theta_1'+\theta_2-\kappa_2)S_{d\bar{d}}^{e\bar{e}}(-2\theta_2)S_{ec}^{fg}(\theta_1-\theta_2)f_{b\bar{b}\bar{e}g}^{\beta/2}(\theta_2'+\ri\pi+\ri 0,-\theta_2'+\ri\pi+\ri 0,\theta_2+\kappa_2,-\theta_1+\kappa_1).
\end{equation}
This yields
\begin{eqnarray}
C_{44}^{2,7}&\equiv&\frac{1}{4}\int_0^\infty\frac{d\theta_1'd\theta_2'}{(2\pi)^2}\,
\bigl|K^{a\bar{a}}(\theta_1')\bigr|^2\,\bigl(K^{b\bar{b}}(\theta_2')\bigr)^*\,
K^{d\bar{d}}(\theta_1')\,e^{2\Delta \ri t(\cosh\theta_1'+\cosh\theta_2'-\cosh(\theta_1'-\kappa_1)-\cosh(\theta_1'+\kappa_2))}\\
&&\qquad\times S_{d\bar{d}}^{e\bar{e}}(-2\theta_1')\,S_{ea}^{af}(-\kappa_1-\kappa_2)\,f_{b\bar{b}\bar{e}f}^{\beta/2}(\theta_2'+\ri\pi+\ri 0,-\theta_2'+\ri\pi+\ri 0,\theta_1'+2\kappa_2,-\theta_1'+2\kappa_1).
\end{eqnarray}
Now shifting the $\theta_2'$-contour to the upper half plane we pick up a pole at $\theta_2'=\theta_1'-2\kappa_1+\ri 0$. Evaluating the corresponding residue and extracting the contribution $\propto t^2$ we find
\begin{equation}
C_{44}^{2,7}\propto \frac{\kappa_1\kappa_2}{2\kappa_1+2\kappa_2-\ri 0}\,\Delta^2t^2+\ldots
\end{equation}
However, multiplying this with $P(\kappa_1)\,P(\kappa_2)$ and performing the integrations over $\kappa_{1,2}$ we find
\begin{equation}
\int d\kappa_1\,d\kappa_2\,\frac{\kappa_1\,\kappa_2}{\kappa_1+\kappa_2-\ri 0}\,
P(\kappa_1)\,P(\kappa_2)\propto\frac{1}{L}\to 0.
\end{equation}
Hence $C_{44}^{2,7}$ contains no contribution $\propto t^2$.

\subsection{Leading long-time behaviour of $\boldsymbol{C_{44}}$}
Taking together \eqref{eq:C444}, \eqref{eq:C443}, and \eqref{eq:C442} we find for the leading long-time behaviour in $\mathcal{O}(K^4)$:
\begin{eqnarray}
C_{44}&=&C_{44}^1+4\,C_{44}^{4,2,a}+\ldots\\
&=&\frac{\mathcal{G}_{\beta/2}\Delta^2t^2}{\pi^2}
\int_0^\infty d\theta_1d\theta_2\,\big|K^{a\bar{a}}(\theta_1)\big|^2\,
\big|K^{b\bar{b}}(\theta_2)\big|^2\,\sinh\theta_1\,\sinh\theta_2\\
&&+\frac{\mathcal{G}_{\beta/2}\Delta^2t^2}{\pi^2}\int_0^\infty d\theta_1d\theta_2\,
\bigl(K^{a\bar{a}}(\theta_1)\bigr)^*\,\bigl(K^{b\bar{b}}(\theta_2)\bigr)^*\,
K^{c\bar{c}}(\theta_1)\,K^{d\bar{d}}(\theta_2)\,T_{abcd}(\theta_1,\theta_2)\,
\sinh\theta_1\,\sinh\theta_2,\qquad\label{eq:Tline}\\
T_{abcd}(\theta_1,\theta_2)&=&
S_{a\bar{b}}^{ef}(-\theta_1-\theta_2)\,S_{c\bar{d}}^{gh}(\theta_1+\theta_2)\,
S_{a\bar{f}}^{gd}(\theta_1-\theta_2)\,S_{c\bar{h}}^{eb}(\theta_2-\theta_1),
\end{eqnarray}
where the dots represent terms that grow at most as $t$ at large times as well as terms $\propto\delta(\kappa_i)$ that have to cancel against the normalisation of the initial state. The term \eqref{eq:Tline} can be further simplified. We multiply the reflection equation \eqref{BYBE} by $S_{b_2b_1}^{d_2d_1}(\theta_2-\theta_1)$ and sum over $b_{1,2}$, followed by multiplication with $S_{a_2d_1}^{e_2e_1}(-\theta_1-\theta_2)$ and summing over $a_2$ and $d_1$. This yields
\begin{equation}
K^{a_1e_1}(\theta_1)K^{e_2d_2}(\theta_2)=
K^{c_1b_1}(\theta_1)K^{c_2c_3}(\theta_2)
S^{b_2c_4}_{c_3c_1}(\theta_1+\theta_2)
S^{a_2a_1}_{c_2c_4}(\theta_1-\theta_2)
S_{b_2b_1}^{d_2d_1}(\theta_2-\theta_1)
S_{a_2d_1}^{e_2e_1}(-\theta_1-\theta_2).
\end{equation}
Now considering initial states satisfying $K^{ab}(\theta)=K^{a\bar{a}}(\theta)\delta_a^{\bar{b}}$ we find 
\begin{equation}
K^{a_1\bar{a}_1}(\theta_1)K^{e_2\bar{e}_2}(\theta_2)\delta_{\bar{a}_1}^{e_1}\delta_{\bar{e}_2}^{d_2}=
K^{c_1\bar{c}_1}(\theta_1)K^{c_2\bar{c}_2}(\theta_2)
S^{b_2c_4}_{\bar{c}_2c_1}(\theta_1+\theta_2)
S^{a_2a_1}_{c_2c_4}(\theta_1-\theta_2)
S_{b_2\bar{c}_1}^{d_2d_1}(\theta_2-\theta_1)
S_{a_2d_1}^{e_2e_1}(-\theta_1-\theta_2).
\end{equation}
This equation in particular holds for $e_1=\bar{a}_1$ and $d_2=\bar{e}_2$, thus after relabeling the indices and using the properties \eqref{eq:Srelations} we arrive at
\begin{equation}
K^{a\bar{a}}(\theta_1)K^{b\bar{b}}(\theta_2)=
K^{c\bar{c}}(\theta_1)K^{d\bar{d}}(\theta_2)T_{abcd}(\theta_1,\theta_2).
\end{equation}
Hence we find for the leading long-time behaviour in $\mathcal{O}(K^2)$
\begin{equation}
C_{44}=\frac{2\mathcal{G}_{\beta/2}\Delta^2t^2}{\pi^2}
\int_0^\infty d\theta_1d\theta_2\,\big|K^{a\bar{a}}(\theta_1)\big|^2\,
\big|K^{b\bar{b}}(\theta_2)\big|^2\,\sinh\theta_1\,\sinh\theta_2+\ldots,
\end{equation}
i.e. the result stated in \eqref{eq:D44asymptotics}.



\begin{thebibliography}{99}
\bibitem{note1}
In terms of integrable quantum quenches such an initial state can be realised by replacing the cosine term in \eqref{H_SG} with $-\lambda\int dx\,\cos\bigl(\beta(\Phi-\Phi_0)\bigr)$ and switching $\lambda\big|_{t<0}=\infty\longrightarrow \lambda\big|_{t>0}=\lambda_+$, $\Phi_0\big|_{t<0}=\Phi_0\longrightarrow \Phi_0\big|_{t>0}=0$. For the special case $\Phi_0=0$, i.e. for quenches of the type \eqref{eq:lambdainftyinitial} one further has $K^{a\bar{a}}(\theta)=K^{\bar{a}a}(\theta)$.
\bibitem{noteF}
This can be argued using an argument similar to the one employed in  Appendix~\ref{app:M>N}.
\end{thebibliography}
\end{document}